\documentclass[12pt]{article}
\usepackage{amsmath,epsf,amssymb,latexsym,enumerate,cite,shadow,array,color}
\usepackage[ps,dvips,matrix,arrow,frame,import,curve,color]{xy}
\UseCrayolaColors

\newtheorem{theorem}{Theorem}
\newtheorem{definition}{Definition}

\newtheorem{conjecture}{Conjecture}

\newif\iffigs\figstrue

\DeclareFontFamily{U}{rsf}{}
\DeclareFontShape{U}{rsf}{m}{n}{
  <5> <6> rsfs5 <7> <8> <9> rsfs7 <10-> rsfs10}{}
\DeclareMathAlphabet\Scr{U}{rsf}{m}{n}

\def\pplogo{\vbox{\kern-\headheight\kern -29pt
\halign{##&##\hfil\cr&{%\sc
\ppnumber}\cr\rule{0pt}{2.5ex}&\ppdate\cr}
}}
\makeatletter
\def\ps@firstpage{\ps@empty \def\@oddhead{\hss\pplogo}%
  \let\@evenhead\@oddhead % in case an article starts on a left-hand page
}
\def\maketitle{\par
 \begingroup
 \def\thefootnote{\fnsymbol{footnote}}
 \def\@makefnmark{\hbox{$^{\@thefnmark}$\hss}}
 \if@twocolumn
 \twocolumn[\@maketitle]
 \else \newpage
 \global\@topnum\z@ \@maketitle \fi\thispagestyle{firstpage}\@thanks
 \endgroup
 \setcounter{footnote}{0}
 \let\maketitle\relax
 \let\@maketitle\relax
 \gdef\@thanks{}\gdef\@author{}\gdef\@title{}\let\thanks\relax}
\makeatother

\def\O{\Scr{O}}
\def\C{{\mathbb C}}
\def\P{{\mathbb P}}
\def\Q{{\mathbb Q}}
\def\R{{\mathbb R}}
\def\Z{{\mathbb Z}}

\def\Im{\operatorname{Im}}
\def\Re{\operatorname{Re}}
\def\id{\operatorname{id}}
\def\obj{\operatorname{obj}}
\def\Hom{\operatorname{Hom}}

\def\sHom{\operatorname{\Scr{H}\!\!\textit{om}}}
\def\Ext{\operatorname{Ext}}
\def\End{\operatorname{End}}

\def\Ker{\operatorname{Ker}}

\def\Vol{\operatorname{Vol}}

\def\Tr{\operatorname{Tr}}

\def\Gl{\operatorname{GL}}
\def\PGl{\operatorname{PGL}}
\def\GO{\operatorname{O{}}}
\def\SU{\operatorname{SU}}
\def\GU{\operatorname{U{}}}

\def\diag{\operatorname{diag}}

\def\Cone{\operatorname{Cone}}
\def\Sym{\operatorname{Sym}}

\def\ch{\operatorname{\mathit{ch}}}
\def\td{\operatorname{\mathit{td}}}

\def\sm{$\sigma$-model}
\def\nlsm{non-linear \sm}

\def\CY{Calabi--Yau}
\def\LG{Landau--Ginzburg}

\def\cM{{\Scr M}}

\def\cA{{\Scr A}}

\def\cK{{\Scr K}}
\def\cD{{\Scr D}}
\def\cI{{\Scr I}}
\def\cH{{\Scr H}}

\def\cT{{\Scr T}}

\def\cE{{\Scr E}}
\def\cF{{\Scr F}}
\def\cX{{\Scr X}}
\def\cG{{\Scr G}}
\def\DC{\mathbf{D}}

\def\ff#1#2{{\textstyle\frac{#1}{#2}}}

\def\mf#1{\mathfrak{#1}}

\def\Lotimes{\mathrel{\mathop\otimes^{\mathbf{L}}}}

\def\csum{\mathord{\looparrowright}}
\def\poso#1{#1\save="x"!LD+<0pt,-0.5mm>;
  "x"!RD+<0pt,-0.5mm>**\dir{.}\restore}

\begin{document}
\setcounter{page}0
\def\ppnumber{\vbox{\baselineskip14pt
\hbox{DUKE-CGTP-04-04}
\hbox{hep-th/0403166}}}
\def\ppdate{March 2004} \date{}

\title{\LARGE D-Branes on \CY\ Manifolds\\[10mm]}
\author{
Paul S.~Aspinwall\\[2mm]
\normalsize Center for Geometry and Theoretical Physics \\
\normalsize Box 90318 \\
\normalsize Duke University \\
\normalsize Durham, NC 27708-0318
}

{\hfuzz=10cm\maketitle}

\def\Large{\large}
\def\LARGE{\large\bf}

\vskip 1cm

\begin{abstract}
In this review we study BPS D-branes on \CY\ threefolds.  Such
D-branes naturally divide into two sets called A-branes and B-branes
which are most easily understood from topological field theory. The
main aim of this paper is to provide a self-contained guide to the
derived category approach to B-branes and the idea of
$\Pi$-stability. We argue that this mathematical machinery is hard to
avoid for a proper understanding of B-branes.  A-branes and B-branes
are related in a very complicated and interesting way which ties in
with the ``homological mirror symmetry'' conjecture of Kontsevich.
We motivate and exploit this form of mirror symmetry.  The examples of
the quintic 3-fold, flops and orbifolds are discussed at some
length. In the latter case we describe the r\^ole of McKay quivers in
the context of D-branes.
\end{abstract}

\vfil\break

%\enlargethispage{2\baselineskip}\thispagestyle{empty}
\tableofcontents
\vfil\eject
%%%%%%%%%%%%%%%%%%%%%%%%%%%%%%%%%%%%%%%%%%%%%%%%%%%%%%%%%%%%%%%%

\section{Introduction}    \label{s:intro}

There can be no doubt that the most important development in string
theory in recent years is the discovery of D-branes. In flat spacetime a
D-brane is regarded as a subspace on which open strings may
end.\footnote{The first reference to such objects that the author is
aware of is, oddly enough, section 4 of \cite{GKRG:Gs}.} Since
string theory modifies classical notions of geometry at short
distances, it is natural to assume that such a simple picture of a
D-brane as a subspace is too na\"\i ve for more general backgrounds.
A more abstract notion of a D-brane is required, one
which coincides with the notion of a subspace when viewed in the
context of only large distances.
The aim of these lectures is to study how this can happen.

It is probably of profound importance in string theory to know a robust
definition of D-branes in the most general space-time background, but
this problem is far too difficult with our present understanding of
string theory. Instead we look for the simplest context in which one
might observe nontrivial D-brane behaviour. We render our model as
simple as possible by the following steps:
\begin{enumerate}
\item Get rid of the enormous complications introduced by time by
  using a compactification model. We will assume our string theory has
  a target space $\R^{1,3}\times X$ for some compact space $X$. We
  focus our attention on $X$.
\item Send the string coupling $g_s$ to zero and consider only quantum
  corrections arising from nonzero $\alpha'$ effects.
\item Use just as much supersymmetry as we can while keeping the
  problem nontrivial. This amounts to an $N=(2,2)$ supersymmetric
  theory on the worldsheet with $X$ a \CY\ threefold.
\item Consider only the ``topological sector'' of the worldsheet
  theory. This results in a finite-dimensional Hilbert space of open
  strings and we remove all oscillator modes.
\end{enumerate}
As we will see, after such dramatic simplifications, a very rich model
remains which requires sophisticated mathematical tools to
analyze. One can only wonder at how abstruse a more realistic D-brane,
with the above assumptions removed, must be!

Much has already been written about D-branes. We refer to
\cite{John:Dbok}, for example, for a review of many aspects of
D-branes. In this paper we chart a slightly different course to usual
to achieve our aims. Firstly we try wherever possible to avoid the
D-brane world-volume approach since this assumes that the D-brane
really is a subspace of the target spacetime. Our ideas are then
planted on the worldsheet which forces us to take the string coupling
to zero. Having put ourselves in the worldsheet, we will avoid much of
the boundary-state formalism that one often employs here. Whether this
is a judicious choice is up to the reader to decide, but it does not
seem to be of great importance in the context of our discussions.

By restricting attention to the topological sector of $N=(2,2)$
worldsheet theories we are in the land of mirror symmetry. In order to
keep these lecture notes a manageable length we will have to assume at
least some familiarity with mirror symmetry for closed strings. We
refer to the TASI 1996 lectures, and in particular \cite{Greene:TASI},
for a review. The more mathematical reader is referred to
\cite{CK:mbook}. We will assume a rudimentary knowledge of the geometry of
\CY\ manifolds. The reviews \cite{Cand:lect,Greene:TASI} should suffice.

These lectures are primarily intended to review the ideas of the
derived category and $\Pi$-stability for B-branes. These subjects have
been reviewed by Douglas in \cite{Douglas:Dlect} from a somewhat
different direction than we employ here. Douglas also has a shorter,
more mathematically-oriented review in the ICM proceedings
\cite{Doug:DICM}. In order to motivate and better understand our
constructions, a good deal of our discussion will also involve mirror
symmetry for open strings. This latter topic has been extensively
studied and reviewed in many places. In particular, the reader can
consult \cite{Hori:bk} and references therein for a very detailed
review of most of the aspects of the subject.

In section \ref{s:wsc} we review the basic ideas one needs from
topological field theory in the context of closed strings. This leads
into mirror symmetry which will be a central tool in our
analysis. Section \ref{s:bound} is then a guide to adding boundaries to
the string worldsheet. By the end of this section we will realize that
there are some difficulties in maintaining mirror symmetry without
broadening our concept of D-branes.

Further analysis requires a degree of mathematical sophistication. We
review the algebraic geometry that we require for further progress in
section \ref{s:maths}. We would like to claim that only the necessary
mathematics has been included here, with no complications introduced
for their own sake. The fact remains however that pretty esoteric
notions in cohomology due to Grothendieck do seem to be directly
applicable to D-brane physics, and so we need to delve fairly deeply
into this abstract world.

In section \ref{s:Bcat} it is then a straight-forward process to apply
the machinery of section \ref{s:maths} to the case of B-branes. We
derive the fact that B-branes are described by the derived category of
coherent sheaves.

The notion of $\Pi$-stability, which is essential in relating the
derived category to ``physical'' D-branes, is reviewed in section
\ref{s:stab}. Much of the motivation for this comes from A-branes and
mirror symmetry which we also discuss at length. Finally in section
\ref{s:app} we give a few examples of the derived category and
$\Pi$-stability. 

%%%%%%%%%%%%%%%%%%%%%%%%%%%%%%%%%%%%%%%%%%%%%%%%%%%%%%%%%%%%%%%%%%%

\section{Worldsheet Models of Closed Strings}   \label{s:wsc}

\subsection{The $N=(2,2)$ \nlsm} \label{ss:nlsm}

Let $\Sigma$ be the string worldsheet. We consider a field theory
based on all possible maps $\phi:\Sigma\to X$, where $X$ is the target
manifold. This \nlsm\ has an action
\begin{equation}
  \frac{i}{8\pi i}\int_\Sigma d^2z\,g_{IJ}(\phi)\frac{\partial\phi^I}
  {\partial z}\frac{\partial\phi^J}{\partial\bar z},
    \label{eq:SCFT}
\end{equation}
where $z$ is a complex coordinate on $\Sigma$ and $\phi^I$ are local
coordinates for the map $\phi$. The letters $I$ and $J$ are associated
with {\em real\/} coordinates here.
The object $g_{IJ}$ may be viewed as a
metric on $X$ but it does not need to be symmetric for the \nlsm\ to
be well-defined. The antisymmetric part of $g_{IJ}$ is usually called
the ``B-field''.

This 2-dimensional field theory only defines string theory to an
extent. We know that nonperturbative effects in the string coupling
are invisible from this point of view. Since the entire content of
these lectures is based on this worldsheet definition of string
theory, one must realize that our results are only completely valid in
the zero string coupling limit.

Assuming $X$ is a K\"ahler manifold, we may construct the $N=(2,2)$
supersymmetric version of the \nlsm\ by adding worldsheet
fermions. We now switch to complex coordinates denoted by $\phi^i$
and its complex conjugate $\phi^{\bar\imath}$. The action
is\footnote{There are many conventions for writing $N=(2,2)$
theories. We are following Witten's notation in \cite{W:AB} where
$\pm$ refers to left and right-moving --- {\em not\/} the sign of the
$\GU(1)$ charge!}
\begin{multline}
  \frac{i}{4\pi i}\int_\Sigma d^2z\Biggl\{g_{i\bar\jmath}\left(
  \frac{\partial\phi^i}{\partial z}
  \frac{\partial\phi^{\bar\jmath}}{\partial\bar z}
  +\frac{\partial\phi^i}{\partial\bar z}
  \frac{\partial\phi^{\bar\jmath}}{\partial z}\right)
  +iB_{i\bar\jmath}\left(
  \frac{\partial\phi^i}{\partial z}
  \frac{\partial\phi^{\bar\jmath}}{\partial\bar z}
  -\frac{\partial\phi^i}{\partial\bar z}
  \frac{\partial\phi^{\bar\jmath}}{\partial z}\right)\\
  +ig_{i\bar\jmath}\psi_-^{\bar\jmath}D\psi_-^i
  +ig_{i\bar\jmath}\psi_+^{\bar\jmath}\bar D\psi_+^i
  +R_{i\bar\imath j\bar\jmath}\psi_+^i\psi_+^{\bar\imath}
               \psi_-^j\psi_-^{\bar\jmath}      
  \Biggr\},
\end{multline}
where $g_{i\bar\jmath}$ is the K\"ahler metric and $B_{i\bar\jmath}$
is a real (1,1)-form encoding the B-field degree of freedom. The
fermions are defined as sections of bundles on $\Sigma$ as follows:
\begin{equation}
\begin{split}
\psi_+^i &\in \Gamma(K^{\frac12}\otimes\phi^*T_X)\\
\psi_+^{\bar\jmath} &\in \Gamma(K^{\frac12}\otimes\phi^*\bar T_X)\\
\psi_-^i &\in \Gamma(\bar K^{\frac12}\otimes\phi^*T_X)\\
\psi_-^{\bar\jmath} &\in \Gamma(\bar K^{\frac12}\otimes\phi^*\bar T_X),
\end{split}
\end{equation}
where $K$ is the canonical bundle on $\Sigma$, i.e., the holomorphic
cotangent bundle\footnote{Note that, since $\Sigma$ is K\"ahler,
$K^{-1}=\bar K$.}, $T_X$ is the holomorphic tangent bundle on $X$ and
bar denotes the corresponding antiholomorphic bundle. $D$ represents
the covariant derivative $D\psi_-^i=\partial\psi_-^i+\partial\phi^j
\Gamma^i_{jk}\psi_-^j$, where $\partial$ is the holomorphic part of
the de Rham differential as usual.

Let $B=\ff i2B_{i\bar\jmath}d\phi^id\phi^{\bar\jmath}$ and assume
$dB=0$.\footnote{Wedge products will often be implicit.} In section
\ref{ss:A} it will become clear that the (continuous) $B$-field degree
of freedom lies in $H^2(X,\R)/H^2(X,\Z)$.

The supersymmetries are given by the following transformations:
\begin{equation}
\begin{split}
  \delta\phi^i &= i\alpha_-\psi_+^i + i\alpha_+\psi_-^i\\
  \delta\phi^{\bar\imath} &= i\tilde\alpha_-\psi_+^{\bar\imath} + 
      i\tilde\alpha_+\psi_-^{\bar\imath}\\
  \delta\psi_+^i &= -\tilde\alpha_-\partial\phi^i
      -i\alpha_+\psi_-^j\Gamma^i_{jk}\psi_+^k\\
  \delta\psi_+^{\bar\imath} &= -\alpha_-\partial\phi^{\bar\imath}
      -i\tilde\alpha_+\psi_-^{\bar\jmath}
      \Gamma^{\bar\imath}_{\bar\jmath\bar k}\psi_+^{\bar k}\\
  \delta\psi_-^i &= -\tilde\alpha_+\bar\partial\phi^i
      -i\alpha_-\psi_+^j\Gamma^i_{jk}\psi_-^k\\
  \delta\psi_-^{\bar\imath} &= -\alpha_+\bar\partial\phi^{\bar\imath}
      -i\tilde\alpha_-\psi_+^{\bar\jmath}
      \Gamma^{\bar\imath}_{\bar\jmath\bar k}\psi_-^{\bar k}
\end{split} \label{eq:N2susy}
\end{equation}
with fermionic parameters $\alpha_-$ and $\tilde\alpha_-$ as sections
of $K^{-\frac12}$ and $\alpha_+$ and $\tilde\alpha_+$ as sections
of $\bar K^{-\frac12}$.

If $X$ is a \CY\ manifold then it is well-known (see, for example,
chapters 3 and 17 of \cite{Pol:books}) that there will be a
metric (close to the Ricci-flat metric if $X$ is large) such that this
supersymmetry is extended to an $N=(2,2)$ superconformal symmetry. We
restrict to this case from now on.

Let us quickly review some basic facts about $N=(2,2)$ superconformal
field theories for \CY\ threefolds in order to fix notation. We urge
the reader to consult other sources (such as
\cite{Dixon:,War:N2lect,Greene:TASI} and chapter 19 of
\cite{Pol:books}) for a fuller account of this important subject if
they are not familiar with it.

A closed string state forms a representation of the superconformal
algebra. This is often encoded in the from of an operator product
relationship between the generators of the algebra and the vertex
operators associated to the closed strings. The generators of the
left-moving algebra are then given by 
\begin{equation}
\begin{split}
  T(z) &= -g_{i\bar\jmath} \frac{\partial\phi^i}{\partial z}
       \frac{\partial\phi^{\bar\jmath}}{\partial z}
       +\ff12g_{i\bar\jmath}\psi_+^i\frac{\partial\psi_+^{\bar\jmath}}
       {\partial z}
       +\ff12g_{i\bar\jmath}\psi_+^{\bar\jmath}\frac{\partial\psi_+^i}
       {\partial z}\\
  G(z) &= \ff12g_{i\bar\jmath} \psi_+^i
         \frac{\partial\phi^{\bar\jmath}}{\partial z}\\
  \tilde G(z) &= \ff12g_{i\bar\jmath} \psi_+^{\bar\jmath}
         \frac{\partial\phi^i}{\partial z}\\
  J(z) &= \ff14g_{i\bar\jmath} \psi_+^i\psi_+^{\bar\jmath}
\end{split}
\end{equation}
with similar expressions for the right-moving $\bar T(\bar z)$, $\bar
G(\bar z)$, $\tilde{\bar G}(\bar z)$ and $\bar J(\bar z)$. 

Two elements of the superconformal algebra are of interest to us. The
first concerns the generator of dilatations of the worldsheet
associated to $T(z)$. The eigenvalue of this operation is the {\em
conformal weight\/} $h$ of a given state. The second is the charge $q$
associated with the $\GO(2)=\GU(1)$ R-symmetry of the superconformal
algebra associated to $J(z)$. Since we have both a left-moving and a
right-moving $N=2$ algebra, we have left-moving weight and charge
which we denote $h$ and $q$, and a right-moving weight and charge
which we denote $\bar h$ and $\bar q$.

The R-symmetry part of the superconformal algebra can essentially be
``factored out'' in the following sense. The $\GU(1)$ currents can be
bosonized using bosons $\varphi$ and $\bar\varphi$:
\begin{equation}
  J(z) = i\sqrt{3}\,\frac{\partial\varphi}{\partial z},\quad
  \bar J(\bar z) = i\sqrt{3}\,\frac{\partial\bar\varphi}{\partial\bar z}
\end{equation}
If we have a vertex operator
in the left-moving sector with charge $q$, then we can essentially
write it as\footnote{Normal ordering is assumed.}
\begin{equation}
  f = f_0\exp(\frac{i}{\sqrt{3}}q\varphi),
\end{equation}
where the operator $f_0$ will have charge 0.

Periodic or anti-periodic boundary conditions on the fermions lead to
the Ramond and Neveu-Schwarz sectors respectively as usual. The NS
sector has $q\in\Z$ whilst the R sector has $q\in\Z+\ff12$.

Naturally there are an infinite number of string states in this theory
but there is a very interesting finite subset which is of central
importance. Unitarity forces certain constraints on the allowed
weights and charges. In the NS sector we have a set of states lying on
the boundary of this set of unitary representations which satisfy
\begin{equation}
  h=|q/2|, \quad q = -3,-2,\ldots,3.
\end{equation}
The operators producing these states from the vacuum are called
``chiral primary'' operators for $q>0$ and ``antichiral primary''
operators for $q<0$. We refer to \cite{BFK:unit,LVW:,Dixon:} for more
details. For simplicity of notation we will usually refer to both the chiral
and antichiral operators as chiral.

The key feature of the chiral operators is that they close nicely
under the operator product to form the ``chiral algebra'' (or, less
precisely, the ``chiral ring''). This finite-dimensional subalgebra of
the full infinite-dimensional algebra of closed string vertex
operators seems to encompass a good deal of information about the full
superconformal field theory. It is best analyzed using methods of
topological field theory as we will see in the following sections. One may
also use methods of ``gauged linear sigma models'' as pioneered in
\cite{W:phase}. Indeed, linear sigma models may be used to analyze
open strings and D-branes as in
\cite{Hori:lsm,HKLM:LS,lsm:multistep}. We will not pursue the linear sigma
model in these lectures.

An operator in the Ramond sector of particular interest is the
``spectral flow operator'' with $q=3/2$:
\begin{equation}
  \Sigma(z) = \exp(i\ff{\sqrt{3}}2\varphi).
\end{equation}
This has an operator product expansion with itself as
\begin{equation}
  \Sigma(z)\Sigma(w) = {(z-w)^{\frac34}}{\Upsilon(z)}+\ldots,
     \label{eq:SSO}
\end{equation}
where
\begin{equation}
\Upsilon(z) = \exp(i\sqrt{3}\varphi)
    = \Omega_{ijk}\psi_+^i\psi_+^j\psi_+^k,
  \label{eq:OmSF}
\end{equation}
is the chiral primary operator with $q=3$
and $\Omega=\Omega_{ijk}d\phi^i d\phi^j d\phi^k$ is the (3,0)-form
on $X$ which is unique up to normalization. The spectral flow operator
is responsible for {\em spacetime\/} supersymmetry. Again we refer the
reader to \cite{Dixon:,LVW:} for details. Note that we have two
spectral flow operators, $\Sigma(z)$ and $\bar\Sigma(\bar z)$, which
give us $N=2$ supersymmetry in the uncompactified spacetime directions.

%%%%%%

\subsection{The A model} \label{ss:A}

The chiral algebra is best studied by passing to a topological field
theory associated to the $N=(2,2)$ superconformal field theory
described in section \ref{ss:nlsm}. There are two topological field
theories that naturally occur this way --- the ``A model'' and the ``B
model'' discovered by Witten \cite{W:tsm,W:AB} which we now
review. We will generally denote the target space for the A-model by
$Y$, and use $X$ as the target space of the B-model.\footnote{This
apparently backwards convention is used since it renders the notation
in some of the sections on algebraic geometry more standard.}

We ``twist'' the superconformal field theory by modifying the bundles
in which the fermions take values. We set
\begin{equation}
\begin{split}
  \chi^i = \psi_+^i &\in \Gamma(\phi^*T_Y)\\
  \chi^{\bar\imath} = \psi_-^{\bar\imath} &\in \Gamma(\phi^*\bar T_Y)\\
  \psi_z^{\bar\imath} = \psi_+^{\bar\imath} 
              &\in\Gamma(K\otimes\phi^*\bar T_Y)\\
  \psi_{\bar z}^i = \psi_-^i 
              &\in\Gamma(\bar K\otimes\phi^*T_Y).
\end{split} \label{eq:Atwist}
\end{equation}
Note that the action (\ref{eq:SCFT}) still makes sense (i.e., it is
invariant under rotations of the worldsheet) with this assignment.

The ``supersymmetry'' (\ref{eq:N2susy}) still holds but notice that
the four $\alpha$ parameters are no longer worldsheet spinors. We consider
a restricted version of this symmetry by setting $\alpha=\alpha_-
=\tilde\alpha_+$ and $\alpha_+=\tilde\alpha_-=0$. That is, we have a
symmetry depending on a single scalar parameter $\alpha$. Let us also
denote the operator which generates this symmetry $Q$. To be precise,
\begin{equation}
  \delta W = -i\alpha\{Q,W\},
\end{equation}
for any operator $W$. It follows that (up to equations of motion)
\begin{equation}
  Q^2=0.
\end{equation}
In other words, $Q$ generates a ``BRST symmetry''. Furthermore, we may
write the action in a simplified form (where $\alpha'$ has been chosen
suitably):
\begin{equation}
  S = \int_\Sigma i\{Q,V\} - 2\pi i\int_\Sigma\phi^*(B+iJ),
    \label{eq:Aact}
\end{equation}
where
\begin{equation}
  V = 2\pi g_{i\bar\jmath}(\psi_z^{\bar\jmath}\,\bar\partial\phi^i +
                      \partial\phi^{\bar\jmath}\,\psi_{\bar z}^i),
   \label{eq:AV}
\end{equation}
and $B+iJ\in H^2(Y,\C)$ is the complexified K\"ahler form.

The next step is to restrict attention only to operators $W$, which
are $Q$-closed, i.e., $\{Q,W\}=0$. The effect of the twisting
(\ref{eq:Atwist}) is to mix the notion of weight $h$ and
$\GU(1)$-charge $q$ from the original untwisted superconformal field
theory. It follows that by restricting to $Q$-closed states we are
effectively restricting attention to the case $h=q/2, \bar h=-\bar
q/2$. That is, to a particular chiral algebra.

Now, suppose we have an operator $W$ which is $Q$-exact in the sense that
$W=\{Q,W'\}$ for some $W'$. By standard methods one can show that any
correlation function involving this operator and other $Q$-closed
operators will vanish. In other words, a $Q$-exact operator is
equivalent to zero in the chiral algebra. 
\begin{center}
 \shabox{This means we are restricting
attention to $Q$-cohomology.}
\end{center}
The reason we have put this statement in a pretentious little box is
that it is the most important mathematical statement in these
lectures. The fact that cohomology is essential will lead to a
proliferation of homological algebra in the later lectures.

Note that the triviality of $Q$-exactness extends to the action
too. That is, under the shift $S\mapsto S+\{Q,S'\}$, correlation
functions are invariant. One can show that a change in the worldsheet
metric leads to a $Q$-exact shift in the action (\ref{eq:Aact}). This
means that the location of the vertex operators on the worldsheet are
not important (assuming the locations are distinct of course) when
computing the correlation functions.

The action (\ref{eq:Aact}) manifestly depends on the complexified
K\"ahler form but any change in complex structure merely changes $V$
and is thus trivial. So the correlation functions in this topological
A-model depend only on the complexified K\"ahler form
$B+iJ$. Furthermore it is manifest from the action that it is only the
cohomology class of $B+iJ$ that is of importance, and that a shift in
$B$ by an element of integral cohomology will not affect the
correlation functions.

Operators will be general functions of the fields $\phi$ and
$\psi$. We first consider ``local operators'' in $\Sigma$, i.e.,
scalars. This means we cannot use $\psi_z^{\bar\imath}$ or $\psi_{\bar
z}^i$ as they are 1-forms on $\Sigma$. A basis for the vector space of
local operators is therefore given by operators of the form
\begin{equation}
W_a = a_{I_1I_2\ldots I_p}\chi^{I_1}\chi^{I_2}\ldots\chi^{I_p},
\end{equation}
where $a=a_{I_1I_2\ldots I_p}d\phi^{I_1}d\phi^{I_2}\ldots d\phi^{I_p}$
is a $p$-form on $Y$. The $I_n$'s represent real indices --- in other
words they may be holomorphic or antiholomorphic. One can then compute
\begin{equation}
  \{Q,W_a\} = -W_{da}.
\end{equation}
That is, for the A-model, {\bf $Q$-cohomology is de Rham cohomology}
and the space of operators is given by $H^*(Y,\C)$.

Let us now address the question of how we might compute a correlation
function between such operators:
\begin{equation}
  \langle W_aW_bW_c\ldots \rangle = \int \cD\phi\cD\chi\cD\psi e^{-S}
   W_aW_bW_c\ldots
\end{equation}
The fact that the action (\ref{eq:Aact}) splits naturally into two
pieces makes life particularly easy when analyzing the space of all
maps $\phi:\Sigma\to Y$. Let us assume $\Sigma$ is a sphere. The space
of all maps then breaks up into connected components corresponding to
elements of $\pi_2(Y)$. On a given component the second term in the
action (\ref{eq:Aact}) is constant and can be pulled out of the path
integral.

The term in the action that remains is $Q$-exact and is therefore
trivial. Although one's first temptation might be to replace something
that is trivial by zero, we do the opposite and rescale it by a factor
that tends to infinity! Then the fact that this integrand is positive
semi-definite means that we effectively restrict the path integral to
maps $\phi$ where this part of the action is zero. These are
``worldsheet instantons''.  In other words, the saddle-point
approximation of instantons is exact for topological field
theories. The worldsheet instantons are given by $V=0$ in
(\ref{eq:AV}). These are {\em holomorphic maps\/}
$\bar\partial\phi^i=0$.

The infinite-dimensional space of all maps $\phi:\Sigma\to Y$ is
therefore replaced by the finite-dimensional space of holomorphic maps
when we perform the path integral. Supersymmetry then cancels the
Pfaffians associated with the fermionic path integral and the
remaining determinants from the $\phi$ integrals. We refer to
\cite{W:tqft} for more details on this cancellation process.

We focus on the $p$-forms for $p$ even since the odd forms do not
directly correspond to operators in the untwisted superconformal field
theory. The 0-form clearly represents the identity operator.
The simplest case is therefore to consider correlation functions
between operators associated to 2-forms. One can show that
\begin{equation}
  \langle W_aW_bW_c\rangle=    \int_Y a \wedge b \wedge c
        + \sum_{\alpha\in I}N^{\alpha}_{abc}e^{2\pi i
          \int_{\Sigma}\phi^*(B+iJ)},
\end{equation}
where $I$ is the set of instantons and $N^{\alpha}_{abc}$ are
integers given by the intersection theory on the moduli space of
rational curves (i.e., holomorphic embeddings of $\Sigma$) in
$Y$, including the possibility of multiple covers \cite{CDGP:,AM:rat}.

The knowledge of these correlation functions between 2-forms is
sufficient to define an algebra (i.e., multiplicative) structure on
$H^{\textrm{even}}(Y,\C)$. This is equivalent to the operator algebra.
In the large radius limit, where $J\to\infty$, this coincides with the
cohomology ring given by the wedge product. At finite volume the
deformed ring is called the ``quantum cohomology ring'' of $Y$. We have
impinged on a vast subject here which we do not have space to explore
more fully. We refer to \cite{CK:mbook} and references therein for a
more detailed account of this important subject together with more
recent developments.

The operator algebra is graded by the degree of the forms. Viewing $Q$
as a generator of a BRST symmetry we can also refer to the grading as
a ``ghost number''. That is, if $a$ is a $p$-form then the operator
$W_a$ has ghost number $p$. The ghost number maps naturally back to
the $\GU(1)$ charges in the untwisted theory. In this case $W_a$ would
map to an operator with $(q,\bar q)=(p/2,-p/2)$.  Note also that the
correlation function of a product of operators is only nonzero if the
total ghost number is 6. This means that the grading of the operator
algebra is preserved under multiplication mod 6.

An important aspect of the A-model for our purposes concerns
deformations of the theory. An operator within the theory may be used
to deform the Lagrangian density if it makes sense to integrate such
an operator over $\Sigma$ to deform the action. To find such operators
we need to look beyond the local operators considered so far. Suppose
$W_a$ is a local operator with ghost number $p$. The operator $dW_a$
(where ``$d$'' is the {\em worldsheet\/} de Rham operator) will have
trivial correlation functions with other operators since the location
of the vertex operator insertions is unimportant. It follows that it
must be $Q$-exact, i.e.,
\begin{equation}
  dW_a = \{Q,W^{(1)}_A\},
\end{equation}
for some operator $W^{(1)}_A$ with ghost number $p-1$. We may repeat
this process again by setting
\begin{equation}
  dW^{(1)}_a = \{Q,W^{(2)}_A\},
\end{equation}
for some operator $W^{(2)}_A$ with ghost number $p-2$. But $W^{(2)}_A$
is a 2-form and so we can naturally integrate it over $\Sigma$. We may
therefore consider a deformation of the theory given by
\begin{equation}
  S \mapsto S+t\int_\Sigma W^{(2)}_A\,d^2z,
\end{equation}
for some infinitesimal $t$. In order to preserve the grading of the
operator algebra given by the ghost number, the deformation of the
action should have ghost number zero, i.e., $p=2$.

It is not hard to see that this deformation of the field theory
corresponds to deforming $B+iJ$ by a 2-form proportional to
$tA$. Since the only dependence of the A-model was on $B+iJ$, we see
that we have described all the deformations of the A-model. (Other
deformations that violate ghost number conservation were considered
in \cite{W:AB}.)

It is important to realize that the topological A-model is a different
quantum field theory to the original $N=(2,2)$ superconformal field
theory. Even though the vector space of primary chiral operators is
naturally a subspace of the infinite-dimensional space of operators in
the untwisted theory, the operator products may be quite
different. There is an exception however. If the worldsheet is flat
then the twisting has no affect. If the $N=(2,2)$ were used to
compactify a heterotic string (rather than the type II strings we
consider in these lectures) then the products which determine the
effective superpotential of the resulting $N=1$ theory in four
dimensions are unchanged in the topological field theory. We refer to
\cite{W:AB} for more details.

We emphasize again that the structure of the operator algebra
depends only upon $B+iJ$ and not the complex structure of $Y$. In
fact, as explored in \cite{W:tsm}, we don't need any complex structure
on $Y$, nor do we require the \CY\ condition. $Y$ can be any
symplectic manifold with a compatible almost complex
structure. Instantons then correspond to pseudo-holomorphic
curves. Since the topological A-model knows about only a small subset
of the data of the untwisted theory, it should not come as a surprise
that it can be applied to a wider class of target spaces.

In this section we considered a {\em fixed\/} worldsheet of genus zero
mapping into $Y$. If higher genera are considered, the A-model becomes
fairly trivial because of ghost number conservation constraints. A
variant of the A-model that is commonly considered consists of
coupling the worldsheet theory to gravity. In other words one includes
all metrics on $\Sigma$ in the path integral. Such a theory now
contains nontrivial information about higher genus worldsheets as
discovered in \cite{BCOV:big}. This ``topological gravity'' is also
important in the ``large N'' duality of \cite{GV:grav,OV:lNd}.

If we were going to do a full treatment of mirror symmetry for open
strings we would certainly have to wade into many of the
technicalities of the A-model coupled to gravity. However, in these
lectures, which focus on the issues of stability, we can get away with
largely ignoring this topic.

%%%%%%%%%%%%%

\subsection{The B model} \label{ss:B}

We may relabel the fermions in the superconformal field theory in a
different way to obtain the ``B-model'' which was also introduced by
Witten \cite{W:AB}. 

Let $\psi^{\bar\jmath}_\pm$ be sections of $\phi^*(\bar T_X)$, while
$\psi^j_+$ is a section of $K\otimes \phi^*(T_X)$ and 
$\psi^j_-$ is a section of $\bar K\otimes \phi^*(T_X)$. Define scalars
\begin{equation}
\begin{split}
  \eta^{\bar\jmath} &= \psi_+^{\bar\jmath} + \psi_-^{\bar\jmath} \\
  \theta_j &= g_{j\bar k}(\psi_+^{\bar k} - \psi_-^{\bar k}),
\end{split}  \label{eq:Btw}
\end{equation}
 and define a 1-form $\rho^j$ with $(1,0)$-form part given by $\psi_+^j$
and $(0,1)$-form part given by $\psi_-^j$.

Now consider a variation corresponding to the original
supersymmetric variation with $\alpha_{\pm}=0$ and
$\tilde\alpha_{\pm}=\alpha$.
As in the A-model, this produces a BRST-like variation $Q$ satisfying
$Q^2=0$ (up to equations of motion).

Now, for a suitable choice of $\alpha'$, we may rewrite the action in
the form
\begin{equation}
  S = i\int \{Q,V\} + U,
\end{equation}
where
\begin{equation}
\begin{split}
  V &= g_{j\bar k}\left(\rho_z^j\bar\partial\phi^{\bar k}+
    \rho_{\bar z}^j\partial\phi^{\bar k}\right)\\
  U &= \int_{\Sigma}\left(-\theta_jD\rho^j-\ff i2R_{j\bar\jmath k\bar k}
     \rho^j\wedge\rho^k\eta^{\bar\jmath}\theta_lg^{l\bar k}\right).
\end{split} \label{eq:SB}
\end{equation}

An additional complication arises in the B-model because the fermions
are twisted in a more asymmetric fashion than in the A-model. For a
general target space $X$ one has a chiral anomaly associated with a
problem properly defining the phase of the Pfaffian associated to the
fermionic path integrals. This anomaly is zero if we require
$c_1(T_X)=0$, i.e., if $X$ is a \CY\ manifold. 

It is not immediately obvious from (\ref{eq:SB}) but $U$ depends only
upon the complex structure of $X$. It is independent of both the
metric on $\Sigma$ and the complexified K\"ahler form on $X$, $B+iJ$.
Thus the correlation functions have a similar independence.

Local observables are now written 
\begin{equation}
  W_A = \eta^{\bar k_1}\ldots \eta^{\bar k_q}\,A_{\bar k_1\ldots\bar
  k_q}^{j_1\ldots j_p}\,\theta_{j_1}\ldots\theta_{j_p},
\end{equation}
where
\begin{equation}
  A = d\bar z^{\bar k_1}\ldots d\bar z^{\bar k_q}\,A_{\bar k_1\ldots\bar
  k_q}^{j_1\ldots j_p}\,\frac{\partial}{\partial z_{j_1}}
  \ldots\frac{\partial}{\partial z_{j_p}},
\end{equation}
is a $(0,q)$-form on $X$ valued in $\bigwedge^p T_X$. One might call
$A$ a ``$(-p,q)$-form''. Note that we can use contraction with the
holomorphic 3-form $\Omega$ to give an isomorphism between the spaces
of $(-p,q)$-forms and $(3-p,q)$-forms. This isomorphism is often used
implicitly and explicitly in discussions of mirror symmetry as we will
see in section \ref{ss:mir}.

Now,
\begin{equation}
  \{Q,W_A\} = -W_{\bar\partial A},
\end{equation}
and so, for the B-model, {\bf $Q$-cohomology is Dolbeault cohomology}
on forms valued in exterior powers of the holomorphic tangent bundle.

The instantons in the B-model are trivial. Setting $V=0$ in
(\ref{eq:SB}) requires $\bar\partial\phi^{\bar k}=\partial\phi^{\bar
k}=0$, i.e., $\phi$ is a constant map mapping $\Sigma$ to a point in
$X$. Thus the correlation functions do not consist of some infinite
sum.

The generators of the operator algebra of interest in the B-model are
given by $(-1,1)$-forms. The three-point functions can be shown to be
\begin{equation}
  \langle W_A W_B W_C\rangle =
    \int_X {\Omega}^{jkl}A_j\wedge B_k \wedge C_k
    \wedge{\Omega},
\end{equation}
where $A=A^j\frac{\partial}{\partial\phi^j}\in
H^1_{\bar\partial}(X,T_X)$ etc. The object $\Omega^{jkl}$ can be
obtained from the antiholomorphic 3-form $\bar\Omega$ using the
K\"ahler metric to raise indices.

A $(-p,p)$-form in the B-model has ghost number $2p$ and maps to an
operator with $(q,\bar q)=(p,p)$ in the untwisted model. Just as in
the A-model we may consider deforming the theory by adding operators
to the Lagrangian density. This time such operators correspond to
$(-1,1)$-forms, i.e., elements of $H^1_{\bar\partial}(X,T_X)$. That
this cohomology group corresponds to deformations of complex structure
of $X$ is well-known (see chapter 15 of \cite{GSW:book} for a nice
account of this).

Note that the B-model does require that $X$ has a complex structure
and that it be \CY. However, it does not require any mention of
$B+iJ$. This means that the B-model requires only ``algebraic''
knowledge of $X$ in the following sense. Suppose that $X$ is an
``algebraic variety'' i.e., a subspace of $\P^N$ defined by the
intersection of various (homogeneous) equations $f_1=f_2=\ldots=0$ in
the homogeneous coordinates. Then the B-model is defined completely by
the equations $f_1,f_2,\ldots$

The fact that the B-model has no instanton corrections together with
the above algebraic nature means that one should think of the B-model
as being the ``easy'' topological field theory and the A-model as the
``difficult'' theory. When we discuss open strings in section
\ref{s:Bcat} the reader may decide that the B-model is not so ``easy''
after all but no one can deny that it is a good deal easier than the
A-model!

%%%%%%%%%%%%%

\subsection{Mirror Symmetry} \label{ss:mir}

There are several definitions of mirror symmetry varying in
strength. We require only a fairly weak definition which asserts that
two \CY\ threefolds $X$ and $Y$ are mirror if the operator
algebra of the A-model with target space $Y$ is isomorphic to the
operator algebra of the B-model with target space $X$.

The original definition is stronger and is a statement concerning
conformal field theories. The strongest definition would be that the
type IIA string compactified on $Y$ yields ``isomorphic'' physics in
four dimensions to the type IIB string compactified on $X$.

A simple analysis of the dimensions of the vector spaces of the
operator algebra yields the simple statement that
$h^{p,q}(Y)=h^{3-p,q}(X)$ and thus $\chi(Y)=-\chi(X)$.

The operator algebra for the A-model on $Y$ depends on a choice of
$B+iJ$ on $Y$ and the operator algebra for the B-model on $X$ depends
on a choice of complex structure for $X$. Thus a precise statement of
mirror symmetry must map the moduli space of $B+iJ$ of $Y$ to the
moduli space of complex structures of $X$. This mapping is called the
``mirror map'' and we now discuss it in some detail for a simple key
example.

Let us introduce the most-studied example of a mirror pair of \CY\
threefolds following \cite{GP:orb,CDGP:}. $Y$ is the ``quintic
threefold, i.e., defined as a hypersurface in $\P^4$ given by the
vanishing of an equation of degree 5 in the homogeneous
coordinates. Since $h^{1,1}(Y)=1$, the moduli space of complexified
K\"ahler classes is only one-dimensional. Let $e$ denote the
positive\footnote{That is, $\int_Y e^3 >0$.} generator of
$H^2(Y,\Z)$. Then, by an abuse of the notation, we will refer to the
cohomology class of the complexified K\"ahler form as $(B+iJ)e$, i.e.,
$B$ and $J$ are real numbers in the context of the quintic.  Basically
we can think of the size of $Y$ (i.e., $J$) being determined purely by
the size of the ambient $\P^4$. 

$Y$ has $h^{2,1}=101$ and thus 101
deformations of complex structure but this is of no interest to us
here.

The mirror $X$ of the quintic is constructed by dividing $Y$ by a
$(\Z_5)^3$ orbifold action. We refer to \cite{Greene:TASI} for a
review of why this orbifold yields the mirror. $Y$ has orbifold
singularities which should be resolved yielding many degrees of
freedom for $B+iJ$. However, all we care about is the complex
structure of $X$ which may be defined by specifying the exact quintic
polynomial used. The most general quintic compatible with the
$(\Z_5)^3$ orbifold action is given by
\begin{equation}
   x_0^5+x_1^5+x_2^5+x_3^5+x_4^5-5\psi x_0x_1x_2x_3x_4.
     \label{eq:Qm}
\end{equation}
Thus the complex structure is determined by the single complex parameter
$\psi$. The mirror map we desire will be a mapping between $B+iJ$ on
the A-model side and $\psi$ on the B-model side. This map turns out to
be quite complicated and is actually a many-to-many mapping.

Because the mirror map is not globally well-defined one generally
starts with a basepoint, which is usually the large radius limit on the
A-model side, and finds the mirror map in some neighbourhood of this
basepoint. One can then try to analytically continue the mirror map to
a larger region.

One may analyze the moduli space intrinsically without any reference
to a specific compactification by studying the general features of
scalar fields in $N=2$ theories of supergravity in four
dimensions. The result is that the moduli space is a so-called
``special K\"ahler manifold'' \cite{WLP:spec,Strom:S,Cand:mir}. For a
nice mathematical treatment of this subject see \cite{Frd:SK}.

The special K\"ahler structure of the moduli space leads to the
existence of favoured (but no uniquely defined) coordinates, the
``special coordinates'' which obey certain flatness constraints. On
the A-model side, the components of $B+iJ$ form such special
coordinates. On the B-model side, the natural complex parameters such
as $\psi$ in (\ref{eq:Qm}) do {\em not\/} form special
coordinates. Instead, the special coordinates are formed from {\em
periods\/} as follows. Let $\alpha_m,\beta^m$ for $m=0\ldots h^{2,1}(Y)$
form a symplectic basis of $H_3(X,\Z)$ in the sense that we have the
following intersection numbers
\begin{equation}
  \alpha_m\cap\alpha_n=0,\quad
  \alpha_m\cap\beta^n=\delta_m^n,\quad
  \beta^m\cap\beta^n=0.
\end{equation}
A ``period'' of the holomorphic 3-form
\begin{equation}
  \varpi_m = \int_{\alpha_m}\Omega,
      \label{eq:per}
\end{equation}
is not intrinsically defined as we may rescale $\Omega$ by a
constant. However, {\em ratios\/} of periods, $\varpi_m/\varpi_0$ for
$m=1\ldots h^{2,1}$ are well-defined. These ratios do form special
coordinates and these are naturally mapped to components of $B+iJ$ by
the mirror map.

To find exactly which periods are mapped to which components of $B+iJ$
one looks at the monodromy of these coordinates around the large
radius limit induced by the symmetry $B\mapsto B+1$. A systematic
method for doing this was analyzed in \cite{Mor:gid}. The criteria we
have described so far almost determines the mirror map uniquely. To
nail down the last constants one really needs to explicitly count some
rational curves on $Y$ and map the correlation functions of the
A-model to that of the B-model directly. Having said that, there is a
conjectured form of the mirror map (which was implicitly used in
\cite{CDGP:}) which appears to work in all known cases. We refer to
\cite{CK:mbook} for more details.

Let's see how all this works for the case of the quintic.  We first
need to find the relationship between the periods of $X$ and the
parameter $\psi$. This relationship is encoded in a differential
equation called the ``Picard--Fuchs'' equation. This is a differential
equation whose solutions are the periods (\ref{eq:per}).  There are
various ways of deriving this equation. A fairly tortuous method was
originally pursued in \cite{CDGP:} with a more direct way discussed in
\cite{Mor:PF}. The nicest method was given in \cite{Bat:var} (see also
\cite{AGM:sd}) in terms of toric geometry.

First introduce a coordinate $z=(5\psi)^{-5}$ on the B-model moduli
space. The method of \cite{Bat:var} yields a differential equation
\begin{equation}
  \left(z\frac{d}{dz}\right)^4\Phi - 5^5z\left(z\frac{d}{dz}+\ff15\right)
\left(z\frac{d}{dz}+\ff25\right)\left(z\frac{d}{dz}+\ff35\right)
\left(z\frac{d}{dz}+\ff45\right)\Phi =0.
   \label{eq:PFq}
\end{equation}
Expanding around $z=0$ we obtain a basis of solutions in
the following form:
\begin{equation}
\begin{split}
\Phi_0 &= \sum_{n=0}^\infty\frac{(5n)!}{n!^5}z^n\\
\Phi_k &= \frac1{(2\pi i)^k}\log(z)^k\Phi_0 + \ldots, \quad k = 1,2,3.
\end{split}
\end{equation}
The monodromy of this set of solutions around $z=0$ is precisely the
right form to be associated with the large radius limit $J=\infty$ on
the A-model side \cite{Mor:gid}. The mirror map is then given
by\footnote{Special geometry and monodromy considerations alone do not
rule out a constant term in this power series.}
\begin{equation}
  B+iJ = \frac{\Phi_1}{\Phi_0} = \frac1{2\pi i}\Bigl(\log(z)
  +770z+717825z^2+\ldots\Bigr).
\end{equation}
 
\iffigs
\begin{figure}[t]
\begin{xy}
  \xyimport(100,100){\epsfxsize=15cm\epsfbox{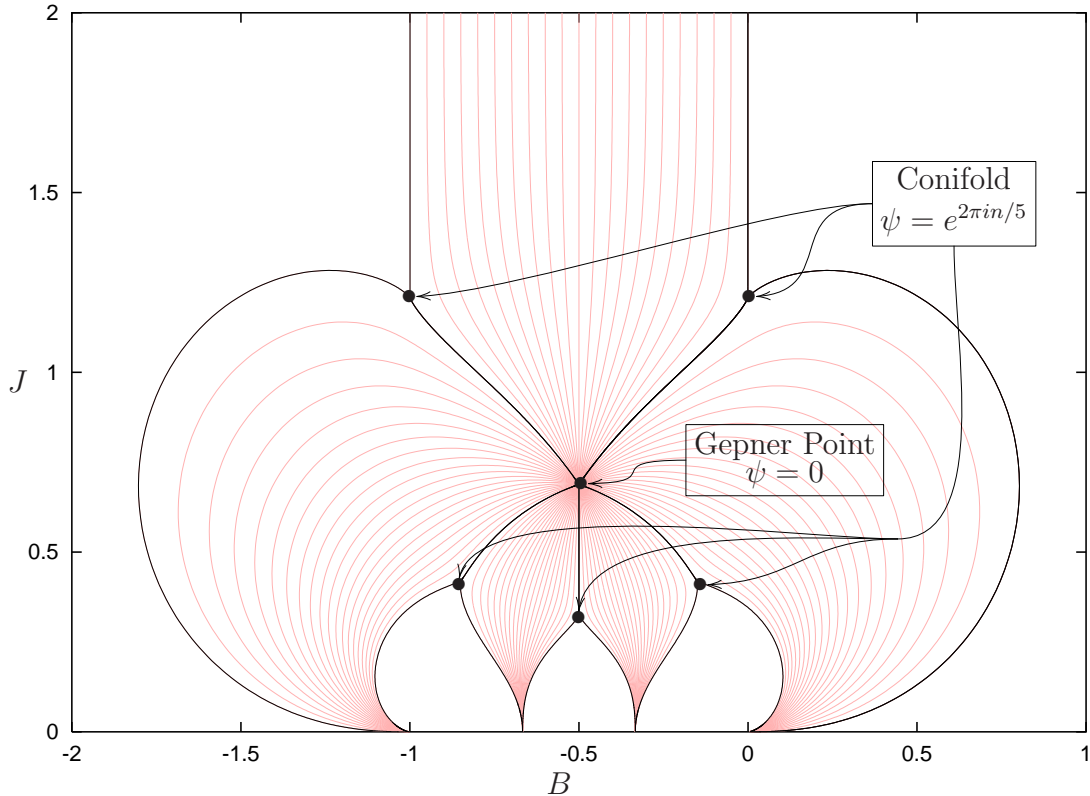}}
  ,(50,-1)*{B}
  ,(2,50)*{J}
  \POS(70,40)*\composite{!<0pt,-1ex>\txt{Gepner Point}
     *!<0pt,1ex>{\psi=0}}*+\frm{-}!CL\ar@(l,r)
  ,(51.9,37)*{\bullet}
  \POS(85,74)*\composite{!<0pt,-1ex>\txt{Conifold}
     *!<0pt,2ex>{\psi=e^{2\pi in/5}}}*+\frm{-}="y"!CL="x"\ar@(l,r)
  ,(66.8,60.7)*{\bullet}
  \POS"x"\ar@(l,r)(36.7,60.7)*{\bullet}
  \POS"y"!CD\ar@{-}@(d,r)(80,30)="y"
  \POS"y"\ar@(l,r)(62.5,24.2)*{\bullet}
  \POS"y"\ar@(l,u)(51.7,20)*{\bullet}
  \POS"y"\ar@(l,u)(41.1,24.2)*{\bullet}
\end{xy}
  \caption{Five fundamental regions for the moduli space of the quintic.}
  \label{f:scorp}
\end{figure}
\fi

There are three points of particular interest in the moduli space
where the Picard-Fuchs equation becomes singular. As we have just
stated, the point $z=0$ corresponds to the large radius limit. The
point $z=\infty$ (or $\psi=0$) corresponds to the ``Gepner model''
\cite{Gep:}. It may also be interpreted as a $\Z_5$-orbifold of the
\LG\ theory \cite{VW:,GVW:,W:phase}. The solutions to the
Picard--Fuchs equation have a branch point of order 5 at this point.
Finally there is a singularity at $z=5^{-5}$ (or $\psi=\exp(2\pi
in/5)$) usually referred to as the ``conifold'' point.

A nice way to visualize the mirror map is to plot fundamental regions
of the moduli space in the $(B+iJ)$-plane. To do this we put branch
cuts along $\psi=R\exp(2\pi in/5)$ for real $R>0$ and $n=0,1$. In
figure \ref{f:scorp} we show the ``scorpion'' diagram from
\cite{CDGP:} which shows five fundamental regions. These consist of the
one containing the large radius limit in the region $-1<B<0$ together
with 4 other fundamental regions obtained by analytically continuing
around the Gepner point.

It is very important to note that the fundamental regions do not
tesselate in general. Monodromy around the large radius limit induces
a shift $B\mapsto B+1$ and it is clear from figure \ref{f:scorp} that
such shifts cause overlaps between fundamental regions.

If we stick to the region containing the large radius limit we see
that the Gepner point represents the ``smallest'' possible quintic
threefold. For further discussion of minimal sizes in this context see
\cite{AGM:sd,me:min-d}.

A more typical example of a mirror pair will require analysis of
moduli spaces of more than one complex dimension. This makes the
problem a good deal more complicated than the quintic but we do not
require any more basic concepts to solve this problem. The
Picard--Fuchs equations are now a set of simultaneous linear partial
differential equations. We refer to \cite{me:min-d}, for example, for
an efficient way of dealing with this situation.

%%%%%%%%%%%%%%%%%%%%%%%%%%%%%%%%%%%%%%%%%%%%%%%%%%%%%%%%%%%%%%%%

\section{Boundaries}    \label{s:bound}

\subsection{The A-model}   \label{ss:bA}

In this section we consider a worldsheet $\Sigma$ with boundaries. A
careful analysis of this gets quite technical quite quickly, taking us
beyond where we need to be for these talks. We refer the reader to
\cite{MPR:bound} (and also \cite{ALZ:bound1,LZ:bound} for the most
thorough treatment. One should also consult \cite{Hori:bk} which is
based on the analysis of \cite{HIV:D-mir}. In the following we will
make rough and ready assumptions which are quite adequate for our
purposes.

\subsubsection{A-branes}  \label{sss:Ab}

As stated earlier one of the main purposes of these
lectures is to demonstrate the existence of D-branes which do not
correspond simply to subspaces. Despite this, we will initially assume
that D-branes {\em are\/} subspaces. Thus we assume that we have a
collection of subspaces $L_a\subset Y$ and that our maps
$\phi:\Sigma\to Y$ obey the condition 
\begin{equation}
  \phi(\partial\Sigma) \subset \bigcup_a L_a,
\end{equation}
i.e., the open strings end on the D-branes $L_a$. We have not yet
constrained the dimensions of the D-branes and one might be free to
consider the case that one of the $L_a$'s fills $Y$ in which case we
have imposed no condition at all. A D-brane that can appear in the
A-model will have to satisfy certain constraints which we now
discuss. Such a D-brane is called an ``A-brane''.

The first step is to apply the variational principle to the
problem. Applying a variation of the fields and then integrating by
parts divides the variation of the action into two parts --- the bulk
and the boundary. Setting the variation of the bulk to zero yields the
Euler--Lagrange equations in the usual way. Demanding that the
variation of the boundary is zero imposes further conditions.

In flat space the vanishing of the variation of the boundary imposes
either Dirichlet or Neumann conditions for the fields $\phi^I$ (see
\cite{PCJ:Dnote} for example). More generally \cite{OOY:Dm,MPR:bound} we set 
\begin{equation}
  \frac{\partial\phi^I}{\partial z} = R^I_J(\phi)
   \frac{\partial\phi^J}{\partial\bar z} + \hbox{fermions},
   \label{eq:Rb}
\end{equation}
where $R$ is a matrix orthogonal with respect to the metric $g_{IJ}$.
Eigenvectors of $R$ with eigenvalue $-1$ give Dirichlet conditions
and are thus associated with directions normal to $L$. To be
completely general, one need not assume that directions tangent to the
D-branes are associated to eigenvectors with eigenvalue $+1$
\cite{OOY:Dm,KO:Azum}, but for our purposes we may make this assumption.

It is impossible to preserve all the $N=(2,2)$ supersymmetry of
section \ref{ss:nlsm} once $\Sigma$ has a boundary. This is because we
must have a reflection condition at the boundary which mixes the
left-moving and right-moving fermions. The best we can do is use the
same reflection matrix as above:
\begin{equation}
  \psi_+^I = R^I_J(\phi)\psi_-^J.
  \label{eq:Rf}
\end{equation}
Now, referring to the A-model twist of (\ref{eq:Atwist}), such a
reflection only really makes sense in the A-model if
$R^i_j=R^{\bar\imath}_{\bar\jmath}=0$ when we use holomorphic
coordinates. That is, only the off-diagonal terms $R^{\bar\imath}_j$ and
$R^i_{\bar\jmath}$ are nonzero.

Now choose a vector $v$ which has eigenvalue $+1$ with
respect to $R$, i.e., a tangent vector in the D-brane. Let us
introduce the almost complex structure $J$, which in holomorphic
coordinates is of the form
\begin{equation}
  J^m_n = i\delta^m_n,\quad J^{\bar m}_{\bar n}=-i\delta^{\bar m}_{\bar n},
\end{equation}
with off-diagonal entries equal to zero. It is then easy to see that
the vector $Jv$ has eigenvector $-1$. Furthermore, $J^2v=-v$, so a
further application of $J$ restores us to the tangent direction.  Thus
$J$ exchanges the directions tangent and normal to the D-brane
$L$. Clearly then $L$ must be of middle dimension, i.e., real
dimension 3.

Note that if $v$ and $w$ are two tangent vectors in $L$ with
eigenvalue $+1$ under $R$, then $w$ is orthogonal to $Jv$ with respect
to the metric $g_{IJ}$. Since, by definition, the K\"ahler form on $Y$
is $\ff12g_{LM}J^M_Nd\phi^Ld\phi^N$, we see that the K\"ahler form
restricted to $L$ is zero.

A middle-dimensional manifold on which the K\"ahler form restricts to
zero is called a {\em Lagrangian submanifold}. Thus we have argued
that the simplest D-branes compatible with the A-model twist appear to
consist of Lagrangian submanifolds. There are further constraints
which we discuss shortly.

A more careful analysis \cite{OOY:Dm,KO:Azum} shows that a \CY\
$n$-fold may have ``coisotropic'' submanifolds of real dimension
$n+2p$ for non-negative integer $p$. Such submanifolds will be of no
interest to us in the case of \CY\ threefolds since $b_5=0$ so long as
the holonomy is not a proper subgroup of $\SU(3)$.

Thus far we have taken care of the analysis of the theory that
pertains to the metric. We should also consider the effect of the
boundary on the $B$-field. 

When $\Sigma$ had no boundary, it was apparent from the A-model action
(\ref{eq:Aact}) that only the cohomology class of $B$ affected any
correlation functions. This is no longer true when $\Sigma$ has a
boundary and so there are more degrees of freedom associated to the
$B$-field than would arise from $H^2(Y,\R)$. Naturally these degrees
of freedom must be associated to the boundary and so should show up in
the D-brane.

This extra freedom may be written in the guise of 1-form $A$ on $Y$
and an addition of a term
\begin{equation}
  S_{\partial\Sigma} = -2\pi i\oint_{\partial\Sigma}\phi^*(A),
   \label{eq:SA}
\end{equation}
to the action. In order to maintain supersymmetry and/or BRST
invariance it is also necessary to add some terms involving fermions
to this boundary contribution.

A shift of the $B$-field by an exact 2-form $d\Lambda$ is then an
invariance of the theory if it is accompanied by a shift $A\mapsto
A-\Lambda$. Thus, with this symmetry understood, the $B$-field is
restored to living in $H^2(Y,\R)$ and we have a new parameter
$A$. Setting $F=dA$, we note that $B+F$ is invariant under the
$\Lambda$-symmetry and thus, unlike $B$ or $F$ alone, can be a
physically meaningful parameter.

It is also important to realize that the theory is no longer invariant under
a lone shift of the $B$-field by an integral 2-form. An invariance is
obtained by accompanying such a shift by a similar shift in $F$. We
will see this effect clearly in section \ref{sss:qmon}.

Typically one thinks of $A$ as the connection on a $\GU(1)$-bundle
associated to the boundary of $\Sigma$ in the form of ``Chan--Paton''
factors. 

Like the $B$-field, the only contribution from $A$ to the correlation
functions will arise from worldsheet instantons. As in section
\ref{ss:A}, worldsheet instantons correspond to holomorphic maps of
$\Sigma$ into $Y$. The action of such an instanton will receive a
contribution from (\ref{eq:SA}) in the form of a line integral of $A$
around the boundary of $\Sigma$ in the D-brane $L$. As this integral
contains only directions tangent to $L$, it is only the projection
of $A$ into the cotangent bundle of $L$ that matters. Thus the
$\GU(1)$-bundle may be considered as living purely on the
D-brane even though we defined $A$ as living in the cotangent bundle
of $Y$.

To derive this notion that $A$ is a connection one should
really follow the ``gerbe'' description of the $B$-field
\cite{Hit:gerb}. We will not attempt to do this here. As we will do
later, one is free to associate larger gauge groups than $\GU(1)$ to
the D-branes. One should always be aware, however, that there is a
natural diagonal $\GU(1)$ in this gauge group which is associated to
the $B$-field by the $\Lambda$ symmetry.

The condition that the BRST symmetry (or supersymmetry of the
untwisted theory) is preserved puts a condition on the connection
$A$. In the case that $B=0$ one can show
\cite{W:CS,OOY:Dm,HIV:D-mir,MPR:bound} that $F=0$, i.e., the
connection must be flat. We will generally restrict to this case.

The statement that an A-brane consists of a Lagrangian submanifold
with a flat bundle is a purely classical statement. Quantum
considerations impose two further constraints. The first arises due to
an anomaly. We would like to preserve the ghost number grading of the
operator product algebra once we include A-branes. It turns out that
an arbitrary Lagrangian submanifold can break this symmetry which, in
physics language, is due to an anomaly.

This anomaly is carefully analyzed in chapter 40 of \cite{Hori:bk} and
is tied to the problem of grading Floer cohomology
\cite{Oh:Fl1}. Since this subject is rather technical we will simply
state the result here. Let us fix a particular choice of a holomorphic
3-form $\Omega$ on $Y$. At any point $p$ on a Lagrangian submanifold
$L$ the volume form of $L$ may be written as a restriction
\begin{equation}
  dV_L = Re^{-i\pi\xi(p)}\Omega|_L,
    \label{eq:ph1}
\end{equation}
where $R$ is a positive real number. $\xi$ gives a map from $L$ to a
circle $\xi:L\to S^1$.  This in turn induces a map on the fundamental
group
\begin{equation}
\xi_*:\pi_1(L)\to\pi_1(S^1)\cong\Z,
   \label{eq:Maz1}
\end{equation}
known as the ``Maslov class'' of $L$.

\iffigs
\begin{figure}
  \centerline{\epsfxsize=9cm\epsfbox{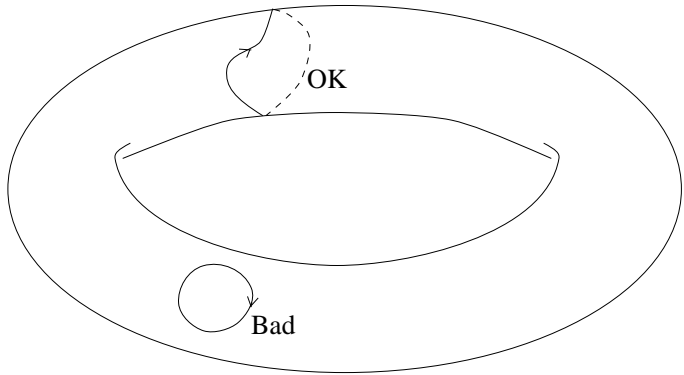}}
  \caption{Loops which do and do not give an anomaly.}
  \label{f:tmaz}
\end{figure}
\fi

The anomaly is absent precisely when the Maslov class of $L$ is
zero. Clearly this is always the case when $\pi_1(L)=0$. For a
nontrivial example we can picture, consider the case where $Y$ is a
one-complex-dimensional torus. Any line of $Y$ is trivially
Lagrangian. As shown in figure~\ref{f:tmaz}, a contractible loop has a
nontrivial Maslov class and so is ruled out as an A-brane.

The second condition on A-branes which arises from quantum effects
concerns destabilizing from open string tadpoles. This requires
knowledge about the open string states to which we now turn.
 
\subsubsection{Open strings for one A-brane}  \label{sss:Aopen1}

The Hilbert space of closed string states in the A-model is unaffected
by the presence of A-branes and is still given by the De Rham
cohomology of $Y$. In addition we need to consider open strings
stretching between two D-branes. Clearly there are two possibilities:
\begin{enumerate}
\item The ends of the open string may lie on the same D-brane.
\item The string stretches between two distinct D-branes.
\end{enumerate}
These two cases are best analyzed differently. The first case was
analyzed by Witten in \cite{W:CS}.

Suppose we have a Lagrangian cycle $L$ with a $\Gl(N,\C)$ vector bundle
$E\to L$. By the usual methods of Chan--Paton factors
\cite{Pol:books}, the open strings will lie in the set $\End(E)$,
i.e., endomorphisms of $E$. In the case of a line bundle we simply
have $\End(E)=\C$. $Q$-invariance allows us to take a scaling limit
\cite{W:CS} effectively taking the string tension to infinity, which
implies that the open string states are only associated with constant
maps $\phi$ and the fermions take values in the tangent bundle of $L$.

Thus, local operators corresponding to the insertion of open string
states on the boundary of $\Sigma$ which maps to $L$ are given by
objects of the form 
\begin{equation}
  a_{I_1I_2\ldots}(\phi)\chi^{I_1}\chi^{I_2}\ldots,
\end{equation}
where $a_{I_1I_2\ldots}(\phi)\in\End(E)$, $\phi\in L$, and
$\chi^{I_k}$ lies in the tangent bundle of $L$. The BRST operator $Q$
acts similarly to section \ref{ss:A} and so the Hilbert space of open
string states is given by the total de Rham cohomology group
\begin{equation}
\bigoplus_{n=0}^3 H^n(L,\End(E)),
\end{equation}
where the ghost number is given by $n$.

The discussion of deformations of the theory induced by operators in
section \ref{ss:A} applies similarly to boundary states. One
difference is that the deforming operator will naturally be integrated
along the one-dimensional $\partial\Sigma$ rather than the
two-dimensional $\Sigma$. Thus we look for ghost number {\em one\/}
boundary operators to give the deformation. These correspond to
elements of $H^1(L,\End(E))$.

It is interesting to explicitly match these deformations coming from
open string vertex operators to the parameters that define the
A-model. First we note that the number of deformations of the
connection on a flat vector bundle are given by $H^1(L,\End(E_{\R}))$
as shown in chapter 15 of \cite{GSW:book} for example. Note that to
count degrees of freedom correctly we restrict to a {\em real form\/} of the
gauge group. Since our open string operators are complex-valued, these
deformations of the connection $A$ account for exactly half of the
degrees of freedom present in $H^1(L,\End(E))$.

So what do the other half of the deformations correspond to? Clearly
deformations of $L$ itself should correspond to deformations of the
A-model. We will now show that such deformations indeed account for the
remaining degrees of freedom.

Let us assume initially the minimal case, i.e., that $E$ is a
line bundle. A deformation of $L$ corresponds to a section of the
normal bundle of $L$. We saw in section \ref{sss:Aopen1} that the
K\"ahler form provides a perfect pairing between vectors normal to $L$
and vectors tangent to $L$. We may thus use the K\"ahler form to
provide a one-to-one mapping between sections of the normal bundle of
$L$ and sections of the cotangent bundle of $L$. Thus a deformation is
given by a 1-form on $L$. Of course, we would like the deformed
submanifold to still be Lagrangian. A simple calculation reveals that
this condition dictates that the 1-form on $L$ be closed.

The result is that Lagrangian deformations of $L$ are in one-to-one
correspondence with closed 1-forms on $L$. In contrast, the degrees of freedom
coming from the A-model consist of {\em cohomology classes\/} of
1-forms on $L$. Thus, if a deformation of $L$ corresponds to a 1-form
which is exact, then it does not affect the A-model. Such a
deformation is called a ``Hamiltonian'' deformation of $L$. Thus we
see that A-branes are really only defined up to Hamiltonian deformation.

If the rank of $E$ is greater than one then some of the deformations
of $L$ corresponding to $H^1(L,\End(E))$ are associated to breaking
$L$ up into a collection of branes with bundles of lower rank. This
all ties together with the picture of enhanced gauge symmetry for
coincident D-branes as discussed in \cite{W:DII}. We should therefore
think of the generic A-brane as comprising of a line bundle $E\to L$
with higher rank bundles obtained by allowing such basic A-branes to
coalesce.

There is one more piece of information about the properties of
A-branes that the open strings in $H^1(L,\End(E))$ can tell us. If an
A-brane background defines a truly stable vacuum for the topological
A-model then the one-point function $\langle W_a\rangle$ will be zero
for any vertex operator associated with $H^1(L,\End(E))$. A nonzero
value, called a ``tadpole'', would force the operator to acquire an
expectation value which would move the D-brane to another location.

We therefore need to know how to compute $\langle W_a\rangle$
exactly. Fortunately Witten \cite{W:CS} discovered a beautiful way of
computing all the correlation functions between open string operators
associated to $H^1(L,\End(E))$.\footnote{Note that Witten's method
applies to the topological field theory {\em coupled to gravity}.}

Without instanton corrections, Witten showed that the correlation
functions could be determined by a Chern--Simons field theory on the
Lagrangian $L$. The effect of instantons is to add an additional term
into the action. At tree level an instanton will consist of a
holomorphic map of a disk into $Y$ with the boundary of the disk
mapped to $L$.

It is this instanton contribution to the effective action that has the
potential to generate tadpoles $\langle W_a\rangle$. Restricting to
the case of a line bundle, the condition that such tadpoles vanish is
that
\begin{equation}
  \sum_{\alpha\in I}\exp\left(2\pi i\int_{D_\alpha}(B+iJ)
+ 2\pi i \oint_{\partial D_\alpha}\!\!\!\!A\right)[\partial D_\alpha]=0
  \quad\hbox{in $H_1(L)$,}
  \label{eq:tadpole}
\end{equation}
where the sum is over all holomorphic disks $D_\alpha$ with $\partial
D_\alpha\subset L$ (including multiple covers).  The notation
$[\partial D_\alpha]$ refers to the cohomology class of $\partial
D_\alpha$. It is an interesting exercise to show that
(\ref{eq:tadpole}) is invariant under Hamiltonian deformations of $L$.

Any Lagrangian violating (\ref{eq:tadpole}) should not be considered
to be an A-brane. This condition on A-branes has been explored in some
cases (in \cite{KKLM:W,KKLM:disk} for example) but a general geometric
understanding appears to be missing. For example, it is not known if
the 3-torus fibrations of SYZ \cite{SYZ:mir} satisfy this condition.
Note that the condition (\ref{eq:tadpole}) depends on $B+iJ$ and the
value of the connection $A$. Thus there can be A-branes which are good
for a specific value of these parameters but will, in general, be
killed by tadpoles. Note also that an $S^3$, which is
simply-connected, always trivially satisfies the vanishing tadpole
condition.

It is worth summarizing the definition of an A-brane that we have finally
settled on:
\begin{center}
\shabox{\parbox{.85\hsize}{An A-brane is an element of the equivalence
    class of Lagrangian 3-manifolds in $Y$ modulo Hamiltonian
    deformations, which satisfies the tadpole cancellation property
    (\ref{eq:tadpole}) and has trivial Maslov class.}}
\end{center}

\vspace{5mm} 
Although we will be doing our best to evade the issue, we
really should mention the $A_\infty$-algebra structure associated to
the A-branes. The correlation functions for the open string vertex
operators which arise from Witten's Chern--Simons theory are not
consistent with an associative algebra in the usual way. Instead one
defines a series of products
\begin{equation}
  m_k(a_1,a_2,\ldots, a_k),
\end{equation}
for which $m_2$ would be the usual product. These higher products are
related in a specific way. We refer to
\cite{Kel:Ainf,LaSh:sh,Fuk:TQFT}, for example, for more details. The
recent paper \cite{HLW:Ainf} explains carefully how the 
$A_\infty$ structure appears directly in the topological field theory.

\subsubsection{Open strings for many A-branes}  \label{sss:Aopen2}

Suppose we have a set of A-branes $L_a$. For simplicity of exposition,
let us initially assume that we just have line bundles over each
brane.  Given a pair of A-branes $L_a$ and $L_b$ we will have a
Hilbert space of open strings beginning on $L_a$ and ending on
$L_b$. This Hilbert space has a grading, which, up to an additive
shift is the ghost number. This additive shift will turn out to be
very important and we will discuss it extensively soon. We use the
following notation for this graded Hilbert space:
\begin{equation}
  \Hom^*(L_a,L_b)=\bigoplus_{m\in\Z} \Hom^m(L_a,L_b).
  \label{eq:Htot}
\end{equation}
We will also denote $\Hom^0(L_a,L_b)$ simply by $\Hom(L_a,L_b)$.

The reason for this notation is that the concept of open strings
between branes fits naturally into the mathematical structure of a
{\em category}. A category is defined as follows (as copied from
\cite{Wei:hom}) 

\begin{definition}
A {\em category\/} $\mathcal{C}$ consists of the following: a
class\footnote{A class is basically the same thing as a set but by
using this language one avoids Russell's paradox.}
$\obj(\mathcal{C})$ of {\em objects}, a set $\Hom_{\mathcal{C}}(A,B)$
of {\em morphisms\/} for every ordered pair $(A,B)$ of objects, an
{\em identity morphism\/} $\id_A\in\Hom_{\mathcal{C}}(A,A)$ for every
object $A$, and a {\em composition function}
\begin{equation}
  \Hom_{\mathcal{C}}(A,B)\times\Hom_{\mathcal{C}}(B,C)
     \to\Hom_{\mathcal{C}}(A,C),
     \label{eq:comp}
\end{equation}
for every ordered triple $(A,B,C)$ of objects. If
$f\in\Hom_{\mathcal{C}}(A,B)$ and $g\in\Hom_{\mathcal{C}}(B,C)$, the
composition is denoted $gf$. The above data is subject to two axioms:
\begin{enumerate}
\item  {\em Associativity axiom:\/} $(hg)f=h(gf)$ for
  $f\in\Hom_{\mathcal{C}}(A,B)$, $g\in\Hom_{\mathcal{C}}(B,C)$ and
  $h\in\Hom_{\mathcal{C}}(C,D)$.
\item {\em Unit axiom:\/} $\id_Bf=f=f\id_A$ for
  $f\in\Hom_{\mathcal{C}}(A,B)$.
\end{enumerate}
\end{definition}

There are many examples of a categories. Some of the obvious ones are
as follows:
\begin{enumerate}
\item Objects are {\em sets\/}, morphisms are {\em maps}.
\item Objects are {\em groups\/}, morphisms are {\em group homomorphisms}.
\item Objects are {\em rings\/} (or modules, etc.), morphisms are {\em
ring homomorphisms} (modules homomorphisms, etc.)
\item Objects are {\em topological spaces}, morphisms are {\em
continuous maps}.
\end{enumerate}
Note that in each case above an object is a set, or some glorified
notion of a set, and so consists of elements. One of the key ideas in
category theory is to phrase things so that you never make any mention
of these elements. There are also categories whose objects are not
composed of elements. The D-brane categories which will be of
particular interest to us are examples of such ``elementless''
categories!

A morphism $f\in\Hom(A,B)$ is often written $f:A\to B$ for obvious
reasons.  As one might guess, we say that two objects, $A$ and $B$, in
a category are {\em isomorphic\/} if there are morphisms $f:A\to B$
and $g:B\to A$ such that $gf=\id_A$ and $fg=\id_B$. That is, there
exists an invertible morphism between $A$ and $B$.

\iffigs
\begin{figure}[ht]
\begin{equation*}
\raisebox{-4cm}{\epsfxsize=5cm\epsfbox{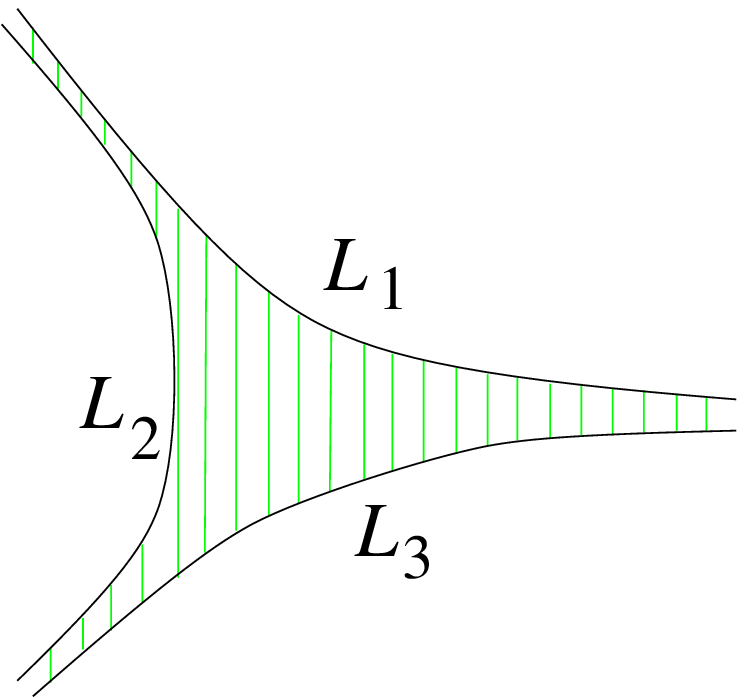}}
\hspace{-2cm}\Hom(L_1,L_2)\otimes\Hom(L_2,L_3)\to\Hom(L_1,L_3)
\end{equation*}
  \caption{Composition of morphisms.}
  \label{f:comp}
\end{figure}
\fi

Clearly we would like to form a category where the A-branes are the
objects and the open strings of the A-model form the morphisms. 
While one is free to define the set of morphisms of the total Hilbert
space as in (\ref{eq:Htot}), we will ultimately see that there is
little difference between this and restricting just to the case of
$\Hom^0(L_1,L_2)$.

The composition function corresponds precisely to the notion of two
open strings joining together as shown in open string diagram of
figure~\ref{f:comp}. The edges of this figure have labels showing to
which D-brane the ends of the open string are attached. This data is
encoded in the correlation functions of the topological field theory.

We see from the previous section that the identity operator $\id_L$
for a given D-brane $L$ is given by the identity operator in
$H^0(L)$.

Just as in the case of a single D-brane, the correlation functions of
the topological A-model coupled to gravity encode a more complicated
product rule than the simple composition (\ref{eq:comp}). Thus the
notion of an $A_\infty$-algebra is generalized to the notion of an
$A_\infty$-category. This structure is very interesting and important
but we do not really need to concern ourselves with it in these
lectures. In particular, if one ignores the higher products, the
category of A-branes that we wish to construct really does satisfy the
axioms of a plain old category specified above.

The information content of the topological A-model with open strings is 
precisely the data associated to the category of A-branes. We already
know exactly what the objects are. We now want to compute the
dimensions of the Hilbert spaces of open strings and the correlation
functions between such states.

To compute the Hilbert space of open strings stretched between two
different D-branes $L_1$ and $L_2$, it is easiest to assume that $L_1$
and $L_2$ intersect transversely. As in section \ref{sss:Aopen1},
the $Q$-invariance of the topological field theory can be used to
argue that open strings can only arise from constant maps
$\phi:\Sigma\to Y$. This means that an open string state is associated
to a point of intersection $L_1$ and $L_2$.

The previous section suggests that locally the Hilbert space should be
given by the De Rham cohomology of this intersection, i.e., the
cohomology of a point. We therefore first guess that there is a
one-dimensional Hilbert space associated with each point of
intersection. Thus the dimension of $\Hom^*(L_1,L_2)$ would be given
by the number of points of intersection between $L_1$ and $L_2$.

This cannot be right. We know the A-model is invariant under
Hamiltonian deformation of $L_1$ or $L_2$ but the number of points of
intersection is not such an invariant. Of course, the {\em oriented\/}
intersection number $\#(L_1\cap L_2)$ is such an invariant as it
depends only on homology classes but this turns out to be too crude
for our purposes.

Let us introduce some notation. Let there be $M$ points of
intersection between $L_1$ and $L_2$ and let the points be labeled
$p_a$, $a=1\ldots M$. Thus we have open string vertex operators
$W_{p_a}$ that create an open string at the point $p_a$.  Our putative
Hilbert space will be denoted $V=\C^M$. Each vertex operator $W_{p_a}$
has a ghost number that we denote $\mu(p_a)$. This leads to a grading
of $V$ by ghost number
\begin{equation}
\begin{split}
  V_i &= \bigoplus_{\mu(p_a)=i}\C\\
  V &= \bigoplus_i V_i.
\end{split}
\end{equation}

The way to determine the true Hilbert space lies in Witten's work on
Morse theory \cite{Witten:mrs} as generalized in the work of Floer
\cite{Flr:sf} and, in particular, by Fukaya \cite{Fuk:mh,Fuk:cat}. We
also refer to chapters 10.5 and 40.4 of \cite{Hori:bk} for a nice
review of this. Because these references are quite thorough, we will
only outline the general picture in the following discussion.

\iffigs
\begin{figure}
  \centerline{\epsfxsize=11cm\epsfbox{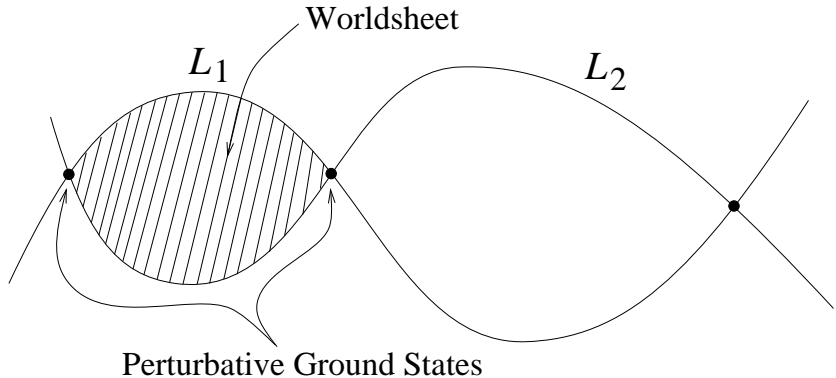}}
  \caption{Instanton Tunneling.}
  \label{f:tun}
\end{figure}
\fi

The basic idea is that an instanton can ``tunnel'' from an open string
state at one point of intersection to an open string at another point
of intersection. The worldsheet of an instanton of such a
tunneling process is shown in figure~\ref{f:tun}. As we saw earlier,
at tree-level these worldsheet instantons are holomorphic disks in $Y$.
These instantons produce a correction to the BRST operator resulting
in:
\begin{equation}
  \{Q,W_{p_a}\} = \sum_b n_{ab} W_{p_b},
\end{equation}
for some coefficients $n_{ab}$ to be determined. Thus the true Hilbert
space will be determined as the $Q$-cohomology of some complex based on
the vector space $V$. Since $Q$ has ghost number one, the complex
looks like
\begin{equation}
\xymatrix@1{
  \ldots\ar[r]^Q&V_{-1}\ar[r]^Q&V_0\ar[r]^Q&V_1\ar[r]^Q&\ldots
}  \label{eq:Flcx}
\end{equation}
We define $\Hom^i(L_a,L_b)$ as the cohomology of this complex at
position $i$.

To compute $n_{ab}$ we must perform an integral of the moduli space of
instantons. This integral must be performed over the fermionic
parameters as well as the obvious bosonic maps $\phi$. By the usual
rules of fermionic integration such an integral vanishes unless the
fermionic parameters cancel in some way, i.e., we have no net
fermionic zero modes. To be more precise, we require that the index of
the Dirac operator for the instanton is equal to the ghost number of
$Q$, i.e., one \cite{Witten:mrs}. 

The index of the Dirac operator also measures the generic (or, to be
precise, virtual) dimension of the moduli space of holomorphic
maps. We refer to \cite{W:matrix} for a nice account of what happens
in the non-generic situation. In the generic case, we thus compute
$n_{ab}$ simply by counting the number of points in the
zero-dimensional instanton moduli space.

For an instanton connecting $p_a$ to $p_b$, the index of the Dirac
operator is given by the difference in ghost numbers
$\mu(p_b)-\mu(p_a)$. Thus we expect that the generic dimension of the
moduli space of instantons is given by
\begin{equation}
  \dim\cM = \mu(p_b)-\mu(p_a) -1.
     \label{eq:FdM}
\end{equation}
We refer the reader to \cite{Flr:mors} for further information on this
point. 

The astute reader should have noticed that we have nowhere specified a
way that one can actually compute $\mu(p_a)$. Given the dimensions of
moduli spaces of instantons, the relation
(\ref{eq:FdM}) only gives enough information to compute the {\em
relative\/} ghost number of two points of intersection of $L_a$ and
$L_b$. Indeed, we have the following very important fact:
\begin{center}
\shabox{\parbox{.85\hsize}{The topological A-model does not contain
enough information to determine the absolute ghost number of an open
string associated to a point of intersection of two A-branes.}}
\end{center}

Just how much ambiguity in the ghost number do we actually have? Given
a pair of D-branes $L_1$ and $L_2$ we are free to shift the ghost
numbers of the open strings from $L_1$ to $L_2$ by some fixed
integer. We also saw in section \ref{sss:Aopen1} that if $L_1=L_2$
then the ghost number was given by the degree of de Rham cohomology
which is perfectly well-defined. Furthermore, we would like to preserve
ghost number in the operator product
\begin{equation}
  \Hom^i(L_1,L_2)\otimes\Hom^j(L_2,L_3) \to \Hom^{i+j}(L_1,L_3).
\end{equation}
The ambiguity in the ghost number can then be accounted for by
assigning a ghost number $\mu(L)$ to each D-brane itself. One then
{\em defines\/} the ghost number of an element of $\Hom^i(L_a,L_b)$ as 
\begin{equation}
  i+\mu(L_b)-\mu(L_a).
  \label{eq:gshift}
\end{equation}
It is easy to see that this definition has all
the properties we desire.

We may restate the above as follows. The topological A-model has a
symmetry which allows us to shift the ghost numbers of the open string
states by assigning arbitrary ghost numbers to the A-branes and
defining the ghost number as in (\ref{eq:gshift}). Note that this idea
of assigning integers to Lagrangian submanifolds to fix this ambiguity
was studied carefully in \cite{Sei:grd}.

\iffigs
\begin{figure}
  \centerline{\epsfxsize=4cm\epsfbox{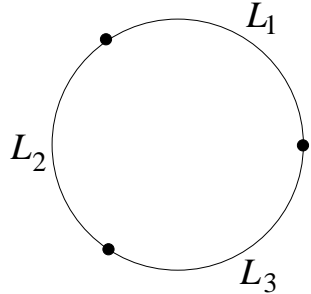}}
  \caption{Disk instanton associated to three-point functions.}
  \label{f:din}
\end{figure}
\fi

We will not give details on how to compute the correlation
functions. It should be clear however that there will be instanton
corrections involved. For example, if we compute the three-point
function associated to figure \ref{f:comp}, at tree-level we will
consider holomorphic disks in $Y$ with boundary conditions shown in
figure \ref{f:din}. The cancellation of fermion zero modes will
enforce ghost number conservation as usual.

In this section we have outlined the definition of the category of
A-branes in the case that the objects $L_a$ and $L_b$ intersect
transversely. Actually one may always use Hamiltonian deformations to
deform {\em any\/} pair of Lagrangian into this case. Thus we actually
have a complete definition of the category of A-branes. 

This category is named after Fukaya who introduced it. The reader
should note that our discussion of the Fukaya category in this section
has omitted a vast number of technical details that have made this
subject the object of a good deal of attention for the past ten
years. We refer to \cite{Fuk:cat,Fuk:TQFT,Fuk:mir1,KS:mir,FOOO:bar},
for example, for more of the gory details. We also refer to
\cite{AP:Cmir,Polsh:Aie} where the Fukaya category (complete with its
$A_\infty$ structure) is determined explicitly for the 2-torus.
Recently, in a remarkable paper \cite{Seidel:quart}, Seidel has
described the Fukaya category for the quartic K3 surface. No other
examples are known.

The generalization of the Fukaya category to the case of higher rank
bundles over each A-brane should be fairly obvious. Rather than
associating $\C$ with each point of intersection, we have a matrix
representing a linear map from the fibre of one bundle to the fibre
of the other over the point of intersection.

We emphasize that nothing in A-model depends on the complex structure
of $Y$. Indeed, the Fukaya category is usually defined purely in terms
of the symplectic geometry of $Y$ thus explicitly removing any
possible dependence on the complex structure. The Fukaya category
depends on $B+iJ$ for both its objects and its composition of
morphisms. The tadpole condition (\ref{eq:tadpole}) has a $B+iJ$
dependence and so certain objects might only exist for particular values
of this parameter. The correlation functions depend on $B+iJ$ through
instanton corrections and so the composition of morphisms are
similarly dependent.

Finally we should point out that worldsheet instantons are generally
expected to adversely affect notions based on the concept of a
spacetime metric. Thus it would be reasonable to expect that the
concept of a Lagrangian submanifold is only really valid at large
radius limit. The composition rules in the Fukaya category are
based on power series associated to instanton effects. Beyond the
radius of convergence of these power series it is reasonable to think
that the Lagrangian submanifold description of A-branes has broken
down. 

%%%%%%%%%%

\subsection{The B-model}   \label{ss:bB}

It should be with relief that we turn attention to the B-model on
$X$. Unfortunately we will see that there is a subtlety concerning the
set of all possible B-branes that will occupy us for most of the
remaining lectures.

\subsubsection{B-branes}  \label{sss:Bb}

We may repeat the analysis of the beginning of section
\ref{sss:Ab}. The difference for the case of B-branes is that the
B-model twist implies that we should impose $R_j^{\bar\imath}
=R_{\bar\jmath}^i=0$ for the reflection matrix in (\ref{eq:Rb}) and
(\ref{eq:Rf}). That is, only the diagonal terms $R_j^i$ and 
$R_{\bar\jmath}^{\bar\imath}$ are nonzero.

This means that the almost complex structure now {\em preserves\/} the
tangent and normal directions to the D-brane, rather than exchanging
them. It follows that the D-brane is a holomorphically embedded
submanifold of $X$. Clearly this forces the dimension of the D-brane
to be even, i.e., 0, 2, 4 or 6.

Although 0, 2 and 4-dimensional B-branes exist, we will at first
restrict attention only to the 6-dimensional case, where the D-brane
fills $X$. That is we put purely Neumann conditions on the open
string. The complexities of B-branes will allow us to deduce the
properties of all the B-branes purely from a knowledge of 
6-branes.\footnote{Note that, in our notation, a $p$-brane is a brane
with $p$-dimensions {\em in the \CY\ directions\/} and any number of
dimensions in the uncompactified part of spacetime.}

As in the A-brane, consideration of the $B$-field forces us to
consider the possibility of a bundle over the B-brane, i.e., a bundle
$E\to X$. Setting the $B$-field equal to zero we may consider the
constraint on this bundle from the requirement that the variation of
the action from the boundary term is zero. In this case, we find that
the curvature, $F$, of the bundle is a 2-form purely of type (1,1)
\cite{W:CS,HIV:D-mir,Hori:bk}. In other words, $E\to X$, is a
{\em holomorphic bundle}. 

We refer again to chapter 15 of \cite{GSW:book} for a very readable
account of holomorphic vector bundles. The basic idea is that the
transition functions for the bundle may be written as holomorphic
functions of the coordinates of $X$. Thus the bundles may be described
very naturally in the language of algebraic geometry. In section
\ref{s:maths} this will allow us to move from the language of bundles
to the language of sheaves which, although alien to most physicists,
is definitely the right language for B-branes.

\subsubsection{Open strings for B-branes}  \label{sss:Bopen}

Since we have chosen purely Neumann boundary conditions on the open
string, we have effectively set the matrix $R$ equal to the identity
in (\ref{eq:Rf}). Thus, from (\ref{eq:Btw}) we have, on the boundary
\begin{equation}
  \theta_j = g_{j\bar k}(\psi_+^{\bar k} - \psi_-^{\bar k})=0,
\end{equation}
and so a local operator will depend only on $\phi$ and
$\eta^{\bar\jmath}$. It follows that local operators look like $(0,q)$-forms.

Suppose we have two B-branes in the form of two bundles $E_1\to X$ and
$E_2\to X$.  The Chan--Paton degrees of freedom are associated with
maps from $E_1$ to $E_2$. We denote the space of such maps as
$\Hom(E_1,E_2)$.

We saw in section \ref{ss:B} that the BRST operator $Q$ looks like the
Dolbeault operator in the B-model. Adding all these ingredients
together, we see that an open string vertex operator for a string
stretching from $E_1$ to $E_2$ is given by the cohomology groups
\begin{equation}
  H^{0,q}_{\bar\partial}(X,\Hom(E_1,E_2)).
   \label{eq:BbH}
\end{equation}

In contrast to the A-brane case, we can choose to declare the ghost
number of an operator in (\ref{eq:BbH}) to be $q$ without ambiguity.

As always, the B-model has no instanton corrections. If
\begin{equation}
\begin{split}
  a &\in H_{\bar\partial}^{0,1}(X,\Hom(E_1,E_2))\\
  b &\in H_{\bar\partial}^{0,1}(X,\Hom(E_2,E_3))\\
  c &\in H_{\bar\partial}^{0,1}(X,\Hom(E_3,E_1)),
\end{split}
\end{equation}
then we may compute the 3-point function exactly from
\begin{equation}
\langle W_aW_bW_c\rangle = \int_X \Tr(a\wedge b\wedge c)\wedge\Omega,
\end{equation}
and deduce an operator product algebra. The $\Hom$ matrices are composed in
the obvious way.

If the B-model is coupled to gravity one may analyze higher $n$-point
functions. In \cite{W:CS} Witten showed that these correlation
functions could be deduced from a ``holomorphic'' Chern--Simons
theory. This $A_\infty$ structure was analyzed more abstractly by
Merkulov \cite{Merk:Dol}. See also \cite{HLW:Ainf} for more discussion
of the $A_\infty$ structure in topological field theories.

\subsubsection{A failure of mirror symmetry}  \label{sss:mf}

Given the dreadful complexities one is forced to endure to define the
Fukaya category (most of which we omitted) the reader is probably
shocked at how easy the B-branes were to analyze.

It would be remarkable if one could now invoke mirror symmetry and say
that the category of A-branes on $Y$ is equivalent to the category of
B-branes on $X$ at this point. Unfortunately this equivalence would be
wrong with our current definition of B-branes. The problem is that we
simply do not have enough B-branes.

Clearly our assumption that B-branes are 6-branes is too strong. The
lower-dimensional branes certainly exist and one might hope that such
branes account for the missing B-branes. Sadly we still fall far short
of the number of objects in the Fukaya category.

There is a lack of symmetry between the A-branes and B-branes which is
key in the failure of mirror symmetry. In section \ref{sss:Aopen2} we
had a real problem when we tried to assign an intrinsic ghost number
to an open string which we solved by labeling the A-branes with a
ghost number. The B-branes did not have this problem. We will
essentially restore mirror symmetry by inflicting the ghost
number ambiguity on the B-branes!

As Kontsevich proposed as far back as 1994 \cite{Kon:mir}, the answer
involves going to the ``derived category'' as we will explain in
section \ref{s:Bcat}.

%%%%%%%%%%%%%%%%%%%%%%%%%%%%%%%%%%%%%%%%%%%%%%%%%%%%%%%%%%%%%%%%%%

\section{Some Mathematical Tools}  \label{s:maths}

Before continuing with the story of B-branes we need some more
mathematical weapons. As these ideas are not familiar to a typical
physicist we will try to be fairly thorough. Most of the ideas in this
section are taken from \cite{Hartshorne:,Wei:hom,KS:shvs}.

As defined in section \ref{ss:bB}, a B-brane is associated to a vector
bundle over $X$. In section \ref{ss:B} we noted that the B-model for a
closed string can be described in purely algebraic terms. In order to
do the same for closed strings we need to replace vector bundles by
something purely algebraic, namely sheaves. This mathematical
construction appears to be unavoidable if one wants to fully
understand B-branes. Anyone who ignores the language of sheaves would
be forced to reinvent it!

In order to discuss sheaves properly we use quite a bit of categorical
language. This will also prove useful later on when we discuss the
derived category.

\subsection{Categories of sheaves}  \label{ss:sheaf}

\subsubsection{Holomorphic functions} \label{sss:olo}

Let us begin with $\P^N$ with homogeneous coordinates
$[z_0,z_1,\ldots,z_N]$. Let $X\subset\P^N$ be an ``algebraic
variety'', i.e., $X$ is defined as the intersection of the zeroes of a
set of polynomials $F_1,F_2,\ldots$ in the homogeneous
coordinates. These polynomials define the space $X$ purely in terms of
algebraic data. Other than $\P^N$ itself, the simplest case consists of
a hypersurface $X\subset\P^N$ defined by a single equation. We never
consider more than one defining equation in these lectures.

How might we put some more ``stuff'' on $X$ that is described purely
in terms of algebraic structures built on the homogeneous coordinates?
The obvious thing to do is to define functions $f:X\to\C$. In terms of
the homogeneous coordinates, the natural way to define such a function
is
\begin{equation}
  f = \frac{g}{h},
   \label{eq:reg}
\end{equation}
where $g$ and $h$ are polynomials in the homogeneous
coordinates. Clearly we require $g$ and $h$ to have the same
homogeneous degree so that the function is well-defined on the
projective space $\P^N$ and thus $X$. Any function $f:X\to\C$ that can
be written in the form (\ref{eq:reg}) in a neighbourhood of a point
$p\in X$, such that $h$ never vanishes in this neighbourhood, is
called {\em regular\/} at $p$. A function is {\em regular on $X$\/} if
it is regular at all the points in $X$.  We will also refer to such
functions on $X$ as {\em holomorphic}.  It is these regular functions
on $X$ which form the prototype of ``extra data'' on $X$ which will be
used to replace the unalgebraic notion of a vector bundle.

\subsubsection{Sheaves} \label{sss:sh}

Let $X$ be a topological space. We make the following\footnote{In
fancy language this makes a presheaf a contravariant functor from the
category of open sets on $X$ to the category of abelian groups.}
\begin{definition}
A {\em presheaf\/} $\cF$ on $X$ consists of the following data
\begin{enumerate}[a)] 
  \item For every open set $U\subset X$ we associate an abelian
  group $\cF(U)$.
  \item If $V\subset U$ are open sets we have a ``restriction''
  homomorphism $\rho_{UV}:\cF(U)\to\cF(V)$.
\end{enumerate}
This data is subject to the conditions
\begin{enumerate} \setcounter{enumi}{-1}
  \item $\cF(\emptyset)=0$.
  \item $\rho_{UU}$ is the identity map.
  \item If $W\subset V\subset U$ then $\rho_{UW}=\rho_{VW}\rho_{UV}$.
\end{enumerate}
\end{definition}
If $\sigma\in\cF(U)$ then we use the notation $\sigma|_V$ for the
restriction $\rho_{UV}(\sigma)$. An element of $\cF(U)$ is called a
{\em section\/} of $\cF$ over $U$. $X$ is an open subset of itself and
an element of $\cF(X)$ is called a {\em global section}.

We then make a more restrictive
\begin{definition}
A {\em sheaf\/} $\cF$ on $X$ is a presheaf satisfying the conditions
\begin{enumerate} \setcounter{enumi}{2}
  \item If $U,V\subset X$ and $\sigma\in\cF(U)$, $\tau\in\cF(V)$ such
  that $\sigma_{U\cap V}=\tau_{U\cap V}$, then there exists $\nu\in
  \cF(U\cup V)$ such that $\nu|_U=\sigma$ and $\nu|_V=\tau$.
  \item If $\sigma\in\cF(U\cup V)$ and $\sigma_U=\sigma_V=0$, then
  $\sigma=0$. 
\end{enumerate}
\end{definition}
This definition makes the data defining a sheaf essentially contained
in very small open sets. That is, the sheaf is defined by ``local''
information.

A simple example of a presheaf is given by associating some fixed
abelian group, such as $\Z$, to every non-empty open set in $X$. The
restriction maps are set equal to the identity. This example is not a
sheaf as it violates condition 3 above when we consider disconnected
open sets. The closest sheaf we can find to this presheaf would be to
associate $\Z^n$ to each open set $U$ where $n$ is the number of
connected components of $U$. This latter sheaf will be useful and we
denote it simply by $\Z$.

Another important sheaf is constructed by making $\cF(U)$ the group
(under addition) of holomorphic functions over $U$. The restriction
map is the obvious restriction map in the usual sense. This ``sheaf of
holomorphic functions'', also known as the {\em structure sheaf}, is
denoted $\O_X$ (or just $\O$).

Yet another example is given by $\O^*$ --- the sheaf of
nonzero\footnote{That is, nowhere zero.} holomorphic functions. This
time the abelian group structure is given by multiplication.

It will be useful to make a category of sheaves. Obviously the sheaves
on $X$ form the objects. We define a morphism $\phi:\cF\to\cG$ of
sheaves as something which associates a homomorphism
$\phi(U):\cF(U) \to\cG(U)$ to each open set $U\subset X$ such that, for
any $V\subset U$, the following diagram commutes:
\begin{equation}
\xymatrix@=15mm{
  \cF(U)\ar[r]^{\phi(U)}\ar[d]^{\rho_{UV}}&\cG(U)\ar[d]^{\rho'_{UV}}\\
  \cF(V)\ar[r]^{\phi(V)}&\cG(V),
}
\end{equation}
where $\rho$ and $\rho'$ are the restriction maps on $\cF$ and $\cG$
respectively. The fact that this forms a category follows
immediately from the properties of homomorphisms of groups.  Note that
the objects, $\cF$, in the category of sheaves are not directly
composed of elements. Having said that, for any open set, $\cF(U)$ is
a group and so is composed of elements. 

We note that two sheaves are said to be {\em isomorphic\/} if there is
an invertible morphism from one to the other.

\subsubsection{Locally free sheaves}  \label{sss:locfree}

Having defined the structure sheaf $\O_X$, we would like to define more
complicated sheaves that are equally algebraic in nature. First recall
the definition of a {\em module\/}:
\begin{definition}
Let $R$ be a ring with a multiplicative identity 1. An $R$-module is
an abelian group $M$ with an $R$-action given by a mapping $R\times
M\to M$ such that
\begin{enumerate}
  \item $r(x+y)=rx+ry$
  \item $(r+s)x = rx+sx$
  \item $(rs)x=r(sx)$
  \item $1x=x$
\end{enumerate}
for any $r,s\in R$ and $x,y\in M$.
\end{definition}

Now, the set of regular functions over $U$, $\O_X(U)$, is an abelian
group under addition as noted earlier, but multiplication of functions
also gives it a ring structure.  This allows us to introduce the
concept of a {\em sheaf of $\O_X$-modules}. That is, let $\cE$ be a
sheaf such that $\cE(U)$ is an $\O_X(U)$-module for any open $U\in
X$. By a common abuse of notation we will refer to a sheaf of
$\O_X$-modules as an $\O_X$-module.

Clearly $\O_X$ is itself an $\O_X$-module. We may also take a sum
of copies 
\begin{equation}
\O_X^{\oplus n}=\underbrace{\O_X\oplus\O_X\oplus\ldots\oplus\O_X}_n
\end{equation}
to give another $\O_X$-module. This is called the {\em free\/}
$\O_X$-module of rank $n$. We call a sheaf $\cE$ {\em locally free\/}
of rank $n$ if there is an open covering $\{U_\alpha\}$ of $X$ such
that $\cE(U_\alpha)\cong\O_X(U_\alpha)^{\oplus n}$ for all $\alpha$.

There is a one-to-one correspondence between holomorphic vector bundles of
rank $n$ on $X$ and locally free sheaves of rank $n$ on $X$. To see
this first consider the trivial complex line bundle over $X$. We may
regard $\O_X(U)$ as the group of {\em holomorphic sections\/} of this
bundle over $U$. Thus, if the covering $\{U_\alpha\}$ trivializes a
vector bundle $E\to X$, then the group of holomorphic sections of $E$
over $U_\alpha$ is given by $\O_X(U_\alpha)^{\oplus n}$.

Conversely, let $\cE$ be a locally free sheaf and let
\begin{equation}
  \phi_\alpha:\cE(U_\alpha)\to\O_X(U_\alpha)^{\oplus n},
\end{equation}
be the explicit isomorphism. On $U_\alpha\cap U_\beta$ we may define
the $n\times n$ matrix of holomorphic functions
\begin{equation}
  \phi_\beta\phi^{-1}_\alpha:\O_X(U_\alpha\cap U_\beta)^{\oplus n}\to
   \O_X(U_\alpha\cap U_\beta)^{\oplus n},
\end{equation}
which defines a holomorphic bundle $E\to X$.

So locally free sheaves are the algebraic way of describing
holomorphic vector bundles. Clearly a trivial bundle corresponds to a
free $\O_X$-module and, as we stated above, the trivial line bundle
corresponds to the structure sheaf $\O_X$.

To get a better feel for locally free sheaves, let us consider some
very simple cases where $X$ is $\P^1$ with homogeneous coordinates
$[z_0,z_1]$. We define an open set $U_0$ by $z_0\neq0$ with an affine
coordinate $y_0=z_1/z_0$. Similarly $U_1$ has $z_1\neq0$ with an
affine coordinate $y_1=z_0/z_1$. Thus $\P^1$ has an open cover
$\{U_0,U_1\}$. 

Consider a holomorphic line bundle on $\P^1$ with a fibre coordinate
$w_i$ over the open set $U_i$. Then the transition function in
$U_0\cap U_1$ may be written in the form
\begin{equation}
  w_1 = y_1^nw_0,
\end{equation}
for some integer $n$.\footnote{This is the first Chern class of the
line bundle.} We define $\O(n)$ as the locally free sheaf
associated to this line bundle. Clearly $\O(0)$ is $\O$.

Now consider an example of a morphism $\O\to\O(n)$. On $U_0$ we use
the identity map 1. To keep the transition function valid, this would
force us to make the morphism look like $g(y_1)\mapsto y_1^n g(y_1)$
on $U_1$ for any $g\in\O(U_1)$. We will write a morphism as
\begin{equation}
\xymatrix@1{
  \O\ar[r]^(0.4){f}&\O(n),
} \label{eq:OfOn}
\end{equation}
where $f$ is a homogeneous function in $[z_0,z_1]$ and we understand
the morphism in $U_\alpha$ to be given by a multiplication by
$f/z_\alpha^n$. Thus, in the case we just described, $f=z_0^n$.
Clearly, so long as $f$ is homogeneous of degree $n$, it will be
compatible with the transition functions of the vector bundle. Indeed,
all morphisms from $\O$ to $\O(n)$ are of this form.

Grothendieck \cite{Grot:P1} proved that {\em any\/} locally free sheaf
of finite rank on $\P^1$ is isomorphic to a sum
$\O(n_1)\oplus\O(n_2)\oplus\ldots$. For $\P^N$ we may define analogous
sheaves $\O(n)$ which will serve as our basic building blocks later
on. Note that Grothendieck's theorem is not valid for $N>1$.

\subsubsection{Kernels and cokernels} \label{sss:kcok}

Given a morphism between two objects in a category we would like to
define the notion of kernel and cokernel of this map.

If we have a map $f:B\to C$ between two groups we would usually define
the kernel of $f$, $\Ker(f)$, to be the subgroup of $B$ which $f$ maps
to the identity of $C$. In the world of categories it is taboo to talk
about elements and so this definition is, in general, no good.

The first thing we need to define is the notion of the identity in
some categorical way. This is done as follows
\begin{definition}
A {\em zero object\/} in a category is an object $0$ such that for any
object $B$ there is precisely one morphism in $\Hom(0,B)$ and
precisely one morphism in $\Hom(B,0)$. If the zero object exists, then
for any pair of objects, $B$ and $C$, we define the {\em zero
morphism\/} (also denoted as $0$) in $\Hom(B,C)$ as the composition
$B\to0\to C$.
\end{definition}
It is easy to show that all zero objects are isomorphic. For the
category of sets the zero object does not exist. For groups, rings,
etc., it is the trivial group, ring, etc. In the category of sheaves
it is the sheaf that associates the trivial group to every set $U$. In
the category of D-branes, it represents the absence of a D-brane!

Secondly, we are going to restrict attention to {\em additive
categories}. This is a category with a zero object and an abelian
group structure (written as addition) on the set of morphisms
$\Hom(B,C)$ such that the distributive law $(f+f')g=fg+f'g$ and
$f(g+g')=fg+fg'$ is true for compositions.\footnote{In addition one
requires the existence of an object $B\oplus C$ for every pair $B$ and
$C$, but this isn't very important in this discussion.} It is easy to
see that the zero morphism is the identity in the group $\Hom(B,C)$.
Clearly the category of abelian groups and the category of sheaves has
an additive structure. In addition, we saw that for D-branes
$\Hom(B,C)$ represents a Hilbert space, thus endowing the category of
D-branes with an additive structure too.

Now we may make the following
\begin{definition}
The {\em kernel\/} of a morphism $f:B\to C$ is a morphism $i:A\to B$ such
that $fi=0$ and which satisfies the following ``universal'' property:
For any morphism $e:A'\to B$ such that $fe=0$, there is a unique morphism
$e':A'\to A$ such that $e=ie'$. That is, the map $e'$ can be
constructed such that the following commutes:\footnote{In these
diagrams the solid lines represent maps which are given, and dotted
lines represent maps to be constructed.}
\begin{equation}
\xymatrix{
  &A'\ar@{-->}[ld]_{e'}\ar[d]^e&\\
  A\ar[r]^i&B\ar[r]^f&C.
}
\end{equation}
\end{definition}

Note that the kernel may not always exist for a general additive category.
We emphasize that, in this categorical language, a kernel is a
morphism --- not an object as one might first think. Of course, since
a morphism must specify the object it is mapping from, we do
intrinsically define a {\em kernel object\/}, $A$, too. If the reader is
unfamiliar with universal properties of maps such as the above, they
should convince themselves that this definition of kernel coincides
with the usual one for, say, the category of groups. To be precise,
the kernel, as defined above, is the {\em inclusion map\/} of the kernel
subgroup into the group itself.

We say $f$ is a {\em monomorphism\/} if the kernel of $f$ is zero.  It
is easy to prove that the uniqueness of $e'$ implies that the kernel
itself is a monomorphism. For groups, etc., a monomorphism is the same
thing as an injective map. If the kernel object exists, it is unique
up to isomorphism.

While we're at it, we can reverse all the arrows in the above to
define the cokernel:
\begin{definition}
The {\em cokernel\/} of a morphism $f:B\to C$ is a morphism $p:C\to D$ such
that $pf=0$ and which satisfies the following ``universal'' property:
For any morphism $g:C\to D'$ such that $gf=0$, there is a unique morphism
$g':D\to D'$ such that $g=g'p$. That is, the map $g'$ can be
constructed such that the following commutes:
\begin{equation}
\xymatrix{
  B\ar[r]^f&C\ar[r]^p\ar[d]^g&D\ar@{-->}[dl]^{g'}\\
  &D'&
}
\end{equation}
\end{definition}
The object $D$ is called the {\em cokernel object}.
Recall that the ``old'' definition of the cokernel of a map $f:B\to C$
is the quotient $C/\Im(f)$. The reader should again check that this
agrees with the categorical definition.

We say $f$ is an {\em epimorphism\/} if the cokernel of $f$ is zero. For
groups, etc., an epimorphism is the same thing as a surjective
map. Again, the cokernel itself is an epimorphism and the cokernel
object is unique, if it exists, up to isomorphism.

So why have we dragged ourselves through all this mathematical
nonsense? The answer is that the category of sheaves does not quite
behave as one might expect when this categorical machinery is
applied to it. One might be forgiven for thinking that given a map
between sheaves, $\phi:\cF\to\cG$, one could apply the old ideas of
kernel and cokernel to the group maps $\phi(U):\cF(U)\to\cG(U)$ to get
a definition of kernel and cokernel of $\phi$. While this works for
the kernel, it can fail for the cokernel. In particular, the cokernel
defined this way, while always a presheaf, need not be a sheaf.

To illustrate what can go wrong, consider the morphism of sheaves
\begin{equation}
  \phi:\O\to\O^*,
\end{equation}
on $\C^*=\C-\{0\}$ by the map $\phi(U)(f)=\exp(2\pi if)$ for any
function $f\in\O(U)$. The coordinate $z$ on the complex plane gives us
a global section $z\in\O^*(\C^*)$. This section is clearly not in the
image of $\phi(\C^*)$. However, if we were to consider a
simply-connected subspace $U\in\C^*$, then $z|_U$ would lie in the
image of $\phi(U)$. We can define a nontrivial presheaf $\cF$ such
that $\cF(U)$ is given by $\O^*(U)/\Im(\phi(U))$, but it violates
property number 4 of the definition of a sheaf in section
\ref{sss:locfree}.

Conversely, the categorical definition of cokernel tells us that
cokernel of $\phi$ in this example is zero. Thus $\phi$ is an epimorphism.

\subsubsection{Abelian categories} \label{sss:abc}

A particular kind of category will be of particular importance to us
--- namely an {\em abelian\/} category. While the D-brane category
itself will turn out not to be abelian, these special categories will
be an essential building block in describing D-branes.

We defined the kernel and cokernel morphisms in section
\ref{sss:kcok}. Now let us further define the {\em image\/} of a
map as the kernel of its cokernel (if it exists) and the {\em
coimage\/} of a map as the cokernel of its kernel. Again, this defines
the image and coimage as morphisms but each has a naturally associated
object too. 

Given any map $f:B\to C$ such that the image and coimage exist,
chasing through the definitions of these various morphisms shows that
we may construct a map $h$ to make the following diagram commute:
\begin{equation}
\xymatrix@=13mm{
  A\ar[r]^{\ker(f)}&B\ar[r]^f\ar[d]_{\operatorname{coim}(f)}&
    C\ar[r]^{\operatorname{coker}(f)}&D\\
  &E\ar@{-->}[r]^h&F\ar[u]_{\operatorname{im}(f)}
} \label{eq:imcoim}
\end{equation}

Now we may define the category of interest:

\begin{definition}
An {\em abelian\/} category is an additive category satisfying the
following axioms:
\begin{enumerate}
\item Every morphism has a kernel and a cokernel (and thus an image
  and coimage).
\item The map $h$ in (\ref{eq:imcoim}) is an isomorphism for any $f$.
\end{enumerate}
\end{definition}

Any category for which the objects are made up of elements, such as
the category of $R$-modules, is abelian.  Since the coimage, etc.,
are defined only up to an isomorphism, in the case of an abelian
category we may assume that $E$ and $F$ are the same objects in
(\ref{eq:imcoim}). That is, every map $f$ factors into its coimage
(which is epic) followed by its image (which is monic).

As usual, an exact sequence may be defined as a sequence of maps
\begin{equation}
\xymatrix@=13mm@1{
  \ldots\ar[r]^{f_{n-2}}&A^{n-1}\ar[r]^{f_{n-1}}&A^{n}\ar[r]^{f_{n}}
  &A^{n+1}\ar[r]^{f_{n+1}}&\ldots,
}
\end{equation}
such that the image of $f_{n-1}$ is the same morphism as the kernel of
$f_n$. 

A short exact sequence is then an exact sequence of the form
\begin{equation}
\xymatrix@1{
   0\ar[r]&A\ar[r]^f&B\ar[r]^g&C\ar[r]&0.
}
\end{equation}
If the reader is finding all this category stuff a bit confusing, they
should check through the above definitions to prove that this short
exact sequence implies that $f$ is the kernel of $g$, and $g$ is the
cokernel of $f$.

Finally in this section we can use the category machinery to define
cohomology abstractly for a complex as follows:
\begin{equation}
\xymatrix@C=2mm{
\ldots\ar[rr]^{f_{n-2}}&&A^{n-1}\ar[rr]^{f_{n-1}}
  &&A^{n}\ar[rr]^{f_{n}}&&A^{n+1}\ar[rr]^{f_{n+1}}&&\ldots\\
&&&L^n\ar[ru]^(0.4){\operatorname{im}(f_{n-1})}\ar@{-->}[rr]^k&&
  K^n\ar[ul]_(0.4){\operatorname{ker}(f_{n})}&&&
}
\end{equation}
If $f_nf_{n-1}=0$, the definition of the kernel of $f_n$ guarantees
the existence of the map $k$ above. The cohomology of this complex at
position $n$ is then defined as the cokernel object of the map $k$. We
generally denote the cohomology $H^n(A^\bullet)$ in the case that it
is an abelian group, or $\cH^n(A^\bullet)$ if it is a sheaf.

There is a remarkable theorem due to Freyd \cite{Frd:embed} that says
that, if the objects in an abelian category can be described as a set,
the category may be embedded in the category of $R$-modules for some
ring $R$. This effectively means that the objects in such a category
can be thought of as consisting of elements.

We will see eventually that the categories we build for A-branes and
B-branes are not abelian --- hence the necessity for the abstractions
of category theory.

\subsubsection{Coherent sheaves} \label{sss:coh}

We saw in section \ref{sss:locfree} that the concept of vector bundles
can be replaced by locally free sheaves in algebraic geometry. We want
to be able to do things like compute cohomology for complexes of these
things so it would be nice if the category of locally free sheaves
were an abelian category. Sadly it is not.

The problem is that, while this category contains all of its kernels,
it does not contain its cokernels. The solution is to start with the
category of locally free sheaves, which is a subcategory of the
category of $\O_X$-modules. Then add in all the cokernel objects
together with all the possible morphisms between these new objects and
the objects we already had. The resulting category is abelian.  Thus
we end up with a minimal abelian full\footnote{A subcategory is full
if, for any pair of objects $A$ and $B$, the set of morphisms
$\Hom(A,B)$ is the same in the subcategory as it was in the original
category.} subcategory of the category of $\O_X$-modules containing
locally free sheaves. This is the category of {\em coherent
sheaves}.\footnote{In these notes we are going to be somewhat careless
about specifying whether ranks of free modules are infinite or
finite. In particular we will make no effort to distinguish between
coherent and {\em quasi-coherent\/} sheaves. \label{fn:qc}}

Let's give a simple example of a coherent sheaf that is not locally
free. We work locally on part of $X$ and pretend it looks like
$\C^3$ with affine coordinates $(x,y,z)$. Consider the following morphism
\begin{equation}
\xymatrix@1@=15mm{
  \O^{\oplus3}\ar[r]^{\left(\begin{smallmatrix}x&y&z
   \end{smallmatrix}\right)}&\O,
} \label{eq:affsky}
\end{equation}
that is, three functions $f_1$, $f_2$, and $f_3$ are mapped to
$xf_1+yf_2+zf_3$ by this morphism. Na\"\i vely speaking, the cokernel
object should consist of functions on $\C$ modded out by functions in
the image of the map. This quotient should kill all functions away
from the origin. We define $\O_p$ as the sheaf such that $\O_p(U)$ is
the trivial group if $U$ does not contain the origin. If $U$ does
contain the origin we set $\O_p(U)=\C$. This is an $\O_X$-module and
is called the {\em skyscraper sheaf\/} of the origin. There is also a
natural map
\begin{equation}
\xymatrix@1{
  \O\ar[r]&\O_p,
} \label{eq:afsk2}
\end{equation}
which takes a function to its value at the origin. The reader can
carefully check through the definitions to see that
(\ref{eq:afsk2}) indeed represents the cokernel of
(\ref{eq:affsky}). That is,
\begin{equation}
\xymatrix@1@=12mm{
  \O^{\oplus3}\ar[r]^{\left(\begin{smallmatrix}x&y&z
   \end{smallmatrix}\right)}&\O\ar[r]&\O_p\ar[r]&0
}
\end{equation}
is an exact sequence.

Thus, the skyscraper sheaf $\O_p$ is a coherent sheaf. It clearly
isn't locally free since its rank appears to jump at the origin. In a
way, it looks like it could be associated to a vector bundle which has
a fibre $\C$ over the origin and has trivial fibre elsewhere. We will
see in section \ref{ss:cohB} that this is a 0-brane on $X$.

We can also compute the image of the map (\ref{eq:affsky}), i.e., the
kernel of the map (\ref{eq:afsk2}). To phrase it a third way, we are
looking for the $\O_X$-module $\cI_p$ that completes the short exact
sequence
\begin{equation}
\xymatrix@1{
0\ar[r]&\cI_p\ar[r]&\O\ar[r]&\O_p\ar[r]&0.
} \label{eq:ideal0}
\end{equation}
One can easily show that $\cI_p$ is a subsheaf of $\O$ and consists of
functions which vanish at the origin. This sheaf is called the {\em
ideal sheaf\/} of the origin.

Again, $\cI_p$ is not locally free. The best we could say is that
somehow it represents a trivial bundle with fibre $\C^3$ everywhere
except at the origin, where it has no fibre. In D-brane language you
might say it is a 6-brane on $X$ with an anti-0-brane glued in at the
origin. We will see later that this isn't such a bad description of
$\cI_p$.

%%%%%%%%%%%%

\subsection{Cohomology}  \label{ss:cohom}

We have replaced the vector bundles of the B-model by coherent
sheaves. We now need to replace the notion of Dolbeault cohomology
with something more algebraic. To do this we use {\em sheaf
cohomology}. Let us emphasize immediately that by sheaf cohomology we
do {\em not\/} mean \v Cech cohomology, which is more topological than
algebraic.\footnote{Unfortunately, many references, such as
\cite{GH:alg}, do mean \v Cech cohomology when they say sheaf
cohomology.} Our sheaf cohomology is the version due to Grothendieck
as reviewed in chapter III of \cite{Hartshorne:} which is much more
suited to the B-model. Having said that, we will start our discussion
with \v Cech cohomology.

\subsubsection{\v Cech cohomology}  \label{sss:cech}

Here we give a lightning review. For more information and examples,
the reader is referred to \cite{BT:}.

Suppose $\mf U=\{U_\alpha\}$ is an open covering of a manifold $X$.
Let us denote $U_{\alpha_0}\cap U_{\alpha_1}\cap\ldots\cap
U_{\alpha_p}$ by $U_{\alpha_0\alpha_1\ldots\alpha_p}$.  Given a sheaf
$\cF$ we define
\begin{equation}
  \check{C}^p(\mf U,\cF)=\prod_{\alpha_0<\alpha_1<\ldots<\alpha_p}
     \!\!\!\!\!\!\cF(U_{\alpha_0\alpha_1\ldots\alpha_p}).
\end{equation}
That is, an element $a\in\check{C}^p(\mf U,\cF)$ is specified by
giving an element $a_{\alpha_0\alpha_1\ldots\alpha_p}\in
\cF(U_{\alpha_0\alpha_1\ldots\alpha_p})$ for each unordered
($p+1$)-tuple of open sets in $\mf U$.

Now define a coboundary $\delta:\check{C}^p\to \check{C}^{p+1}$ by
\begin{equation}
  (\delta a)_{\alpha_0\alpha_1\ldots\alpha_p\alpha_{p+1}} =
    \sum_{k=0}^{p+1}(-1)^k a_{\alpha_0\alpha_1\ldots\hat\alpha_k
    \ldots\alpha_{p+1}}, \label{eq:ccb}
\end{equation}
where the notation $\hat\alpha_k$ means omit $\alpha_k$.

It is easy to prove that $\delta^2=0$ and so we can define \v Cech
cohomology, $\check{H}^p(\mf U,\cF)$, as the cohomology of the complex
\begin{equation}
\xymatrix@1{
  0\ar[r]&\check{C}^0(\mf U,\cF)\ar[r]^\delta
  &\check{C}^1(\mf U,\cF)\ar[r]^\delta
  &\check{C}^2(\mf U,\cF)\ar[r]^\delta&\ldots
}
\end{equation}
This definition of \v Cech cohomology depends on the open covering $\mf
U$ but one can show that, as the covering becomes finer and finer, one
approaches a well-defined limit which we call $\check{H}^p(X,\cF)$. An
open cover which yields $\check{H}^p(\mf
U,\cF)\cong\check{H}^p(X,\cF)$ is a so-called {\em good cover\/} where
each finite intersection $U_{\alpha_0}\cap U_{\alpha_1}\cap\ldots\cap
U_{\alpha_p}$ is diffeomorphic to $\R^n$. We refer to \cite{BT:} for
more details.

As is well-known, given a short exact sequence of sheaves
\begin{equation}
\xymatrix@1{
0\ar[r]&\cE\ar[r]&\cF\ar[r]&\cG\ar[r]&0,
}
\end{equation}
we have an associated long exact sequence of cohomology:
\begin{equation}
\xymatrix@1{
  \ldots\ar[r]&\check H^k(X,\cE)\ar[r]&\check H^k(X,\cF)\ar[r]&
  \check H^k(X,\cG)\ar[r]&\check H^{k+1}(X,\cE)\ar[r]&\ldots
} \label{eq:les}
\end{equation}
Also, if $X$ is an $n$-dimensional manifold, then $\check H^k(X,\cF)=0$ for
any $\cF$ if $k<0$ or $k>n$.

It will prove useful to know all the cohomology groups $\check
H^k(\P^n,\O(m))$ for the locally-free sheaves of rank one introduced
in section \ref{sss:locfree}. The zeroth \v Cech cohomology consists
of sections defined over each $U_\alpha$ such that the differences of
these sections over the pair-wise intersections vanishes. In other
words, the zeroth \v Cech cohomology consists of {\em global
sections}.

The global sections of a sheaf $\cF$ which is an $\O_X$-module are in
one-to-one correspondence with morphisms $\O_X\to\cF$. This important
fact will be used many times below. The morphism may be produced from
the section simply by multiplication by the section. Given the
morphism, the definition of morphisms between $\O_X$-modules forces it
to be given by multiplication by a global section.

The analysis around (\ref{eq:OfOn}) therefore shows that global
sections of $\O(m)$ are given by homogeneous functions of degree $m$
in the homogeneous coordinates $[z_0,z_1,\ldots,z_n]$. Thus,
\begin{equation}
  \dim \check H^0(\P^n,\O(m)) = \binom{n+m}{m},
\end{equation}
where the binomial coefficient is defined to be zero if either of its
entries are negative.

The method of \v Cech cohomology may be applied to compute the higher
cohomologies. This is tedious and there are other ways to do the
computation (see \cite{Hartshorne:,GH:alg} for example). The result is that
\begin{equation}
\begin{split}
  \dim \check H^k(\P^n,\O(m)) &= 0\quad\hbox{for $k\neq0,n$}\\
  \dim \check H^n(\P^n,\O(m)) &= \binom{-m-1}{-n-m-1}.
\end{split}
\end{equation}

Now suppose $X\subset\P^n$ is an algebraic variety of dimension $n-1$
corresponding to the zeroes of a polynomial $f$ of homogeneous degree
$d$. We denote the embedding $i:X\hookrightarrow\P^n$.

Let us introduce a little notation. Suppose we
have a map $f:X\to Y$ between two algebraic varieties and a sheaf
$\cF$ on $X$. We may define the sheaf $f_*\cF$ on $Y$ by $f_*\cF(U)=
\cF(f^{-1}U)$. More specifically, suppose we have an embedding
$i:X\hookrightarrow\P^n$, and we are given a sheaf $\cE$ on $X$. The
sheaf $i_*\cE$ is therefore given by $i_*\cE(U)=\cE(U\cap\P^n)$ for
all open subsets $U\subset\P^n$. This naturally embeds the set of
sheaves on $X$ into the sheaves on $\P^n$.

The structure sheaf $\O_X$ may thus be pushed forward into a sheaf
$i_*\O_X$ on $\P^4$. By a modest abuse of notation we refer to this
latter sheaf as $\O_X$. Clearly $\O_X$ is then a quotient sheaf of
$\O$. In fact, it is not hard to see that we have a short exact sequence
of sheaves
\begin{equation}
\xymatrix@1{
  0\ar[r]&\O(-d)\ar[r]^-f&\O\ar[r]&\O_X\ar[r]&0.
} \label{eq:OX}
\end{equation}
Using (\ref{eq:les}), and the fact that $H^k(X,\cF)=H^k(\P^n,i_*\cF)$,
this allows the computation of the cohomology of $\O_X$. Similarly we
may tensor (\ref{eq:OX}) by $\O(m)$ to compute the cohomology of
$\O_X(m)=\O_X\otimes\O(m)$.

\subsubsection{Spectral sequences}  \label{sss:sseq}

We would like to compare \v Cech cohomology with something we already
know about such as Dolbeault cohomology. A very quick way of doing
this is to use ``spectral sequences'' which we now review.  The idea
of spectral sequences will also be a recurring topic in the rest of
lectures.  Again, a comprehensive review of this important subject is
beyond the scope of these lectures and we recommend the interested
reader consult \cite{BT:,McL:ss}, for more information. Here we focus
on how a spectral sequence is used rather than why it works.

Suppose we are given a {\em double complex}, $E_0^{p,q}$,
\begin{equation}
\begin{xy}
\xymatrix@C=20mm{
  \vdots&\vdots&\vdots&\\
  E_0^{0,2}\ar[r]^\delta\ar[u]^d&E_0^{1,2}\ar[r]^\delta
           \ar[u]^d&E_0^{2,2}\ar[r]^\delta\ar[u]^d&\ldots\\
  E_0^{0,1}\ar[r]^\delta\ar[u]^d&E_0^{1,1}\ar[r]^\delta
           \ar[u]^d&E_0^{2,1}\ar[r]^\delta\ar[u]^d&\ldots\\
  E_0^{0,0}\ar[r]^\delta\ar[u]^d&E_0^{1,0}\ar[r]^\delta
           \ar[u]^d&E_0^{2,0}\ar[r]^\delta\ar[u]^d&\ldots\\
} 
% +<0pt,0pt> collapses box.
\save="x"!LD+<-3mm,0pt>;"x"!RD+<0pt,0pt>**\dir{-}?>*\dir{>}\restore
\save="x"!LD+<0pt,-3mm>;"x"!LU+<0pt,-2mm>**\dir{-}?>*\dir{>}\restore
\save!CD+<0mm,-4mm>*{p}\restore
\save!CL+<-3mm,0mm>*{q}\restore
\end{xy}  \label{eq:dcx}
\end{equation}
where each row and each column forms a complex. We assume that the
two derivatives anticommute $d\delta+\delta d=0$.

Clearly we may make a new complex
\begin{equation}
\xymatrix@1{
  0\ar[r]&E^0\ar[r]^D&E^1\ar[r]^D&E^2\ar[r]^D&\ldots,
}  \label{eq:Dcx}
\end{equation}
where
\begin{equation}
  E^n = \bigoplus_{p+q=n} E_0^{p,q},
\end{equation}
and $D=d+\delta$. The big question is, how to we compute the
cohomology of (\ref{eq:Dcx})?

The spectral sequence method is to inductively form a sequence of
stages $E_r^{p,q}$.  $E_{r+1}^{p,q}$ is defined as the cohomology of
$E_r^{p,q}$ with respect to a differential
\begin{equation}
  d_r: E_r^{p,q} \to E_r^{p+r,q-r+1},
\end{equation}
where $d_0$ is given by $d$ in figure (\ref{eq:dcx}) and $d_1$ is
given by the image of $\delta$ in $E_1^{p,q}$. We won't need to know
how to compute $d_r$ for $r\geq2$ since any problem that requires such
knowledge is essentially too difficult for these lectures!

For large enough $r$ ($r\geq2$ if we're lucky) the differentials $d_r$
are all zero. This means that, for large $r$, the $E_r^{p,q}$'s become
independent of $r$ and so are written $E_\infty^{p,q}$. The cohomology
of (\ref{eq:Dcx}) is then given as\footnote{This is not quite right
for torsion subgroups but we do not consider such things in these
lectures.}
\begin{equation}
  H_D^n = \bigoplus_{p+q=n} E^{p,q}_\infty.
    \label{eq:ssT}
\end{equation}

\subsubsection{Dolbeault cohomology}  \label{sss:dolb}

Let's try the spectral sequence construction to prove that Dolbeault
cohomology is equal to a particular example of \v Cech cohomology. We
consider the sheaf $\cA^{m,n}$ of $(m,n)$-forms on $X$. That is,
$\cA^{m,n}(U)$ is the group of differentiable (but not necessarily
holomorphic) sections of the bundle
$\bigwedge^mT_X\otimes\bigwedge^n\bar T_X$ over the open set $U\subset
X$.

Now, build the double complex
\begin{equation}
E^{p,q}_0 = \check C^p(\mf U,\cA^{m,q}),
\end{equation}
where $\mf U$ is a good cover of $X$. The ``horizontal'' $\delta$
operator in (\ref{eq:dcx}) is given by the \v Cech coboundary
(\ref{eq:ccb}) and the ``vertical'' $d$ operator is given by the
Dolbeault operator $\bar\partial:\cA^{m,n}\to\cA^{m,n+1}$.

The first step is to take the cohomology under the vertical map
$d_0=\bar\partial$ to obtain the $E_1^{p,q}$'s. On $\R^N$ any
$(m,n)$-form which is $\bar\partial$-closed must be
$\bar\partial$-exact. Thus all the cohomology groups vanish except the
bottom row given by $q=0$. The groups $E_1^{p,0}$ are given by
$(p,0)$-forms which are killed by $\bar\partial$. In other words, the
$E_1^{p,q}$ stage looks like
\begin{equation}
\begin{xy}
 \xymatrix{
  \vdots&\vdots&\vdots&\\
  0\ar[r]^\delta&0\ar[r]^\delta&0\ar[r]^\delta&\ldots\\
  0\ar[r]^\delta&0\ar[r]^\delta&0\ar[r]^\delta&\ldots\\
  \check C^0(\mf U,\Omega^m)\ar[r]^\delta
    &\check C^1(\mf U,\Omega^m)\ar[r]^\delta
    &\check C^2(\mf U,\Omega^m)\ar[r]^\delta&\ldots
 }
\save="x"!LD+<-3mm,0pt>;"x"!RD+<0pt,0pt>**\dir{-}?>*\dir{>}\restore
\save="x"!LD+<0pt,-3mm>;"x"!LU+<0pt,-2mm>**\dir{-}?>*\dir{>}\restore
\save!CD+<0mm,-4mm>*{p}\restore
\save!CL+<-3mm,0mm>*{q}\restore
\end{xy}
\end{equation}
where $\Omega^m$ is the sheaf of holomorphic $m$-forms. That is,
$\Omega^m(U)$ is the group of holomorphic sections of $\bigwedge^mT_X$
over $U$.

Now we take the $\delta$-cohomology to form the $E_2^{p,q}$
stage. Clearly we end up with $E_2^{p,q}=\check H^p(X,\Omega^m)$ for
$q=0$ and $E_2^{p,q}=0$ for $q>0$.

For the next stage we note that $d_2$ cannot map between nonzero
entries and therefore must be zero. Similarly $d_r=0$ for $r>2$. Thus,
in this case, $E_2^{p,q}=E_\infty^{p,q}$. Applying (\ref{eq:ssT}) we
see
\begin{equation}
  H_D^n = \check H^n(X,\Omega^m). \label{eq:HD1}
\end{equation}

The alert reader will be wondering why, when the definition of the
double complex in (\ref{eq:dcx}) is symmetric under interchange of $d$
and $\delta$, the spectral sequence procedure clearly treated them
differently. We can exchange the r\^oles of rows and columns to define
another spectral sequence $\tilde E_r^{p,q}$ with $\tilde E_0^{p,q}=
E_0^{p,q}$ and differential
\begin{equation}
  \tilde d_r: \tilde E_r^{p,q} \to \tilde E_r^{p-r+1,q+r},
\end{equation}
with $\tilde d_0=\delta$. This spectral sequence may also be used to
compute the total cohomology $H_D^n$.

Under such a reversal we compute the \v Cech cohomology first. This
may be done by using the fact that the following ``Mayer--Vietoris
sequence'' is exact:
\begin{equation}
\xymatrix@1{
0\ar[r]&\cA^{m,n}(X)\ar[r]&\check C^0(\mf U,\cA^{m,n})\ar[r]^\delta
  &\check C^1(\mf U,\cA^{m,n})\ar[r]^\delta
  &\check C^2(\mf U,\cA^{m,n})\ar[r]^-\delta&\ldots.
} \label{eq:MV}
\end{equation}
Elements of the group $\cA^{m,n}(X)$ are {\em globally\/} defined
$(m,n)$-forms. The exactness of the start of this sequence should be
pretty clear. For the other terms one uses a trick using a partition
of unity. We refer to page 94 of \cite{BT:} for the proof in an
essentially identical situation.

The exactness of (\ref{eq:MV}) implies that the $\tilde E_1^{p,q}$
stage of the spectral sequence looks like
\begin{equation}
\begin{xy}
 \xymatrix@!C=15mm{
  \vdots&\vdots&\vdots&\\
  \cA^{m,2}(X)\ar[u]^{\bar\partial}&0\ar[u]^{\bar\partial}&
       0\ar[u]^{\bar\partial}&\ldots\\
  \cA^{m,1}(X)\ar[u]^{\bar\partial}&0\ar[u]^{\bar\partial}&
       0\ar[u]^{\bar\partial}&\ldots\\
  \cA^{m,0}(X)\ar[u]^{\bar\partial}&0\ar[u]^{\bar\partial}&
       0\ar[u]^{\bar\partial}&\ldots
 }
\save="x"!LD+<-3mm,0pt>;"x"!RD+<0pt,0pt>**\dir{-}?>*\dir{>}\restore
\save="x"!LD+<0pt,-3mm>;"x"!LU+<0pt,-2mm>**\dir{-}?>*\dir{>}\restore
\save!CD+<0mm,-4mm>*{p}\restore
\save!CL+<-3mm,0mm>*{q}\restore
\end{xy}
\end{equation}
The first column is nothing but the usual Dolbeault complex. Thus
$\tilde E_2^{0,q}=H_{\bar\partial}^{m,q}(X)$ and 
$H_D^n=H_{\bar\partial}^{m,n}(X)$. Comparing to (\ref{eq:HD1}) and
relabeling a little, we obtain Dolbeault's theorem:
\begin{equation}
  H_{\bar\partial}^{p,q}(X) = \check H^{q}(X,\Omega^p).
\end{equation}
Thus Dolbeault cohomology can be rewritten as \v Cech cohomology. 

Suppose $E$ is a holomorphic vector bundle over $X$. The above argument can be
generalized to
\begin{equation}
  H_{\bar\partial}^{p,q}(X,E)=\check H^{q}(X,\Omega^p\otimes\cE),
   \label{eq:DckE}   
\end{equation}
where $\cE$ is the locally-free sheaf associated to $E$ and the tensor
product ``$\otimes$'' is defined as a sheaf of tensor products of
$\O_X$-modules.

\subsubsection{Sheaf cohomology}  \label{sss:shco}

Now we give a definition of cohomology which is couched purely in
terms of category language. An object $I$ in a category is called {\em
injective} if, given a monomorphism $f:A\to B$ and any map
$g:A\to I$, we may construct a map $g':B\to I$ such that
$g'f=g$. This may be pictured as the following diagram:
\begin{equation}
\xymatrix{
  0\ar[r]&A\ar[r]^f\ar[d]^g&B\ar@{-->}[dl]^{g'}\\
  &I&
}
\end{equation}
The reader might like to check that $\GU(1)$ is injective in the
category of abelian groups for example.

In the category of $\O_X$-modules, injective objects have an
interesting property as follows. Given an open set $U\subset X$ we
define $\O_U$ as the sheaf $\O_X$ restricted to $U$ and then
``extended by zero'' outside $U$. Roughly speaking, this is the sheaf of
holomorphic functions on $U$. We refer to page 68 of
\cite{Hartshorne:} for the precise definition of extending by zero.
Using a similar argument to that in section
\ref{sss:cech}, for any $\O_X$-module $\cF$ one may argue that
\begin{equation}
  \Hom(\O_U,\cF) = \cF(U).  \label{eq:flb1}
\end{equation}
Now, if $V\subset U$ then $\O_V$ is a subsheaf of $\O_U$, i.e.,
there is a monomorphism $\O_V\to \O_U$. If $\cI$ is an injective
$\O_X$-module, then, by the definition above, we have a surjective map
$\Hom(\O_U,\cI)\to\Hom(\O_V,\cI)$. That is, from (\ref{eq:flb1}), the
restriction map $\rho_{UV}:\cI(U)\to\cI(V)$ is surjective.

A sheaf whose restriction maps are all surjective is called {\em
flabby}.\footnote{Or, in French, {\em flasque}.} Thus we have shown
that injective $\O_X$-modules are flabby sheaves.

Given any object $A$ in an abelian category, an {\em injective
resolution\/} of $A$ is a long exact sequence of the form
\begin{equation}
\xymatrix@1{
  0\ar[r]&A\ar[r]&I^0\ar[r]&I^1\ar[r]&I^2\ar[r]&\ldots,
} \label{eq:injres}
\end{equation}
where the $I^k$ are all injective objects. Injective resolutions need
not exist. In the category of $\O_X$-modules, injective resolutions do
always exist (see page 207 of \cite{Hartshorne:} for example).

At this point we need to introduce the notion of a {\em functor\/} in
a category. In our categories of D-branes we will have no direct
physical manifestation of a functor but the concept still proves
valuable.

\begin{definition}
  A {\em functor\/} $F:\mathcal{C}\to\mathcal{D}$ is a rule that
associates an object $F(C)$ of $\mathcal{D}$ to every object $C$ of
$\mathcal{C}$ and a morphism $F(f):F(C_1)\to F(C_2)$ in $\mathcal{D}$
to every morphism $f:C_1\to C_2$ in $\mathcal{C}$. It satisfies
\begin{enumerate}
\item $F(\id_{\mathcal C})=\id_{\mathcal D}$
\item $F(fg) = F(f)F(g)$.
\end{enumerate}
\end{definition}

Suppose we have a short exact sequence 
\begin{equation}
\xymatrix@1{
  0\ar[r]&A\ar[r]^f&B\ar[r]^g&C\ar[r]&0,
}
\end{equation}
in an abelian category. A functor $F$ is said to be {\em left-exact\/} if
\begin{equation}
\xymatrix@1{
  0\ar[r]&F(A)\ar[r]^{F(f)}&F(B)\ar[r]^{F(g)}&F(C),
}
\end{equation}
is also exact, and {\em right-exact\/} if the following is exact:
\begin{equation}
\xymatrix@1{
  F(A)\ar[r]^{F(f)}&F(B)\ar[r]^{F(g)}&F(C)\ar[r]&0.
}
\end{equation}

An example of a functor from the category of abelian groups to itself
is $\Hom(G,-)$ for some fixed group $G$. This maps a group $A$ to
$\Hom(G,A)$. If $f\in\Hom(G,A)$, then any
homomorphism $h:A\to B$ yields a map $hf:G\to B$ thus inducing the
required map $\Hom(G,A)\to\Hom(G,B)$. The functor $\Hom(G,-)$ is
easily shown to be left-exact but not right-exact.

Given an injective resolution (\ref{eq:injres}) and a left-exact
functor $F$, we may construct a complex\footnote{Note we omitted the
first term from (\ref{eq:injres}).}
\begin{equation}
\xymatrix@1{
0\ar[r]&F(I^0)\ar[r]&F(I^1)\ar[r]&F(I^2)\ar[r]&\ldots
}
\end{equation}
The cohomology of this complex at position $n$ is defined as the $n$th
{\em right-derived functor\/} of $A$ and is denoted
$\mathbf{R}^nF(A)$.
The reader is invited to check that the left-exactness of $F$ means
that $\mathbf{R}^0F(A)=F(A)$. While it is not obvious at first sight,
these derived functors do not depend on the choice of an injective
resolution. 

In the category of $\O_X$-modules, the functor of interest is
$\Hom(\O_X,-)$. As we saw in (\ref{eq:flb1}), $\Hom(\O_X,\cF)=\cF(X)$,
i.e., the group of global sections of $\cF$. Thus we may also view
$\Hom(\O_X,-)$ as the {\em global section functor}. We may then define
{\em sheaf cohomology\/} for $\O_X$-modules as the right-derived
functors of the global section functor. That is,
\begin{equation}
  H^n(X,\cF) = \mathbf{R}^n \Hom(\O_X,-) (\cF).
\end{equation}
Since $\mathbf{R}^0F(A)=F(A)$, $H^0(X,\cF)$ corresponds to the group
of global sections of $\cF$ --- just like \v Cech cohomology.

OK, so this is all pretty abstract! At this point the reader would
probably like some examples to work through to get a feel for sheaf
cohomology. The truth is that this definition of cohomology is awful
for practical calculations. The best one can generally do is show that
it is equivalent to some other form of cohomology that can be computed
realistically. The reason we have bothered to introduce sheaf
cohomology is that its definition is very powerful in the context of
the B-model as we see later.

Sheaf cohomology is equivalent to \v Cech cohomology as can be seen as
follows. Given an $\O_X$-module $\cF$, construct an injective
resolution
\begin{equation}
\xymatrix@1{
  0\ar[r]&\cF\ar[r]&\cI^0\ar[r]&\cI^1\ar[r]&\cI^2\ar[r]&\ldots
} \label{eq:Fres}
\end{equation}
Consider the double complex given by
\begin{equation}
  E^{p,q}_0 = \check{C}^p(\mf U,\cI^q).
\end{equation}
It follows from (\ref{eq:Fres}) that at the first stage of the
spectral sequence we have that $E_1^{p,q}=\check{C}^p(\mf U,\cF)$ for
$q=0$ and is zero otherwise. Thus
$E_2^{p,q}=E_\infty^{p,q}=\check{H}^p(X,\cF)$ for $q=0$ and is
zero otherwise. This yields $H_D^n=\check{H}^n(X,\cF)$.

Applying the spectral sequence the other way requires us to compute
the \v Cech cohomology of injective $\O_X$-modules. We saw above that
an injective $\O_X$-module is a flabby sheaf. It can be shown (e.g.,
page 221 of \cite{Hartshorne:}) that if $\cF$ is flabby, then the \v
Cech cohomology groups $\check{H}^n(X,\cF)$ are zero for $n>0$. As
always, $\check{H}^0(X,\cF)$ is given by the global sections $\cF(X)$.

This means that the $\tilde E_1^{p,q}$ stage of the spectral
sequence looks like
\begin{equation}
\begin{xy}
 \xymatrix@!C=15mm{
  \vdots&\vdots&\vdots&\\
  \cI^2(X)\ar[u]&0\ar[u]&
       0\ar[u]&\ldots\\
  \cI^1(X)\ar[u]&0\ar[u]&
       0\ar[u]&\ldots\\
  \cI^0(X)\ar[u]&0\ar[u]&
       0\ar[u]&\ldots
 }
\save="x"!LD+<-3mm,0pt>;"x"!RD+<0pt,0pt>**\dir{-}?>*\dir{>}\restore
\save="x"!LD+<0pt,-3mm>;"x"!LU+<0pt,-2mm>**\dir{-}?>*\dir{>}\restore
\save!CD+<0mm,-4mm>*{p}\restore
\save!CL+<-3mm,0mm>*{q}\restore
\end{xy}
\end{equation}
and $\tilde E_2^{p,q}$ computes the sheaf cohomology of $\cF$. Thus
$H_D^n=H^n(X,\cF)$ and we obtain the equivalence
\begin{equation}
  H^n(X,\cF) = \check H^n(X,\cF).
\end{equation}
Note that we have used the language of $\O_X$-modules in the
section. Everything works equally well if we restrict attention to
coherent sheaves since we may always form injective resolutions using
coherent sheaves.\footnote{Pedants should remind themselves of the
footnote on page \pageref{fn:qc}.}

We may take the definition of sheaf cohomology a little
further. The functor $\Hom(\cE,-)$ is left-exact for any $\O_X$-module
$\cE$. We denote its right derived functors by ``Ext'':
\begin{equation}
  \mathbf{R}^n \Hom(\cE,-)(\cF) = \Ext^n(\cE,\cF).
     \label{eq:Ext}
\end{equation}
Let us note the following obvious statements:
\begin{equation}
\begin{split}
  \Ext^0(\cE,\cF) &= \Hom(\cE,\cF)\\
  \Ext^n(\O_X,\cF) &= H^n(X,\cF).
\end{split}
\end{equation}

A very useful fact about these $\Ext$ groups is that they satisfy
``Serre duality''. We refer to \cite{Hartshorne:} for more details. In
the case of a smooth \CY\ $m$-fold, Serre duality states that
\begin{equation}
  \Ext^n(\cE,\cF) \cong \Ext^{m-n}(\cF,\cE).
     \label{eq:Serre}
\end{equation}

Now suppose we have two vector bundles $E$ and $F$ over $X$. The bundle
$\Hom(E,F)$ is also a vector bundle as we saw in section
\ref{sss:Bopen}. We may associate locally free sheaves $\cE$ and $\cF$
to the bundles $E$ and $F$ respectively. The locally-free sheaf we
associate to $\Hom(E,F)$ is denoted $\sHom(\cE,\cF)$. We emphasize
that $\sHom(\cE,\cF)$, which is a sheaf, should not be confused with
$\Hom(\cE,\cF)$ which is the abelian group of morphisms from $\cE$ to
$\cF$. $\Hom(\cE,\cF)$ is actually the group of {\em global sections\/} of
$\sHom(\cE,\cF)$.

From (\ref{eq:DckE}) it follows that
\begin{equation}
\begin{split}
  H^{0,q}(X,\Hom(E,F)) &= \check H^q(X,\sHom(\cE,\cF))\\
   &= H^q(X,\sHom(\cE,\cF)).
\end{split}
\end{equation}
Since sheaf cohomology is the right-derived functor of global section,
and a global section of $\sHom(\cE,\cF)$ is given by $\Hom(\cE,\cF)$,
we may further deduce that
\begin{equation}
  H^q(X,\sHom(\cE,\cF)) = \Ext^q(\cE,\cF).
\end{equation}

Thus we have achieved our goal. We have converted the Dolbeault
cohomology language of differential geometry into purely algebraic
ideas. The statement in section \ref{sss:Bopen} that an open string
from a B-brane $E\to X$ to a B-brane $F\to X$ is given by an element
of the Dolbeault cohomology group $H^{0,q}_{\bar\partial}(X,\Hom(E,F))$
is now restated in the form 
\begin{center}
\shabox{\parbox{.85\hsize}{An open string from the B-brane associated
 to the locally-free sheaf $\cE$ to another B-brane given by the
 locally-free sheaf $\cF$ is given by an element of the group
 $\Ext^q(\cE,\cF)$.}}
\end{center}

The reader is probably thoroughly unimpressed at this point
given the lengths of abstraction we went to. Hopefully the later
lectures will convince the reader that it is all worthwhile!

%%%%%%%%%%%%%%%%%%%%%%%%%%%%%%%%%%%%%%%%%%%%%%%%%%%%%%%%%%%%%%%%%%

\section{The Category of B-branes}  \label{s:Bcat}

\subsection{Deformations and complexes}  \label{ss:Bdef}

The problem with the B-model we have thus far is that it doesn't
contain enough B-branes. The first thing to try to do is to see if we
can deform the B-branes we already know about into something new.

Looking at the topological field theory, we already saw that we could
use vertex operators as deformations. The closed string operators are
required to have ghost number two and correspond to $H^1(X,T_X)$. These
give the expected deformations of complex structure.

As in section \ref{sss:Aopen1} the open strings vertex operators which
deform the theory must be ghost number one.  For a theory with a
single D-brane given by the locally free sheaf $\cE$ these correspond
to $\Ext^1(\cE,\cE)$. Ignoring potential obstructions in the moduli
space this agrees with the expected deformations of the sheaf
$\cE$.\footnote{If $\cE$ is associated to a vector bundle $E$ then
$\Ext^1(\cE,\cE)=H^1(X,\End(E))$. See section 15.7.3 of
\cite{GSW:book} for one way of seeing why this latter group
corresponds to deformations of $E$.}

What about the ghost number one open strings stretched between two
distinct B-branes $\cE$ and $\cF$? The first guess would be to look at 
vertex operators in $\Ext^1(\cE,\cF)$. One can do this but it turns
out that one still doesn't generate enough B-branes. To get the right
answer we need to be more general.

Our experience with the A-model in section \ref{sss:Aopen2} tells us
that assigning a ghost number to an open string stretched between two
distinct D-branes is a little troublesome. The B-model has no right to
be so unambiguous in its knowledge of the ghost number and so we
should inflict the same ignorance on it. That is, let us label a
B-brane $\cF$ with some ghost number $\mu(\cF)$. An open string
stretching from $\cE$ to $\cF$ in the group $\Ext^q(\cE,\cF)$ is then
given a ghost number 
\begin{equation}
  q + \mu(\cF) - \mu(\cE), \label{eq:qFE}
\end{equation}
in agreement with (\ref{eq:gshift}). We are certainly free to attach
such ghost number labels to the B-branes without any effect on the
B-model. We will also see in section \ref{s:stab} that associating a ghost
number to the B-branes themselves is essential if we want to
understand the untwisted superconformal field theory.

We may construct a general collection of D-branes in terms of a
locally-free sheaf $\cE$ in which we have a decomposition:
\begin{equation}
  \cE = \bigoplus_{n\in\Z} \cE^n,
\end{equation}
where $\cE^n$ is a B-brane with ghost number $n$. The ghost number one
operators in this B-model are therefore elements of
$\Ext^k(\cE^n,\cE^{n-k+1})$ for any $n$ and $k$.

We have already noted that the case $k=1$ corresponds to deformations
of the sheaves that we already know about. The case $k=0$
concerns open strings
\begin{equation}
\begin{split}
  d &= \sum_n d_n\\
  d_n&\in\Ext^0(\cE^n,\cE^{n+1}) = \Hom(\cE^n,\cE^{n+1}),
\end{split} \label{eq:ddef}
\end{equation}
i.e., morphisms $d_n:\cE^n\to\cE^{n+1}$ of locally-free sheaves.
It will turn out that, by just studying this case, we will obtain the
deformations for the other values of $k$ for free.

Let $W^{(1)}_d$ be the operator obeying
\begin{equation}
  \{Q,W^{(1)}_d\}=d_\Sigma d,
\end{equation}
where $d_\Sigma$ is the worldsheet de Rham operator. We deform the
action by
\begin{equation}
  S = S_0 + \oint_{\partial\Sigma} W^{(1)}_d.
\end{equation}
Following the usual Noether method, we may show that this results in a
change in the BRST charge
\begin{equation}
  Q = Q_0 + d.
\end{equation}
So, to maintain the relation $Q^2=0$, we are required to impose
\begin{equation}
  \{Q_0,d\} + d^2=0.   \label{eq:CaMa}
\end{equation}
We are assuming that $d$ was an open string vertex operator in our
original theory before deformation and so we assume
$\{Q_0,d\}=0$. Naturally this can only be justified if the deformation
of the theory is infinitesimal in some way. Indeed the analysis we
perform below only really describes the tangent space for the
deformations. There can be obstructions to these
deformations.

Anyway, with the assumption $\{Q_0,d\}=0$, we require $d^2=0$. What
does this mean exactly? $d$ is an open string vertex operator given by
a sum in (\ref{eq:ddef}). Multiplication of $d$ by itself means we use
the operator product algebra for open strings. The first thing to note
is that the boundary conditions must make sense in order to obtain a
nonzero result: an open string $A\to B$ can only combine with an open
string $C\to D$ to produce a string $A\to D$ if $B$ and $C$ represent
the same D-brane. Secondly, we stated in section \ref{ss:B} that the
operator product was given by simple wedge product between forms. Here
we are simply multiplying zero forms valued in a group of
homomorphisms. Thus the operator product is simply a composition of
these homomorphisms. The result is that $d^2=0$ implies that
\begin{equation}
  d_{n+1}d_n=0\quad\hbox{for all $n$.}
\end{equation}
In other words we have a {\em complex}
\begin{equation}
\xymatrix@1{
  \ldots\ar[r]^{d_{n-1}}&\cE^n\ar[r]^{d_n}&\cE^{n+1}\ar[r]^{d_{n+1}}&
   \cE^{n+2}\ar[r]^{d_{n+2}}&\ldots,
} \label{eq:Ecx}
\end{equation}
which we denote $\cE^\bullet$ for short.

A B-brane is therefore more generally represented by a complex of
locally-free sheaves. The maps in the complex represent a deformation
from the initial simple collection of sheaves. Note that a sheaf $\cE$
itself is a complex in a rather trivial way:
\begin{equation}
\xymatrix@1{
  \ldots\ar[r]^-0&0\ar[r]^0&\cE\ar[r]^0&
   0\ar[r]^-0&\ldots
} \label{eq:E0}
\end{equation}
Our convention, in this context, will be to assume that $\cE$ is in
position 0 of the complex.

So, a contender for the objects in our category of B-branes appears to
be complexes of locally-free sheaves. This will turn out to be the
correct answer but we will need to quotient out by a large set of
equivalences. That is, two different complexes may represent the same
B-brane. In the language of categories this means that two complexes
are related by an invertible morphism, i.e., they are
isomorphic. Thus, it is by analyzing the morphisms, i.e., open
strings, that we will know if two complexes represent the ``same''
B-brane.

%%%%%%%%%%%%%

\subsection{Open strings} \label{ss:Bop}

The deformation above will also affect the spectrum of open strings
between the B-branes. In this section we compute the corresponding
Hilbert spaces of open string states.

Initially let us assume that the open strings come from a B-brane that
is a complex and goes to a B-brane that is just a locally-free sheaf.
To be more precise, suppose, for simplicity, we have a collection of
locally free sheaves $\cE^0,\cE^1,\ldots$ and another locally-free
sheaf $\cF$. Let $\cE^n$ have ghost number $n$ and $\cF$ have ghost
number 0. Now deform the theory by turning the collection of $\cE$'s
into a complex (\ref{eq:Ecx}) with boundary maps $d_n^{\cE}$. We want
to consider the open strings $\cE^\bullet\to\cF$.

Suppose $\cF$ has an injective resolution
\begin{equation}
\xymatrix@1{
  0\ar[r]&\cF\ar[r]&\cI^0\ar[r]&\cI^1\ar[r]&\cI^2\ar[r]&\ldots
} \label{eq:Fres2}
\end{equation}
We now construct the double complex $E_0^{p,q}=\Hom(\cE^{-p},\cI^q)$:
\begin{equation}
\begin{xy}
\xymatrix{
  &\vdots&\vdots&\vdots\\
  \ldots\ar[r]^(0.3){d_2^{\cE}}&
      \Hom(\cE^2,\cI^2)\ar[r]^{d_1^{\cE}}\ar[u]^{Q_0}&
      \Hom(\cE^1,\cI^2)\ar[r]^{d_0^{\cE}}\ar[u]^{Q_0}&
      \Hom(\cE^0,\cI^2)\ar[u]^{Q_0}\\
  \ldots\ar[r]^(0.3){d_2^{\cE}}&
      \Hom(\cE^2,\cI^1)\ar[r]^{d_1^{\cE}}\ar[u]^{Q_0}&
      \Hom(\cE^1,\cI^1)\ar[r]^{d_0^{\cE}}\ar[u]^{Q_0}&
      \Hom(\cE^0,\cI^1)\ar[u]^{Q_0}\\
  \ldots\ar[r]^(0.3){d_2^{\cE}}&
      \Hom(\cE^2,\cI^0)\ar[r]^{d_1^{\cE}}\ar[u]^{Q_0}&
      \Hom(\cE^1,\cI^0)\ar[r]^{d_0^{\cE}}\ar[u]^{Q_0}&
      \Hom(\cE^0,\cI^0)\ar[u]^{Q_0}
} 
% +<0pt,0pt> collapses box.
\save="x"!RD+<3mm,0pt>;"x"!LD+<0pt,0pt>**\dir{-}?>*\dir{>}\restore
\save="x"!RD+<0pt,-3mm>;"x"!RU+<0pt,-2mm>**\dir{-}?>*\dir{>}\restore
\save!CD+<0mm,-4mm>*{p}\restore
\save!CR+<3mm,0mm>*{q}\restore
\end{xy}  \label{eq:Bbd1}
\end{equation}

The maps in this sequence are the obvious ones induced by the
complexes (\ref{eq:Ecx}) and (\ref{eq:Fres2}). We label the vertical
maps $Q_0$ since we know, by the definition of $\Ext$ in section
\ref{sss:shco}, that cohomology in this direction produces the open
string Hilbert spaces before the deformation $d$ is turned on.

Clearly we have something that looks just like the spectral sequence
construction of section \ref{sss:sseq}. The only difference is that we
have made $p$ negative to make the horizontal maps point to the
right. Note that nothing in the spectral sequence construction in
section \ref{sss:sseq} depended on the positivity of $p$ and $q$.
The required anti-commutivity of the differentials is given by $Q^2=0$
as we saw above.

Thus, a spectral sequence construction applied to (\ref{eq:Bbd1})
yields the total cohomology, i.e., the cohomology of $Q=Q_0+d$, which
is exactly what we are after! In keeping with the notation before the
deformation, we denote the Hilbert space of open strings of ghost
number $q$ by $\Ext^q(\cE^\bullet,\cF)$.

Note that there is no reason why we couldn't include $\cE^n$ for $n<0$
in the complex. It just made the diagram a little easier to draw.

We need to work a little harder in the case that the string starts on
a sheaf $\cE$ and ends on a complex $\cF^\bullet$ given by
\begin{equation}
\xymatrix@1{
  \ldots\ar[r]&\cF^0\ar[r]^{d_0^{\cF}}&\cF^1\ar[r]^{d_1^{\cF}}&
      \cF^2\ar[r]^{d_2^{\cF}}&\ldots,
}
\end{equation}
where each $\cF^p$ has an injective resolution
\begin{equation}
\xymatrix@1{
  0\ar[r]&\cF^p\ar[r]&\cI^{p,0}\ar[r]&\cI^{p,1}\ar[r]&\cI^{p,2}\ar[r]&\ldots
} \label{eq:Eres}
\end{equation}
From the definition of an injective object in section \ref{sss:coh} we
may prove the following generalization
\begin{theorem}
Given any maps $f:A\to B$, $g:A\to I$ in an abelian category with I an
injective object, a map $g'$ can be constructed to make the following
commutative:
\begin{equation}
\xymatrix{
  A\ar[r]^f\ar[d]^g&B\ar@{-->}[dl]^{g'}\\
  I&
}
\end{equation}
so long as $g\ker(f)=0$.
\end{theorem} %Ex?
From this, we may prove that any map $d^{\cF}_p:\cF^p\to\cF^{p+1}$ may
be extended to a map between resolutions
\begin{equation}
\xymatrix{
  0\ar[r]&\cF^p\ar[r]\ar[d]^{d^{\cF}_p}&\cI^{p,0}\ar[r]\ar@{-->}[d]&
      \cI^{p,1}\ar[r]\ar@{-->}[d]&\cI^{p,2}\ar[r]\ar@{-->}[d]&\ldots\\
  0\ar[r]&\cF^{p+1}\ar[r]&\cI^{p+1,0}\ar[r]&\cI^{p+1,1}\ar[r]&
    \cI^{p+1,2}\ar[r]&\ldots,
} \label{eq:Irm}
\end{equation}
where the vertical set of maps form a {\em chain map\/}, i.e., every
square commutes.

We may build a double complex from $\cI^{p,q}$ if we switch the sign
of every vertical map in (\ref{eq:Irm}) to make the squares
anticommute. So we build a single complex of injective objects
\begin{equation}
\xymatrix@1{
  \ldots\ar[r]&\cI^{n-1}\ar[r]&\cI^{n}\ar[r]&\cI^{n+1}\ar[r]&\ldots,
} \label{eq:Ising}
\end{equation}
where
\begin{equation}
  \cI^n = \bigoplus_{p+q=n} \cI^{p,q}.
\end{equation}
Applying the $\Hom(\cE,-)$ functor, we obtain
$\Ext^n(\cE,\cF^\bullet)$ as the cohomology of the induced complex
\begin{equation}
\xymatrix@1{
  \ldots\ar[r]&\Hom(\cE,\cI^{n-1})\ar[r]&\Hom(\cE,\cI^{n})\ar[r]&
     \Hom(\cE,\cI^{n+1})\ar[r]&\ldots
}
\end{equation}

Clearly the general case, $\Ext^n(\cE^\bullet,\cF^\bullet)$, must be
computed by a {\em triple\/} complex
$E_0^{p,q,s}=\Hom(\cE^p,\cI^{q,s})$.  The Hilbert space of open
strings is then given by the cohomology of this with respect to
$Q=Q_0+d^{\cE}+d^{\cF}$. 

To compute this we collapse the double complex $\cI^{q,s}$ into a
single complex $\cI^q$ as in (\ref{eq:Ising}). Now we have a double
complex given by $E_0^{p,q}=\Hom(\cE^p,\cI^q)$ from which
$\Ext^n(\cE^\bullet,\cF^\bullet)$ may be found.

Note that this Hilbert space $\Ext^n(\cE^\bullet,\cF^\bullet)$ of open
strings of ghost number $n$ from $\cE^\bullet$ to $\cF^\bullet$ occurs
commonly in homological algebra and is known as the {\em hyperext\/}
group (see chapter 10 of \cite{Wei:hom} for example).

Let us introduce the useful notion of shifting for complexes. Let
$\cF^\bullet[n]$ denote the complex $\cF^\bullet$ shifted $n$ places to the
left.\footnote{Beware! All sane people define the shift as being to
the {\em left\/} but that doesn't include everyone.} Thus, if the
$q$th position of $\cF^\bullet$ contains $\cF^q$, the $q$th position
of $\cF^\bullet[n]$ contains $\cF^{q+n}$. It is then easy to convince
oneself that
\begin{equation}
  \Ext^q(\cE^\bullet[m],\cF^\bullet[n]) =
  \Ext^{q-m+n}(\cE^\bullet,\cF^\bullet),
\end{equation}
i.e., these shift operators just change the ghost number of the B-branes.

It would seem sensible to define
$\Hom(\cE^\bullet,\cF^\bullet)=\Ext^0(\cE^\bullet,\cF^\bullet)$. Doing
this actually {\em defines\/} the category of B-branes. We know what
the objects are, namely complexes of locally-free sheaves, and now
we've defined the morphisms. We should check, of course, that the morphisms
satisfy the axioms of a category. This is not hard to do and we leave
it as an exercise for the reader.

%%%%%%%%%%%%%%%%%%%%%%%%%%%%%%

\subsection{The derived category}  \label{ss:DC}

Logically speaking, we have achieved our goal. Section \ref{ss:Bop}
completely defined a category of B-branes. Practically speaking,
however, we need to analyze the mathematical structure of this
category in order to extract useful information about it. In
particular, we would like a more intrinsic description of it.

We constructed the category of B-branes by using the right-derived
functor $\Ext$. A category in which the morphisms are obtained from the
derived functors of some other category, as above, is called a
{\em derived category\/} for obvious reasons.

The definition of the derived category proceeds as follows. We begin
with an {\em abelian\/} category $\mathcal{C}$. The derived category
of $\mathcal{C}$, denoted $\mathbf{D}(\mathcal{C})$ has objects
consisting of {\em complexes\/} of objects of $\mathcal{C}$.  We will
denote these chain complexes $\cE^\bullet$, etc., as in the last section.  If
we were being careful, we would distinguish the case where these
complexes had finite length and call it the ``bounded derived
category''. As it is, we will be sloppy and implicitly assume this
finiteness condition most of the time.

We will build up the set of morphisms in two stages.
As we saw in section \ref{ss:Bop}, a chain map is defined as a map
between complexes such that all squares commute. Given two chain maps
$f,g:\cE^\bullet\to\cF^\bullet$ we define a {\em chain homotopy\/}
from $f$ to $g$ as a set of maps $\{h_n\}$ such that we have a diagram
\begin{equation}
\xymatrix@!=16mm{
  \ldots\ar[r]^{d^{\cE}_{n-2}}&
    \cE^{n-1}\ar[r]^{d^{\cE}_{n-1}}\ar[d]^(0.3){f_{n-1},g_{n-1}}
               \ar[dl]^{h_{n-1}}&
    \cE^{n}\ar[r]^{d^{\cE}_{n}}\ar[d]^(0.3){f_n,g_n}\ar[dl]^{h_{n}}&
    \cE^{n+1}\ar[r]^{d^{\cE}_{n+1}}\ar[d]^(0.3){f_{n+1},g_{n+1}}
               \ar[dl]^{h_{n+1}}
          &\ldots\ar[dl]^{h_{n+2}}\\
  \ldots\ar[r]^{d^{\cF}_{n-2}}&
    \cF^{n-1}\ar[r]^{d^{\cF}_{n-1}}&
    \cF^{n}\ar[r]^{d^{\cF}_{n}}&
    \cF^{n+1}\ar[r]^{d^{\cF}_{n+1}}&\ldots,
}
\end{equation}
with $f_n-g_n=d^{\cF}_{n-1}h_n+h_{n+1}d^{\cE}_n$ for all $n$. 

The first set of morphisms that we include in the derived category
consists of the set of chain maps modulo chain homotopies. 

Comparing to section \ref{ss:Bop}, the reader might like to check
that, given a general map between two chains $\cE^\bullet$ and
$\cF^\bullet$, a map that is a chain map will be $Q$-closed, and that
two chain maps differing by a chain homotopy differ by something
$Q$-exact. Thus these morphisms do give {\em part\/} of the set of
morphisms in the category of B-branes.

Given a chain map $f:\cE^\bullet\to\cF^\bullet$, we induce a map
$f^n_*:\cH^n(\cE^\bullet)\to\cH^n(\cF^\bullet)$ between the {\em
cohomologies\/} of the complexes. We should emphasize that we do {\em
not\/} mean sheaf cohomology, but rather the cohomology in the sense
of abelian categories in section \ref{sss:abc}. Thus, for the category
of $\O_X$-modules, the objects $\cH^n(\cE^\bullet)$ are sheaves ---
hence the notation.

A chain map is called a {\em quasi-isomorphism\/} if the induced
morphisms $f^n_*$ are isomorphisms in the category $\mathcal{C}$ for
all $n$. If the morphism $f$ is a quasi-isomorphism, we add another morphism
$f^{-1}$ to the derived category which composes with $f$ to give the
identity. Adding in all these inverse morphisms finally constructs
the derived category $\mathbf{D}(\mathcal{C})$. Thus the derived
category looks somewhat like
\begin{equation}
\xymatrix{
A\ar@/^/[rr]^{f\approx}&&B\ar@/^/[ll]^{f^{-1}}\ar[dr]^g&&
       C\ar@/^/[dl]^{k^{-1}}\\
&D\ar[ur]_h&&E\ar@/^/[ur]^{k\approx}
}
\end{equation}
where $A,B,\ldots$ are chain complexes; $f,g,\ldots$ are equivalence
classes of chain maps modulo homotopy; and $\approx$ denotes a
quasi-isomorphism.

Adding in these inverses makes a lot of objects in
$\mathbf{D}(\mathcal{C})$ isomorphic. To be precise, any two objects
are isomorphic if the complexes are related by a sequence of
quasi-isomorphisms $\cE^\bullet_1\to\cE^\bullet_2
\leftarrow\cE^\bullet_2\leftarrow\cE^\bullet_3\to\ldots\leftarrow
\cE^\bullet_m$, where the arrows may point in either direction.

Note that any complex $\cE^\bullet$ obviously has the same cohomology
as the sequence given by the cohomology itself, i.e., the following
complex with zero morphisms:
\begin{equation}
\xymatrix@1{
  \ldots\ar[r]^-0&\cH^0(\cE^\bullet)\ar[r]^0&\cH^1(\cE^\bullet)\ar[r]^0
    &\cH^2(\cE^\bullet)\ar[r]^-0&\ldots,
}  \label{eq:ccom}
\end{equation}
However, it is not necessarily true that there is a chain map in either
direction between $\cE^\bullet$ and (\ref{eq:ccom}). Thus, in general,
a complex need not be quasi-isomorphic to its cohomology. This very
important fact leads to the complicated structure of the derived
category. 

Suppose $\mathcal{C}$ is the abelian category with objects
corresponding to complex linear vector spaces and morphisms
corresponding to linear maps. In this case, there {\em is\/} always a
quasi-isomorphism between a complex and its cohomology and thus the
derived category takes on a simple form. Every isomorphism class of
objects is determined by its cohomology. We emphasize again though
that this simplification does not happen in a more general case, such
as sheaves.

Let's see what these quasi-isomorphisms do in the B-brane
category. The first thing we should note is that the category of
locally-free sheaves is not abelian. In order to compute the
cohomology of a complex we need a larger abelian category in which
the category of locally-free sheaves is embedded. We may take this to
be $\O_X$-modules for example.

Consider the double complex $E_0^{p,q}=\cI^{p,q}$ that we constructed
in (\ref{eq:Irm}) for the complex $\cF^\bullet$. Applying the
spectral sequence construction we obtain $E_1^{p,0}=\cF^p$ and
$E_1^{p,q}=0$ for $q>0$. It follows that
$E_2^{p,0}=E_\infty^{p,0}=\cH^p(\cF^\bullet)$
and thus the total cohomology of this complex is given by
$\cH^n(\cF^\bullet)$.
Of course, the total cohomology of this double complex is, by
construction, the cohomology of the single combined complex given in
(\ref{eq:Ising}). The means that the injective resolution in
(\ref{eq:Irm}) is equivalent to the statement that
\begin{equation}
\xymatrix{
  \ldots\ar[r]&\cF^{n}\ar[r]\ar[d]&\cF^{n+1}\ar[r]\ar[d]
          &\cF^{n+2}\ar[r]\ar[d]&\ldots\\
  \ldots\ar[r]&\cI^{n}\ar[r]&\cI^{n+1}\ar[r]&\cI^{n+2}\ar[r]&\ldots
}
\end{equation}
is a quasi-isomorphism. We also say that this quasi-isomorphism
represents an injective resolution of the {\em complex\/} $\cF^\bullet$.

Now, if we have a quasi-isomorphism $\cE^\bullet\to\cF^\bullet$, we
may compose this chain map with the quasi-isomorphism (injective
resolution) $\cF^\bullet\to\cI^\bullet$ to obtain another
quasi-isomorphism $\cE^\bullet\to\cI^\bullet$ --- but this is clearly
an injective resolution again. Thus $\cI^\bullet$ represents an
injective resolution of both $\cE^\bullet$ and $\cF^\bullet$!

Referring back to all the computations in section \ref{ss:Bop}, where
we used these injective resolutions, it should now be fairly clear
that any two complexes related by a quasi-isomorphism are isomorphic
objects in the category of B-branes. If $\cE^\bullet$ and
$\cF^\bullet$ are quasi-isomorphic, the construction of
$\Hom(\cE^\bullet,\cF^\bullet)=\Ext^0(\cE^\bullet,\cF^\bullet)$
is identical to the construction of $\Hom(\cF^\bullet,\cF^\bullet)$
and thus 
contains a natural ``identity'' element, as does
$\Hom(\cF^\bullet,\cE^\bullet)$ and these elements are naturally
inverses to each other. Thus the quasi-isomorphisms are invertible ---
just like the derived category.

% Charlotte born here.

Given two objects $\cE^\bullet$ and $\cF^\bullet$ in the derived
category, how might we go about computing the set of morphisms
$\Hom(\cE^\bullet,\cF^\bullet)$? We can chase the inverted
quasi-isomorphisms as follows. Suppose we have a third object
$\cE_1^\bullet$ with the following chain maps
\begin{equation}
\xymatrix{
  \cE_1^\bullet\ar[r]^f\ar[d]^{\approx}&\cF^\bullet\\
  \cE^\bullet
}
\end{equation}
where $\approx$ denotes a quasi-isomorphism. These maps do not imply
the existence of a chain map $\cE^\bullet\to\cF^\bullet$, but in the
derived category the map $f$ {\em will\/} contribute to
$\Hom(\cE^\bullet,\cF^\bullet)$ since it may be composed with the
inverse of the quasi-isomorphism.

Thus to compute $\Hom(\cE^\bullet,\cF^\bullet)$ we need to look at
chain maps between all objects quasi-isomorphic to $\cE^\bullet$ and
$\cF^\bullet$. To actually carry this process out is hopelessly
impractical. Thankfully there is a often a better way.

Suppose the abelian category $\mathcal{C}$ is such that all objects
have an injective resolution. Thus, for any complex $\cF^\bullet$, we
have a quasi-isomorphism $\cF^\bullet\to\cI^\bullet$, where
$\cI^\bullet$ is a complex of injective objects. With a bit of effort,
one may then show (see sections 3.10 and 3.11 of \cite{GM:Hom} or
chapter 10 of \cite{Wei:hom}) that $\Hom(\cE^\bullet,\cF^\bullet)$ is
equal to the set of chain maps from $\cE^\bullet$ to $\cI^\bullet$
modulo chain homotopies.

But wait! This is {\em exactly\/} how we were computing the Hilbert
space of open strings $\Hom(\cE^\bullet,\cF^\bullet)$ in section
\ref{ss:Bop}. Thus, to be almost precise, {\bf the category of B-branes
is the derived category of locally-free sheaves.}

%%%%%%%%%%%%%%%%%%%%%%%%%%%%%%%%%%

\subsection{Coherent sheaves}  \label{ss:cohB}

So what's wrong with the last statement in the previous section? The
objects in the category of B-branes are indeed complexes of
locally-free sheaves and the morphisms are computed exactly in the
manner of the derived category. 

The only problem is that the way we defined the derived category, we
had to begin with an {\em abelian\/} category. This was necessary so
that we could take the cohomology of the complex and thus define the
notion of a quasi-isomorphism. Locally-free sheaves do not form an
abelian category since they do not contain their own cokernels. The
way we defined quasi-isomorphisms was to embed the category of
locally-free sheaves into the category of $\O_X$-modules where the
cohomology was defined.

This is only really a cosmetic problem. To be pedantic we should
replace the category of locally-free sheaves by the minimal abelian
full subcategory of $\O_X$-modules containing locally-free sheaves. In
section \ref{sss:coh} we saw that this is the category of {\em coherent
sheaves}. We have thus proven that\footnote{If we kept track of
finiteness of complexes, we would assert that it is the {\em bounded\/}
derived category.}
\begin{center}
\shabox{The category of B-branes is the
derived category of coherent sheaves $\DC(X)$.}
\end{center}
This was first conjectured by Kontsevich \cite{Kon:mir}. This proof is
an improved version of an argument in \cite{AL:DC} which, in turn, was
based on ideas by Douglas \cite{Doug:DC}.

We should emphasize that we have added nothing by going from
locally-free sheaves to coherent sheaves. On a smooth space, any
coherent sheaf $\cA$ has a locally-free resolution, i.e., an exact
sequence\footnote{The maximal length of this resolution is given by
the dimension of $X$ \cite{Hartshorne:}, and thus we know we need go
no further than $\cF^{-3}$.}
\begin{equation}
\xymatrix@1{
  0\ar[r]&\cF^{-3}\ar[r]&\cF^{-2}\ar[r]&\cF^{-1}\ar[r]&\cF^0\ar[r]&\cA\ar[r]&0,
}
\end{equation}
where each $\cF^k$ is locally free. This is nothing but a
quasi-isomorphism, $\cF^\bullet\to\cA$, between a complex of
locally-free sheaves and a coherent sheaf. Similarly, any complex of
coherent sheaves is quasi-isomorphic to a complex of locally-free
sheaves.

We saw in section \ref{sss:coh} that an example of a coherent sheaf
looked a lot like a 0-brane. Since we have now shown that all coherent
sheaves are B-branes, we will assert that it really is the 0-brane.

Suppose we have an embedding $i:S\hookrightarrow X$, and we are given
a sheaf $\cE$ on $S$. In section \ref{sss:cech} we defined a sheaf
$i_*\cE$ on $X$. This naturally embeds the set of sheaves on $S$ into
the sheaves on $X$.

One might be forgiven for thinking that, if $\cE$ is a locally-free
sheaf associated to a vector bundle $E$, then $i_*\cE$ would represent
a B-brane wrapping the cycle $S$ with vector bundle $E\to S$. It turns
out that this is not true. This may be traced to the Freed--Witten
anomaly \cite{FW:D}. To get the correct answer requires an explicit
analysis of the vertex operators in the topological field theory for
the 2-cycles and 4-cycles as was done by Katz and Sharpe
\cite{KS:Ext,Shrp:Extlect}. The sheaf $i_*\cE$ corresponds to a
``bundle'' $E\otimes K_S^{-\frac12}$, where $K_S$ is the 
canonical line bundle of $S$. Note that if $S$ does not admit a spin
structure, then $E\otimes K_S^{-\frac12}$ is a ``twisted bundle'' in
the sense that its first Chern class is not integral. This is in
agreement with \cite{FW:D}.

To recap, we only needed to consider 6-branes in order to find the
correct category for all B-branes. That said, the precise
identification of which sheaves correspond to 2-branes and 4-branes
requires the further analysis of \cite{KS:Ext,Shrp:Extlect}.

It should be noted that there are many many more coherent sheaves on
$X$ than these wrapped branes $i_*\cE$. Indeed, the derived category
$\DC(X)$ itself is a vast thing encompassing a good deal more than one would
expect for B-branes. This is because we have yet to analyze the {\em
stability\/} of the B-branes --- something the B-model knows nothing
about. A physical D-brane in the untwisted theory will only correspond
to stable objects in some sense and this condition will rule out the
vast majority of objects in $\DC(X)$.

In this proof of B-branes being described by the derived category
$\DC(X)$ we should note we assumed that B-branes really are described
by a category. In particular, we assumed that two B-branes which are
isomorphic in the category are the ``same'' B-brane. It is probably a
deep philosophical question as to when two abstractly-defined D-branes
are the ``same'' in a strict sense. All we can say is that, within the
language of the data of topological field theory, two B-branes which
are isomorphic in $\DC(X)$ are indistinguishable. If someone wishes to
add extra data beyond the topological field theory, then it could be
that two quasi-isomorphic complexes represent different D-branes.

%%%%%%%%%%%%%%%

\subsection{More deformations}  \label{ss:mdefs}

In section \ref{ss:Bdef} we only considered deformations arising from
$\Hom(\cE^n,\cE^{n+1})$. The obvious question to ask is whether there
are any more deformations which can take us outside the derived category.

For example, we could turn on some open strings corresponding to
$g_n\in\Ext^3(\cE^n,\cE^{n-2})$ to produce a more complicated
``complex'':
\begin{equation}
\xymatrix@1{
  \ldots\ar[r]^{d_{n-1}}
  &\cE^n\ar[r]^{d_n}&
  \cE^{n+1}\ar[r]^{d_{n+1}}\ar@/^7mm/[ll]_{g_{n+1}}&
   \cE^{n+2}\ar[r]^{d_{n+2}}\ar@/_7mm/[ll]_{g_{n+2}}&
   \ldots\ar@/^7mm/[ll]_{g_{n+3}}
} \label{eq:Ecx2}
\end{equation}
This actually produces nothing new. To see this first replace the
complex $\cE^\bullet$ by a quasi-isomorphic complex $\cI^\bullet$ of injective
sheaves. Now use the definition of $\Ext$ in section \ref{sss:shco}
and we see that the strings $g_n$ are converted into maps
$g_n:\cI^n\to\cI^{n+1}$ returning us to the case considered in section
\ref{ss:Bdef}. 

About the most general deformation we may consider is as
follows. Suppose we have two D-branes given by complexes $\cE^\bullet$
and $\cF^\bullet$. Assuming the ghost numbers of the components were
not affected by turning on the differentials, an open string
corresponding to $f\in\Hom(\cE^\bullet[-1],\cF^\bullet)$ will have ghost
number one. Thus we may consider a deformation given by $f$. It is
easy to see that this produces a new combined complex:
\begin{equation}
\xymatrix@C+10mm{
\ldots\ar[r]&{\begin{matrix}\cE^{-1}\\ \oplus\\\cF^{-1}\end{matrix}}
\ar[r]^{\left(\begin{smallmatrix}d^\cE&0\\f&d^\cF\end{smallmatrix}\right)}
&{\begin{matrix}\cE^0\\ \oplus\\\cF^0\end{matrix}}
\ar[r]^{\left(\begin{smallmatrix}d^\cE&0\\f&d^\cF\end{smallmatrix}\right)}
&{\begin{matrix}\cE^1\\ \oplus\\\cF^1\end{matrix}}\ar[r]
&\ldots
} \label{eq:Cone}
\end{equation}
This construction is well-known in the context of the derived category
and is known as the {\em mapping cone\/} of $f$. We refer to
\cite{Thom:DCg} for a nice account of why it has this name. We denote
the new complex in (\ref{eq:Cone}) as
$\Cone(f:\cE^\bullet[-1]\to\cF^\bullet)$ or just $\Cone(f)$.
  
The cone construction encompasses almost all the deformations we can
consider. For example, a complex itself can be considered an iterated
cone:
\begin{equation}
  \cE^\bullet = \ldots\Cone(d_2:\Cone(d_1:\Cone(d_0:\cE^0\to\cE^1)
   \to\cE^2)\to\cE^3)\ldots,
\end{equation}
where we think of a sheaf as a complex with a single entry.

The only exception to this rule is the case of deformations given by
$\Ext^1(\cE^n,\cE^n)$. Adding such a vertex operator to the action
simply deforms $\cE^n$ itself. This is not quite the same thing as
forming $\Cone(f:\cE^\bullet\to\cE^\bullet[1])$ in the derived
category although the concepts are very closely related.  In the
latter case we are turning on a string between $\cE^\bullet$ and a
second copy of this D-brane whereas in the former case there was an
open string beginning and ending on the same D-Brane. What
$\Cone(f:\cE^\bullet\to\cE^\bullet[1])$ actually represents is a
family of infinitesimal deformations of $\cE^\bullet$ rather than the
deformed $\cE^\bullet$. We refer to \cite{MK:cxman} for more details
on the theory of deformations.

There is an interesting feature of the deformations we are considering
which is worth discussing. Suppose we turn on a nonzero map
$f:\cE^0\to\cE^1$. Clearly we can rescale this map by a nonzero number
$c\in\C$. There is now a quasi-isomorphism
\begin{equation}
\xymatrix{
\ldots\ar[r]&0\ar[r]&\cE^0\ar[r]^f\ar[d]^1&
   \cE^1\ar[r]\ar[d]^c&0\ar[r]&\ldots\\
\ldots\ar[r]&0\ar[r]&\cE^0\ar[r]^{cf}&\cE^1\ar[r]&0\ar[r]&\ldots,
}
\end{equation}
and so the deformations $f$ and $cf$ represent the same
B-brane.\footnote{Note that this is not true if $c=0$.} We
may use this feature to justify our assumption that the maps $d$ in
section \ref{ss:Bdef} were infinitesimal since their scale doesn't
matter at all! 

Thus, as the deformation $f$ is turned on from zero, we suddenly
obtain the new D-brane and any increase in $f$ makes no
difference. Such discontinuous behaviour is common in algebraic
geometry and therefore it should come as no surprise that B-brane
exhibit such behaviour. In the untwisted theory one might expect more
continuous behaviour and in section \ref{sss:Pi} we will see that this is so.

%%%%%%%%%%%%%%%

\subsection{Anti-branes and K-Theory}  \label{ss:K}

One of the key steps in arriving at the derived category picture was
associating a ghost number with each B-brane in section
\ref{ss:Bdef}. How physical is this? That is, is there much of a
difference between a B-brane associated to the complex $\cE^\bullet$
and a B-brane associated to the shifted complex $\cE^\bullet[n]$ for
some $n$?

Relative shifts certainly matter. We have
\begin{equation}
  \Hom(\cE^\bullet,\cF^\bullet) \neq
  \Hom(\cE^\bullet,\cF^\bullet[n]),
\end{equation}
for generic $\cE^\bullet,\cF^\bullet$ and nonzero $n$.

If we shift all the complexes by the same $n$ then there is no
change in the physics. That is,
\begin{equation}
  \Hom(\cE^\bullet,\cF^\bullet) =
  \Hom(\cE^\bullet[n],\cF^\bullet[n]),
\end{equation}
and there is no change in any of the operator products. Thus, it looks
like there is a gauge symmetry of the theory associated to a global
shift of the complexes by any integer.

While this is essentially correct, there is a subtlety used in the
language of D-branes that makes it preferable to state the gauge
symmetry in a different way. Consider a complex as follows:
\begin{equation}
\xymatrix@1{
  \ldots\ar[r]&0\ar[r]&\cE\ar[r]^c&\cE\ar[r]&0\ar[r]&\ldots,
} \label{eq:cancel}
\end{equation}
where the nontrivial map is given by multiplication by $c\in\C$.
If $c\neq0$, the complex (\ref{eq:cancel}) is quasi-isomorphic to
zero. That is, the two $\cE$'s in (\ref{eq:cancel}) cancel
out. In other words, the $\cE$ on the left is the ``anti-brane'' of
the $\cE$ on the right. Turning on $c$ must represent a ``tachyon
condensate'' in the sense of Sen \cite{Sen:dbd} which performs the
cancellation. We will have much more to say about such tachyons in
section \ref{sss:tachy} and \ref{sss:Pi}.

The generalization of this cancellation is that the mapping cone of
the identity map $\Cone(\id:\cE^\bullet\to\cE^\bullet)$ is
quasi-isomorphic to zero for any complex $\cE^\bullet$. From section
\ref{ss:mdefs} it follows that $\cE^\bullet[1]$ represents the
anti-brane to $\cE^\bullet$. The gauge symmetry is therefore stated as
follows \cite{Doug:DC}
\begin{center}
\shabox{\parbox{.85\hsize}{The B-model is subject to a gauge symmetry
generated by simultaneously shifting {\em all\/} the B-brane complexes one
place to the right (or left) and exchanging the notion of D-brane and
anti-D-brane.}}
\end{center}

We should warn that this anti-brane language is a little crude
and can lead to misleading statements. For example $\cE^\bullet[1]$ is
the anti-brane to $\cE^\bullet$, and $\cE^\bullet[2]$ is the
anti-brane to $\cE^\bullet[1]$ in the above sense. It does {\em not\/}
follow that $\cE^\bullet$ and $\cE^\bullet[2]$ are the same D-brane,
however, since $\Hom(\cE^\bullet,\cF^\bullet)$ is not generically
equal to $\Hom(\cE^\bullet[2],\cF^\bullet)$.

This gauge symmetry means that any intrinsic physical property
associated to a D-brane $\cE^\bullet$ is also given to
$\cE^\bullet[n]$ for any $n$. This would include mass, stability (to
be discussed later), etc. Thus, if a particular D-brane $\cE^\bullet$
becomes massless at a given point in moduli space, all the D-branes
$\cE^\bullet[n]$ become massless. However, it does not mean that an
infinite number of D-branes has become massless in a physically
meaningful way, since all these D-branes are gauge equivalent. For
counting purposes the collection $\{\cE^\bullet[n]:n\in\Z\}$ is one
D-brane!

The fact that D-brane/anti-D-brane annihilation is built into the
derived category descriptions means that we can map the derived
category to Witten's K-theory language for D-branes \cite{W:K}. To do
this we basically disregard all the information contained in the
morphisms. We saw above that we could deform two D-branes
$\cE^\bullet[-1]$ and $\cF^\bullet$ into a single D-brane represented
by the cone of a morphism $f:\cE^\bullet[-1]\to\cF^\bullet$. 
Thus the B-brane $\Cone(f)$ is composed of $\cE^\bullet[-1]$ and
$\cF^\bullet$, where $\cE^\bullet[-1]$ is an anti-$\cE^\bullet$.
We may therefore assert that
\begin{equation}
\begin{split}
  [\Cone(f)] &= [\cF^\bullet] + [\cE^\bullet[-1]] \\
     &= [\cF^\bullet] - [\cE^\bullet],
\end{split} \label{eq:ConG}
\end{equation}
where $[\quad]$ represents some kind of ``class'' of a
D-brane. We may define an abelian group $\cK(X)$ which is generated by all the
objects in $\DC(X)$ and we divide out by all relationships of the form 
(\ref{eq:ConG}) for all possible mapping cones. This group $\cK(X)$ is called
the ``Grothendieck group'' of $X$ (see page 77 of \cite{KS:shvs} for
more details). The Grothendieck group was also discussed in
\cite{Shrp:DC} in the context of D-branes.

We may naturally map the derived category to K-theory as
follows. Using locally-free resolutions we may replace any complex by
a quasi-isomorphic complex $\cE^\bullet$ of locally-free sheaves. We
may then construct the K-theory object
\begin{equation}
  \ldots\ominus E^{-1}\oplus E^0\ominus E^1\oplus E^2\ominus\ldots,
\end{equation}
where $E^i$ is the holomorphic vector bundle associated to $\cE^i$.
One can show that this leads to a well-defined map $\cK(X)\to
K(X)$. Note that this map need not be surjective. $K(X)$ is generated
by all vector bundles whereas we have restricted attention to
holomorphic vector bundles. This is because we have focused only on
B-branes, which are essentially BPS. The full K-theory might require
some non-BPS branes in order to generate all possible classes.

Anyway, we should emphasize that K-theory contains much less
information than the derived category. For example, all 0-branes on
$X$ would be represented by the same K-theory element. In contrast,
two 0-branes corresponding to distinct points in $X$ are associated to
non-isomorphic objects in $\DC(X)$. We like to think of K-theory as a
``poor man's derived category'' that knows only about D-brane charge.

A more precise notion of D-brane charge may be defined from the
world-volume of the D-brane. This may be computed by anomaly inflow
arguments following \cite{GHM:inf,CY:1/2,MM:K}. This subject is
covered by Harvey's lectures at this TASI meeting and so we may refer
to \cite{Harv:anom} and be brief here. A D-brane corresponding to a
vector bundle $E$ is given a charge
\begin{equation}
  Q(E) = \ch(E)\sqrt{\td(X)},
    \label{eq:QD}
\end{equation}
where $\ch(E)$ is the Chern character of $E$ and $\td(X)$ is the Todd
class of the tangent bundle of $X$. Note that $Q(E)$ is
an element of $H^{\textrm{even}}(X,\Q)$. It follows from above that
this extends to the derived category by
\begin{equation}
  Q(\cE^\bullet) = \ch(\cE^\bullet)\sqrt{\td(X)},
\end{equation}
where
\begin{equation}
  \ch(\cE^\bullet) = \ldots-\ch(E^{-1})+\ch(E^0)-\ch(E^1)+\ch(E^2)-\ldots
\end{equation}

In section \ref{ss:cohB} we argued that a D-brane wrapped on $S$ was
given by a coherent sheaf $i_*\cE$.  The charge of such a D-brane can
be computed using the {\em Grothendieck--Riemann--Roch theorem}.  In
the special case that we have an embedding $i:S\to X$, this asserts
that
\begin{equation}
  \ch(i_*\cE)\td(X) = i_!(\ch(\cE)\td(S)),
     \label{eq:GRR}
\end{equation}
where $i_!$ is defined on cohomology as $P\cdot i_*\cdot P^{-1}$,
where $P$ is Poincar\'e duality and $i_*$ in this latter context is
the natural map induced by $i$ on homology. It follows that, for any
$C\in H^{\textrm{even}}(X)$, we have the following
formula
\begin{equation}
  \int_X C\cdot Q(i_*\cE) = \int_S\ch(\cE)
      \sqrt{\frac{\td(S)}{\td(N)}}\cdot i^*C,
      \label{eq:sub0}
\end{equation}
where $N$ is the normal bundle of $S$ in $X$. 

That said, in section \ref{ss:cohB} we also saw that $i_*\cE$
corresponds to a B-brane given by a (twisted) bundle $E'=E\otimes
K_S^{-\frac12}$ over $S$, where $E$ is the bundle associated to
$\cE$. Using the relation $\td=\exp(\ff12c_1)\hat A$ and the fact that
$c_1(X)=0$, we may therefore rewrite (\ref{eq:sub0}) as
\begin{equation}
  \int_X C\cdot Q(i_*\cE) = \int_S\ch(E')
      \sqrt{\frac{\hat A(S)}{\hat A(N)}}\cdot i^*C,
      \label{eq:sub}
\end{equation}
which is the formula one would arrive at via anomaly consideration
\cite{Harv:anom}.

%%%%%%%%%%%%%%%

\subsection{Mirror symmetry restored?}  \label{ss:m2}

If the A-model on $Y$ is ``the same'' as the B-model on $X$, then it
would appear that we have motivated the proposal that the Fukaya
category on $Y$ is equivalent to $\DC(X)$, the derived category on
$X$. That was Kontsevich's original proposal. It turns out that there
is still a small fly in the ointment, as we discuss in section
\ref{sss:mirN}, but this proposal seems to be very close to the truth.
In this section we note a few miscellaneous features of how well this
mirror symmetry works.

The D-brane charge of an A-brane is simply given by its homology class
in $[L_i]\in H_3$ multiplied by the rank of the bundle over the
Lagrangian. Let us denote the Poincar\'e dual of $[L_i]$ by $l_i\in H^3(X)$.
Note that we have a natural symplectic inner product on
these charges given by the oriented intersection number $\#([L_1]\cap
[L_2])$. It can be shown \cite{Flr:mors} that the orientations of the
points of intersection in figure \ref{f:tun} are opposite if the
difference in their ghost numbers is odd. Thus the intersection number
is given by the Euler characteristic of the complex
(\ref{eq:Flcx}). That is,
\begin{equation}
\begin{split}
  \#([L_1]\cap[L_2]) &= \int_Y l_1\cdot l_2\\
         &= \sum_i(-1)^i\dim\Hom^i(L_1,L_2).
\end{split}    \label{eq:Aint}
\end{equation}

If the Lagrangian $L_i$ is mirror to a complex $\cE_i^\bullet$ then
the right-hand-side of (\ref{eq:Aint}) is clearly mirror to the
alternating sum of $\dim\Ext^i(\cE_1^\bullet,\cE_2^\bullet)$. The
Hirzebruch--Riemann-Roch theorem says that this is given by
\begin{equation}
\begin{split}
  \sum_i(-1)^i\dim\Ext^i(\cE_1^\bullet,\cE_2^\bullet)
      &= \int_X \ch(\cE_1^\bullet)^\vee \cdot \ch(\cE_2^\bullet)
           \cdot\td(X)\\
   &= \int_X Q(\cE_1^\bullet)^\vee \cdot Q(\cE_2^\bullet),
\end{split} \label{eq:Bint}
\end{equation}
where, if $\omega$ is a $2p$-form, then
$\omega^\vee=(-1)^p\omega$. Thus we see a very nice agreement for
the pairing between A-brane charges and B-branes charges. Note that
the ``${}^\vee$'' is necessary in (\ref{eq:Bint}) to get a symplectic
inner product.

The tadpole cancellation condition in section \ref{sss:Aopen2}
produces two interesting aspects of the moduli space of A-branes:
\begin{enumerate}
\item First-order deformations of the Lagrangian, which correspond to
$H^1(L)$, may be obstructed and do not lead to genuine A-brane deformations.
\item Some A-branes may depend on very special values for $B+iJ$ and
disappear completely for generic $B+iJ$.
\end{enumerate}
The mirror statements in the B-model are both true:
\begin{enumerate}
\item The first-order deformations of coherent sheaves, which
correspond to $\Ext^1(\cE,\cE)$ can be obstructed. We refer to
\cite{Thom:obs} for examples.
\item There are some sheaves which only exist for special values of
complex structure. An example of this is given by 2-branes wrapped
around an algebraic curve of high genus in $X$ \cite{Wil:Kc}.
\end{enumerate}
This subject was also analyzed in \cite{KKLM:W,KKLM:disk}. Note that
this is a typical example of mirror symmetry in that instanton effects
(i.e., tadpoles) in the A-model are mapped to effects in the B-model
that can be understood from classical geometry.

\begin{table}
\renewcommand{\arraystretch}{1.3}
\begin{tabular}{|l||l|l|}
  \hline
  &A-model&B-model\\
  \hline
  Geometry&Symplectic (no complex structure)&Algebraic (no metric)\\
  Category&Fukaya category&Derived category\\
  D-branes&Lagrangians&Complexes of coherent sheaves\\
  Open strings&Floer cohomology&$\Ext$'s\\
  Dependence&$B+iJ$&complex structure\\
  Charges&$l_i\in H^3$&$\ch(\cE)\sqrt{\td(X)}\in H^{\textrm{even}}(X)$
  or $K(X)$\\
  \hline
\end{tabular}
\caption{Mirror symmetry for A-branes and B-branes.}
\label{tab:mir1}
\end{table}

In table \ref{tab:mir1} we review the picture of mirror symmetry that
we have obtained so far. The reader might be a little disappointed to
note that we haven't actually used the more exotic elements of the
derived category in this discussion of mirror symmetry --- everything
was done for coherent sheaves. We will give a very explicit example
that requires a nontrivial complex in section \ref{sss:qex}.

%%%%%%%%%%%%%%%%%%%%%%%%%%%%%%%%%%%%%%%%%%%%%%%%%%%%%%%%%%%%%%%%%%

\section{Stability}   \label{s:stab}

So far we have dealt with D-branes in the context of topological field
theory. This was sufficient to understand the origins of the Fukaya
category in the case of A-branes and the derived category in the case
of B-branes. In the untwisted theory it is the D-branes that
correspond to BPS states that descend to D-branes in the topological
field theory. Having said that, the BPS condition is stronger than
that imposed on branes in the topological field theories. In order for
an A-brane or a B-brane to correspond to a BPS state in the untwisted
theory we need to impose a further condition --- namely ``stability''.

The purpose of studying stability is two-fold. As we just said, in
order to make contact with the ``real world'', i.e., the untwisted
theory, a D-brane must be stable. In addition, stability makes us
study a mathematical structure on the categories of D-branes, i.e.,
``distinguished triangles'', that provides further insight into the
intrinsic structure of the D-brane categories.

Stability of D-Branes was also studied in \cite{Dnf:Dstab,DGR:Dquin}
using a quite different method than we employ here.

%%%%%%%%%%%%%%%%

\subsection{A-Branes}  \label{ss:Astab}

\subsubsection{Special Lagrangians}  \label{sss:sLag}

The spacetime supersymmetry arises from the spectral flow operators
discussed at the end of section \ref{ss:nlsm}. These are associated
with the holomorphic 3-form $\Omega$ on $X$ as in (\ref{eq:OmSF}). The
$N=2$ spacetime supersymmetry arises because we have a spectral flow
operator in the left-moving and right-moving sector. The boundary
conditions on the open string destroys the independence of these
sectors and the best we can do is to preserve an $N=1$ supersymmetry. To
do this we can set
\begin{equation}
  \Sigma = \exp(-i\pi\xi)\bar\Sigma,
\end{equation}
on the ends of the string, where $0\leq\xi<2$. The parameter $\xi$
measures ``which'' $N=1$ spacetime supersymmetry is preserved from the
original $N=2$.\footnote{$N$ spacetime supersymmetries give a
$\GU(N)$ R-symmetry. One might therefore expect a $\GU(2)/\GU(1)$
choice of $N=1$ supersymmetries in $N=2$. However, the spectral flow picture of
$N=2$ spacetime supersymmetry only sees a $\GU(1)\times\GU(1)$
subgroup of the R-symmetry so the parameter only lives in
$\GU(1)$.}

This boundary conditions given by $R^i_{\bar\jmath}$ in section
\ref{sss:Ab} imply that this is given by
\begin{equation}
  \Omega|_L = \exp(-2i\pi\xi)\bar\Omega|_L,
\end{equation}
on the A-brane $L$. In section \ref{sss:Ab} we noted that $\Omega|_L$ was
equivalent to the real volume form on $L$ up to some complex
constant. This means that our choice of the real parameter $\xi$
coincides with that of (\ref{eq:ph1}). That is,
\begin{equation}
  \xi = \frac1\pi\arg\frac{\Omega|_L}{dV_L}.
    \label{eq:xiA1}
\end{equation}

The key issue is that the parameter $\xi$ must be the same at all
points on the Lagrangian $L$ in order for the same spacetime
supersymmetry to be preserved everywhere. A Lagrangian for which $\xi$
is a constant is called a {\em special Lagrangian}. Thus it is the
special Lagrangians which correspond to BPS states as first observed
in \cite{BBS:5b}.

Note that if $\xi$ is a constant, then we may rewrite (\ref{eq:xiA1})
as
\begin{equation}
  \xi = \frac1\pi\arg\int_L\Omega.
\end{equation}
Note also that $\Omega$ is only defined up to a complex constant so
the value of $\xi$ might appear somewhat meaningless. Indeed, the
standard definition of a special Lagrangian is to put $\xi=0$ and so
assert that the real part of $\Omega|_L$ is zero. We will need the
idea of comparing values of $\xi$ between different special
Lagrangians and so we retain the notion here, although one should
always bear in mind that only relative values of $\xi$ have any
meaning. 

In section \ref{sss:Aopen1} we gave a very specific definition of an
A-brane. In addition to being a Lagrangian it had to satisfy two extra
condition. It should be obvious that the map $\xi_*$ in
(\ref{eq:Maz1}) is trivial and thus the Maslov class condition is
automatically satisfied for a special Lagrangian. The second condition
concerned the tadpole cancellation. This, in general, is not
automatically satisfied in the special Lagrangian case and so remains
an extra condition to be imposed.

It is very easy to motivate the idea that special Lagrangians are BPS
states. A special Lagrangian is a {\em calibrated submanifold\/} in
the sense of Harvey and Lawson \cite{HL:calib}. The details of this
definition need not concern us but the useful fact is that any
calibrated submanifold automatically minimizes the volume of any
manifold in its homology class. Thus, if we think of a D-brane as
some kind of membrane with a tension, a D-brane that wraps a special
Lagrangian submanifold is clearly stable. Note that there is no reason
to suppose that all minimal 3-manifolds in a \CY\ are special
Lagrangians, reflecting the fact that not all stable D-branes are
necessarily BPS.

\subsubsection{A geometrical decay} \label{sss:Adec}

Tadpoles aside, in section \ref{sss:Aopen1} we saw that an A-brane
(with a line bundle) has a moduli space given by $H^1(L)$. It can be
shown \cite{McL:sLag} that the moduli space of special Lagrangians
is also given by $H^1(L)$. Thus, locally, the moduli space of special
Lagrangians agrees with the moduli space of Lagrangians modulo
Hamiltonian deformation. At first sight, this might suggest that in
each equivalence class of Lagrangians modulo Hamiltonian deformation
there is a unique special Lagrangian.

This is not actually true. It turns out the vast majority of
Lagrangians have no special Lagrangian equivalent to them by
Hamiltonian deformation. From our perspective, the best way to see
this is to consider how special Lagrangians can ``disappear'', or
``decay'', as the complex structure of the target space $Y$ is
deformed. Note that a Lagrangian submanifold is defined purely in
terms of the symplectic structure of $Y$ induced by the K\"ahler form
and so has no dependence on the complex structure. Adding the
``special'' in special Lagrangian does introduce a dependence on the
complex structure.

A quite explicit picture for the decay of special Lagrangians was
given by Joyce \cite{Jc:VsLag} which we follow here. %pg39

Let us first consider special Lagrangian planes $\R^m\subset\C^m$. We
may specify such a plane by
\begin{equation}
  \Pi^\phi = \{(e^{i\phi_1}x_1,e^{i\phi_2}x_2,\ldots,e^{i\phi_m}x_m):
       x_j\in\R\}.
\end{equation}
This plane is determined by the real numbers $\phi_j$. Using the
standard holomorphic $m$-form $\Omega=dz_1\wedge
dz_2\wedge\ldots\wedge dz_m$ we obtain
\begin{equation}
\begin{split}
  \xi(\Pi^\phi)&=\frac1\pi\arg\int_{\Pi^\phi} \Omega \\
   &= \frac1\pi\sum_{j=1}^m \phi_j\pmod2.
\end{split} \label{eq:ximod}
\end{equation}

If we reverse the orientation of $\Pi^\phi$ we
shift $\xi(\Pi^\phi)$ by one. Such a reversal of orientation may be
viewed as replacing an A-brane by an anti-A-brane. If we forget about
the orientation we are free to restrict the $\phi_j$'s to the range
$0\leq\phi_j<\pi$. In this case we have
\begin{equation}
  \sum_{j=1}^m \phi_j = k\pi,
\end{equation}
for $0\leq k<m$. We say such an intersection of planes is of {\em
type\/} $k$.

Let us denote by $\Pi^0$ the plane for which
$\phi_1=\phi_2=\ldots=\phi_m=0$. Now consider two D-branes
intersecting transversely at the origin in the form
$\Pi^0\cup\Pi^\phi$. The transverse condition amounts to $\phi_j>0$
for all $j$.  One may therefore interpret these two D-branes as {\em
one\/} singular special Lagrangian so long as the type of the intersection
is integral. The fact that such a configuration is a BPS state
was first noted in \cite{BDL:angles}

If the type of the intersection is 1, this singular A-brane can be written
as a limit in a family of smooth special Lagrangians following Lawlor
\cite{Law:neck}. A smooth member of this family is called a ``Lawlor
neck'' and can be described as follows \cite{Harv:calib,Jc:VsLag}. Let
\begin{equation}
  P(x) = \frac{\prod_{j=1}^m(1+a_jx^2)-1}{x^2}.
\end{equation}
Now fix a positive real number $A$. The positive real numbers
$a_1,\ldots,a_m$ are then implicitly and uniquely determined by $A$
and $\phi_1,\phi_2,\ldots,\phi_m$ by the following equations:
\begin{equation}
\begin{split}
  \phi_j &= a_j\int_{-\infty}^\infty\frac{dx}{(1+a_jx^2)\sqrt{P(x)}}\\
  A &= \frac{\omega_m}{\sqrt{a_1\cdots a_m}},
\end{split}
\end{equation}
where $\omega_m$ is the volume of a unit sphere in $\R^m$.  These
equations impose the condition $\sum\phi_j=\pi$, i.e., type 1. Note
that if we include orientations, since the type is odd, we would say
that we have unbroken spacetime supersymmetry if one of the planes is
viewed as a D-brane and the other plane is viewed as an anti-D-brane.

Now define functions $\eta_j:\R\to\C$ by
\begin{equation} 
  \eta_j(y)=\exp\left(ia_j\int_{-\infty}^y\frac{dx}{(1+a_jx^2)\sqrt{P(x)}}
     \right)\sqrt{\frac{1}{a_j}+y^2}.
\end{equation}
This allows the Lawlor neck to be defined as
\begin{equation}
  L^{\phi,A}=\{(\eta_1(y)x_1,\eta_2(y)x_2,\ldots,\eta_m(y)x_m):
    y\in\R, x_j\in\R, x_1^2+\ldots+x_m^2=1\}.
\end{equation}
One can then show that this is a smooth special Lagrangian submanifold
of $\C^m$ which approaches $\Pi^0\cup\Pi^\phi$ as $A\to 0$. This
submanifold also asymptotically approaches $\Pi^0\cup\Pi^\phi$ as one
moves far from the origin. For the precise analytical details of this
we refer again to \cite{Jc:VsLag}. Topologically this space looks like
a cylinder $S^{m-1}\times\R$. It is impossible to sketch this space
completely realistically since, for $m=1$, the condition
$\sum\phi_j=0$ makes the case trivial, and for $m>1$ we are in at
least four dimensions. The case $m=2$ is shown roughly in figure
\ref{f:ln}.

\iffigs
\begin{figure}
  \centerline{\epsfxsize=8cm\epsfbox{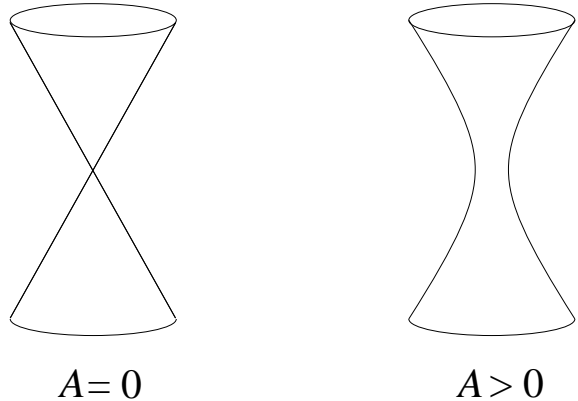}}
  \caption{The Lawlor Neck.}
  \label{f:ln}
\end{figure}
\fi

Now suppose we have a \CY\ $m$-fold $Y$ containing two A-branes $L_1$
and $L_2$ which intersect transversely at a point $p$. We may use a
$\GU(m)$ transformation to rotate the tangent plane of $L_1$ into the
standard plane $\Pi^0$ as above. This same $\GU(m)$ matrix rotates the
tangent plane of $L_2$ into $\Pi^\phi$ thus defining a type for the
intersection as above.  Again assume that the type of intersection is
equal to 1.

Let us consider a small deformation of complex structure of $Y$. Joyce
\cite{Jc:VsLag} proved that, close to the point of intersection of
$L_1$ and $L_2$, the local geometry could produce a Lawlor neck to
smooth out the neighbourhood of the point of intersection. Let us now
spell out in details exactly what happens.

Suppose $L_1$ and $L_2$ are smooth submanifolds of $Y$. As we deform
the complex structure of $Y$ we would like to find special
Lagrangian submanifolds of the deformed $Y$ which are the deformed
versions of $L_1$ and $L_2$. That is, we would like to be able to
follow $L_1$ and $L_2$ as we deform $Y$ without these D-branes
disappearing for some reason. As long as we consider
sufficiently small deformations, we are guaranteed to be able to do
this \cite{Mrsh:sLag,Jc:VsLag}. This means that we may define
$\xi(L_1)$ and $\xi(L_2)$ as real numbers which vary continuously over
the moduli space for small deformations of $Y$. To be precise, fix the mod 2
ambiguity in $\xi(L_1)$ arbitrarily, then we set $\xi(L_2)=\xi(L_1)+1$
to reflect the fact that the intersection is type 1. Now $\xi(L_1)$
and $\xi(L_2)$ are defined over the moduli space of $Y$ at least in
some neighbourhood of the starting point. We will refer to the value
of $\xi$ defined in $\R$ as the ``grading'' of a special Lagrangian.

A wall divides the moduli space into $\cM^+$
and $\cM^-$ corresponding to the sign of $\xi(L_1)-\xi(L_2)+1$. We
begin at a point in the wall. As we deform the complex structure to
the $\cM^+$ side of this wall, the point of intersection is smoothed
out by a Lawlor neck for $A>0$. On the other side, $\cM^-$, no
smoothing occurs.

On the $\cM^+$ side of the wall we use the notation $L_1\csum L_2$ to
denote the smoothed new special Lagrangian. Very close to the wall,
$L_1\csum L_2$ looks asymptotically like $L_1\cup L_2$ away from the
point of intersection and the geometry near the point of intersection
is replaced by a Lawlor neck.  The notation is intentionally
asymmetric since $L_1\csum L_2$ is quite different from $L_2\csum
L_1$.  Note that away from the wall, $L_1\cup L_2$ is no longer a
special Lagrangian since the two components have a different value of
grading $\xi$. The smooth space $L_1\csum L_2$ is the smooth space homological
to $L_1\cup L_2$ which minimizes the volume, i.e., energy of a D-brane
wrapped around these cycles. It follows that
\begin{equation}
  \left|\int_{L_1\csum L_2}\!\!\!\!\!\Omega\right|
   < \left|\int_{L_1}\Omega\right| 
   + \left|\int_{L_2}\Omega\right|,
\end{equation}
and we may choose
\begin{equation}
  \xi(L_2) < \xi(L_1\csum L_2) < \xi(L_1)+1. 
         \label{eq:xid1}
\end{equation}
 
In $\cM^-$ there is no smooth special Lagrangian minimizing the volume
of $L_1\cup L_2$ and $L_1\cup L_2$ itself is not a BPS state. Thus 
spacetime supersymmetry is broken in $\cM^-$.

\iffigs
\begin{figure}
\begin{tabular}{|m{9mm}|m{38mm}m{55mm}m{40mm}|}
  \hline
  $\cM_+$&
  \begin{xy}
     \xyimport(100,100){\epsfxsize=20mm\epsfbox{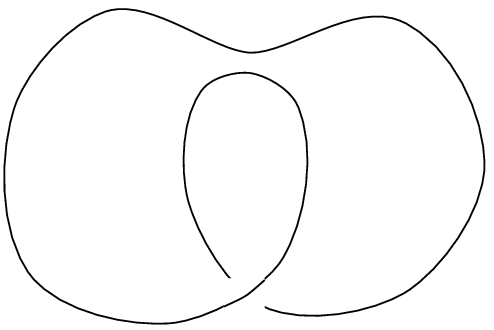}}
     !CU(1.5)!CL(1.2)
     *\frm{}
     \POS(130,20)*{L_1\csum L_2}
  \end{xy}&
     $\xi(L_2)<\xi(L_1\csum L_2) < \xi(L_1)+1$&
     $L_1\csum L_2$ is stable.\\
  \hline
  Wall&
  \begin{xy}
     \xyimport(100,100){\epsfxsize=20mm\epsfbox{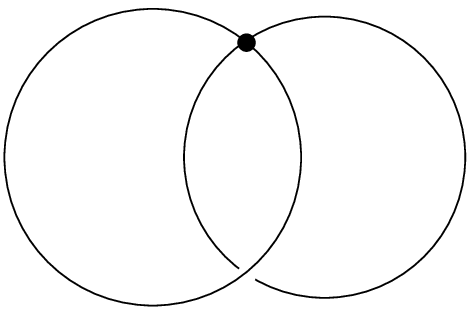}}
     !CU(1.5)
     *\frm{}
     \POS(-10,30)*{L_1}
     \POS(110,80)*{L_2}
  \end{xy}&
     $\xi(L_2)=\xi(L_1\csum L_2) = \xi(L_1)+1$&
     $L_1\csum L_2$ is marginally stable.\\
  \hline
  $\cM_-$&
  \begin{xy}
     \xyimport(100,100){\epsfxsize=20mm\epsfbox{dlect-fda.ps}}
     !CU(1.5)
     *\frm{}
     \POS(-10,30)*{L_1}
     \POS(110,80)*{L_2}
  \end{xy}&
     $\xi(L_2)> \xi(L_1)+1$&
     $L_1\csum L_2$ is unstable.\\
  \hline
\end{tabular}
  \caption{A-Brane Decay.}
  \label{f:Adecay}
\end{figure}
\fi

To recap, in $\cM^+$ we have a BPS state $L_1\csum L_2$. We also have
BPS states $L_1$ and $L_2$ but the mass of $L_1\csum L_2$ is less than
the sum of the masses of $L_1$ and $L_2$. As we hit the wall,
$L_1\csum L_2$ becomes $L_1\cup L_2$. Beyond the wall in $\cM^-$ we
only have BPS states $L_1$ and $L_2$ which together break
supersymmetry. What we have just described is a {\em decay\/} of a BPS
state $L_1\csum L_2$ into its products $L_1$ and $L_2$ as we pass from
$\cM^+$ into $\cM^-$. In $\cM^+$ we view $L_1\csum L_2$ as a {\em
bound state\/} of $L_1$ and $L_2$. We depict this story in figure
\ref{f:Adecay}. 

This is familiar from the standard properties of BPS states in $N=2$
theories in four dimensions as studied, for example, by Seiberg and
Witten \cite{SW:I}. We refer to \cite{Lc:SWintro} for background in
this subject.

In the above discussion, the {\em period\/} of the holomorphic 3-form
is playing the r\^ole of the {\em central charge\/}, Z, of the BPS
state. That is
\begin{equation}
  Z(L) = \int_L\Omega.
    \label{eq:Zper}
\end{equation}
This relationship may also be derived from the fact that both $Z$ and
the periods are subject to the same rules of special geometry
\cite{Strom:S,SW:I,Frd:SK}.\footnote{One might also try to derive this
relationship directly from (\ref{eq:SSO}). See also section 19.3 of
\cite{Hori:bk}.}

\subsubsection{Tachyon condensates} \label{sss:tachy}

In the previous section we saw a fairly concrete geometrical picture
for A-brane decay. The same decay process can be motivated from a
different perspective by using the idea of tachyon condensation from
\cite{Sen:dbd} (see also \cite{TZ:tasi01} and references therein).

Consider an intersection of two special Lagrangian planes $\Pi^0$ and
$\Pi^\phi$ in $\C^m$ as in section \ref{sss:Adec}. One may analyze the
masses of the open strings which begin on $\Pi^0$ and end on
$\Pi^\phi$ following \cite{BDL:angles} or section 13.4 of
\cite{Pol:books} (see also \cite{Hori:bk}). The result is that there
are R sector strings which are always massless and NS sector scalar
fields which have a mass 
\begin{equation}
  M^2 = \frac1{2\pi}\left(\sum_{j=1}^m\phi_j-\pi\right).
    \label{eq:Mang}
\end{equation}
These scalars fields are not projected out by the GSO process if one
of the D-branes is viewed as an anti-D-brane. 

We now propose an equation which generalizes (\ref{eq:ximod}) to
remove the mod 2 ambiguity. In section \ref{sss:Adec} we gave a way of
defining $\xi$ to be valued in $\R$ by demanding continuity over the
moduli space of complex structures at least as long as the associated
A-brane did not decay. If $L_1$ and $L_2$ intersect at a point $p$
with Floer index (i.e., ghost number) $\mu(p)$ then
\begin{equation}
  \xi(L_2)-\xi(L_1)+\mu(p) = \frac1\pi\sum_{j=1}^m\phi_j.
    \label{eq:xis}
\end{equation}
This follows from continuity and the following fact. Suppose $L_1$ and
$L_2$ intersect at two points $p_1$ and $p_2$ and that each Lagrangian
has a trivial Maslov class as in section \ref{sss:Ab}. Using the
arguments of \cite{Flr:mors} one can show that the difference in
$\mu(p_1)$ and $\mu(p_2)$ is equal to the difference in 
$\sum\phi_j$ for each point.

The equation (\ref{eq:xis}) ties the ambiguity in defining the ghost
number $\mu(p)$, which we discussed in section \ref{sss:Aopen2}, to
the ambiguity in the definition of the grading $\xi$. The ambiguity in
$\mu(p)$ was fixed by labeling each A-brane $L$ with some integral
ghost number $\mu(L)$. Borrowing some notation from the derived
category, let $L[n]$ be exactly the same A-brane as $L$ except that we
have increased its ghost number by $n$. It follows from
(\ref{eq:gshift}) and (\ref{eq:xis}) that
\begin{equation}
  \xi(L[n]) = \xi(L) + n.
     \label{eq:xisft}
\end{equation}

Restricting attention to open strings with ghost number 0, i.e.,
$\Hom(L_1,L_2)$ in the Fukaya category, we see that
\begin{equation}
  2M^2 = \xi(L_2)-\xi(L_1)-1.
     \label{eq:FukM}
\end{equation}
Thus, comparing to the last section, in $\cM^+$ we have $M^2<0$ and so
the open string in $\Hom(L_1,L_2)$ is tachyonic. This is entirely
consistent with the fact that there is a ground state $L_1\csum L_2$
lower in energy than $L_1\cup L_2$. This tachyon condenses to form 
$L_1\csum L_2$. In $\cM^-$ the open string is not tachyonic and no
condensation occurs.

The tachyonic condensation picture therefore gives a very simple
description of the hard analysis performed by Joyce reviewed in
section \ref{sss:Adec}. What we would like to conjecture is that this
tachyon picture gives a complete criterion for how A-branes decay as
one moves in the moduli space of complex structures. This is certainly
well-motivated from a physics point of view but the differential
geometry required to make such a statement rigorous is
difficult. Progress has been made in this direction in
\cite{Thms:mmm,ThmY} for example.

Note that since $0\leq \phi_j<\pi$, it follows from (\ref{eq:xis})
that, if $Y$ is a \CY\ $m$-fold,
\begin{equation}
  0\leq\xi(L_2)-\xi(L_1)+\mu(p)<m.
    \label{eq:unit}
\end{equation}
This relation is nicely consistent with the unitarity of
representations of the superconformal algebra as studied in
\cite{BFK:unit}. Any open string vertex operator in the topological
field theory corresponds to a chiral primary field (in the NS sector)
in the untwisted theory in the sense of \cite{LVW:}. This field must have
conformal weight $h$ between 0 and $c/6$, where $c$ is the central
charge. For a \nlsm, $c/3=m$. Finally, a vertex operator for a primary
chiral field of conformal weight $h$ is associated to a mass of
$M^2=h-\ff12$. Thus, comparing to (\ref{eq:Mang}) and
(\ref{eq:xis}), we see agreement.

We emphasize that the A-brane decay process occurs due to deformations
of complex structure of $Y$. This makes it essentially invisible to
the topological A-brane since the latter depends only upon
$B+iJ$. This is consistent with the fact that, as far as the A-model
is concerned, A-branes are Lagrangian with no special condition
applied. Without the special condition there is no decay process.

Finally in this section we recall from standard string theory analysis
that there are open string states in the NS sector corresponding to
vector particles in the uncompactified dimensions. These have mass
\begin{equation}
  2M^2 = \xi(L_2)-\xi(L_1).
\end{equation}
These are therefore always massless when $L_1=L_2$. In other words we
have vectors associated to $L_1$ given by $\Hom(L_1,L_1)$. These are
the vectors associated to gauge group present in the D-brane --- to be
precise, $\Hom(L_1,L_1)$ is the complexification of the gauge
algebra. In the case of a single irreducible D-brane we expect a
$\GU(1)$ gauge group and thus $\Hom(L_1,L_1)=\C$. If the gauge group
is enhanced, either because we have two distinct D-branes, or because
we have coincident D-branes, the gauge group, and thus $\Hom(L_1,L_1)$
will be bigger.  The fact that the irreducibility of D-brane is
equivalent to $\Hom(L_1,L_1)=\C$ may be viewed as a version of Schur's
Lemma in representation theory.

\subsection{B-Branes}  \label{ss:Bstab}

The message we keep repeating in these lectures is that the B-model
should be easier to analyze than the A-model. While this is true, it
doesn't necessarily mean that the B-model is easier to picture in
classical terms. This is particularly true for D-brane decay. In the
case of the A-branes of section \ref{ss:Astab} we have a direct
picture of how special Lagrangians decay as the complex structure is
varied. It should be pointed out however that this picture is very
difficult to make explicit in concrete cases. Conversely we will see
in this section that B-brane decay is not classical at all, thanks
essentially due to nonperturbative $\alpha'$ corrections. This makes
it hard to picture and we are forced to introduce more mathematics not
familiar to the typical physicist. Having said that, we can give
fairly explicit examples of B-brane decay.

\begin{table}
\renewcommand{\arraystretch}{1.3}
\begin{tabular}{|l||l|l|}
  \hline
  &A-model&B-model\\
  \hline
  A/B-branes&Lagrangians&Complexes of coherent sheaves\\
  BPS A/B-branes&Special Lagrangians&$\Pi$-stable complexes\\
  Dependence of corr.\ funcs.&$B+iJ$&complex structure\\
  Dependence of stability&complex structure&$B+iJ$\\
  Bound state&$A\csum B$&$\Cone(A\to B)$\\
  \hline
\end{tabular}
\caption{Mirror symmetry for BPS A-branes and B-branes.}
\label{tab:mir2}
\end{table}

In table \ref{tab:mir2} we review how mirror symmetry relates the
ideas of stability between the A-model and B-model. Notice in
particular that the r\^oles of complex structure and $B+iJ$ are
exchanged between the topological field theory dependence and the
stability criteria.

\subsubsection{Triangles}  \label{sss:trian}

Just like the A-model, the B-model itself should not know about any
stability issue. What we do demand from the B-model though is some
criterion of whether a given object in the derived category can {\em
potentially\/} decay into two other objects. 

The discussion of A-brane decay via tachyon condensates in section
\ref{sss:tachy} showed that, when we were on the wall of marginal
stability, the open string was massless. Thus it acts like a marginal
(but probably not truly marginal) operator in the conformal field
theory. In this sense a decay (or binding) process looks like a
deformation. Thus it is the mapping cone of (\ref{eq:Cone}) which
defines a potential bound state of two D-branes.

The mapping cone construction in the derived category gives rise to a
{\em triangulated\/} structure on the category. This mathematical
structure turns out to be central to the notion of D-brane
stability. The fact we ignored it in section \ref{ss:Astab} turns out
to be a problem as we will see later. So let us now turn to the
definition of this triangulated structure.

A triangulated category $\mathcal{C}$ is an additive category with two further
ingredients:
\begin{enumerate}
\item A {\em translation\/} functor $T:\mathcal{C}\to\mathcal{C}$
  which is an isomorphism. If $A$ is an object (or morphism) in
  $\mathcal{C}$ we will denote $T^n(A)$ by $A[n]$.
\item A collection of {\em distinguished triangles}. A triangle is a set of
  three objects and three morphisms in the form
\begin{equation}
\xymatrix{
&C\ar[dl]|{[1]}_(0.3)c&\\
A\ar[rr]^a&&B\ar[ul]_b,\\
} \label{eq:tri}
\end{equation}
where the ``[1]'' on the arrow denotes that $c$ is a map from $C$ to
$A[1]$. 
\end{enumerate}
A triangle may also be written as
\begin{equation}
\xymatrix@1{
A\ar[r]^a&B\ar[r]^b&C\ar[r]^-c&A[1].
}
\end{equation}
A morphism between two triangles is simply a commutative diagram of
the form
\begin{equation}
\xymatrix@!C{
A\ar[r]^a\ar[d]^f&B\ar[r]^b\ar[d]^g&C\ar[r]^-c\ar[d]^h&A[1]\ar[d]^{f[1]}\\
A'\ar[r]^{a'}&B'\ar[r]^{b'}&C'\ar[r]^-{c'}&A'[1].
} \label{eq:trimor}
\end{equation}

This data is subject to the following axioms:
\begin{enumerate}[{\bf TR1:}]
\item
\begin{enumerate}[a)]
\item For any object $A$, the following triangle is distinguished:
\begin{equation}
\xymatrix{
&0\ar[dl]|{[1]}&\\
A\ar[rr]^{\textrm{id}}&&A\ar[ul],\\
}
\end{equation}
\item If a triangle is isomorphic to a distinguished triangle then, it
  too, is distinguished.
\item Any morphism $a:A\to B$ can be completed to a distinguished
  triangle of the form (\ref{eq:tri}).
\end{enumerate}
\item The triangle (\ref{eq:tri}) is distinguished if and only if 
\begin{equation}
\xymatrix{
&C\ar[dl]_c&\\
A[1]\ar[rr]|(0.4){[1]}^(0.6){-a[1]}&&B\ar[ul]_b,\\
} 
\end{equation}
is also distinguished. That is, we may shuffle the edge containing
``[1]'' around the triangle translating the objects and morphisms accordingly.
\item Given two triangles and the vertical maps $f$ and $g$ in
(\ref{eq:trimor}), we may construct a morphism $h$ to complete
(\ref{eq:trimor}).
\item {\em The Octahedral Axiom:}
\begin{equation} 
\xymatrix@C=16mm{
&B\ar[lddd]\ar[rd]&\\
D\ar@{-->}[dd]|{[1]}\ar[ur]|{[1]}&&E
        \ar[ll]\ar@{-->}[dddl]\\
&&\\
C\ar[rr]|{[1]}\ar@{-->}[dr]&&A\ar[uuul]\ar@{-->}[uu]\\
&{F}\ar@{-->}[ur]|{[1]}\ar@{-->}[uuul]&
} \label{eq:oct}
\end{equation}
Four faces of the octahedron are distinguished
triangles and the other four faces commute. The relative orientations
of the arrows obviously specify which is which.

The octahedral axiom specifies that, given $A,B,C,D,E$ and the solid
arrows in the octahedron, there is an object $F$ such that the
octahedron may be completed with the dashed arrows. The pairs of maps
that combine to form maps between $B$ and $F$ also commute.

\end{enumerate}

For any abelian category $\mathcal{C}$, the derived category
$\DC(\mathcal{C})$ is a triangulated category. The translation functor
is the same as the shift functor that we introduced earlier of
course. The distinguished triangles are provided by the mapping cone
--- any vertex of a distinguished triangle is isomorphic to the
mapping cone of the opposite edge when the ``[1]'' is shuffled around
to the appropriate edge. We refer to \cite{Wei:hom,GM:Hom} for the
proof that $\DC(\mathcal{C})$ satisfies the above axioms.

Given a short exact sequence 
\begin{equation}
\xymatrix@1{
0\ar[r]&A\ar[r]^a&B\ar[r]^b&C\ar[r]&0,
}
\end{equation}
in $\mathcal{C}$, we induce a distinguished triangle (\ref{eq:tri}) in
$\DC(\mathcal{C})$ for the corresponding single-entry complexes with
$a$ and $b$ induced in the obvious way.  This short exact sequence can
be also phrased ``$B$ is an extension of $C$ by $A$''. The group of
extensions of $C$ by $A$ is given by $\Ext(C,A)=\Ext^1(C,A)$ (see
\cite{MacL:hom} for example). Thus, the short exact sequence
determines an element $c$ which is a morphism in $\Ext^1(C,A)$ which
equals $\Hom(C,A[1])$ in $\DC(\mathcal{C})$. This is the map $c$ in
(\ref{eq:tri}).  The derived category is not an abelian category since
kernels and cokernels do not always exist. Thus one cannot define
short exact sequences in $\DC(\mathcal{C})$. In a sense, distinguished
triangles are a weaker notion of short exact sequences that are ``the
best one can do'' for the derived category.

This definition of a triangulated category was invented by Verdier
\cite{Verd:tri} for completely abstract reasons of course, but it
turns out to be precisely what is needed for the rules of D-brane
decay. The basic triangle (\ref{eq:tri}) should be read as {\em the
D-branes $A$ and $C$ may bind via the potentially tachyonic open
string $c$ to form $B$.}  We may then go through each axiom in turn
and say what it means:
\begin{enumerate}[{\bf TR1:}]
\item
\begin{enumerate}[a)]
\item $A$ binds with 0 (the empty brane) to produce $A$.
\item We consider two objects in $\DC(X)$ which are isomorphic to be
  the same D-brane. Thus this rule is required for consistency.
\item The existence of an open string from $A$ to $B$ means that $B$
can potentially decay into $A$ and some other decay product $C$. This
is not obvious but this axiom may be rephrased after the following.
\end{enumerate}
\item If $B$ can potentially decay into $A$ and $C$, then $C$ can
  potentially decay into $A[1]$ and $B$. This is consistent with
  the observation in section \ref{ss:K} that $A[1]$ could be
  interpreted as an anti-$A$.

Note that using this axiom we may now rephrase {\bf TR1:} c) as
follows. Given an open string from $A$ to $B$ we may potentially form
a bound state of these two D-branes.
\item Given open strings between D-branes $A$ and $A'$ and between
$B$ and $B'$, we may construct open strings between the corresponding
  bound states.
\item This formidable looking axiom is little more than a statement of
  associativity in the rules for combining D-branes. If we crudely
  write addition to represent rules for combining, the distinguished
  triangles in (\ref{eq:oct}) can be read (using {\bf TR2}) as
\begin{equation}
\begin{split}
  C &= A[1] + B  \\
    &= A[1] + (E + D[-1]) \\
    &= (A[1] + E) + D[-1] \\
    &= F + D[-1].
\end{split}
\end{equation}
\end{enumerate}
One may choose to regard these rules for D-brane decay as mainly
self-evident, or as proven since we have proven that the category of
B-branes is the derived category and therefore triangulated.

The triangulated structure encodes the long exact sequences associated
to cohomology as follows. A functor between two triangulated
categories is exact if it preserves triangles. An example of such an
exact functor is $\Hom(M,-)$ for some fixed object $M$. This is a
functor from an arbitrary category to the category of vector spaces. As
mentioned in section \ref{ss:DC}, the derived category of vector
spaces is rather trivial in the sense that every complex is
quasi-isomorphic to a complex where all the differential maps are
zero. Using this fact, let us write the object $\Hom(M,A)$ in the
derived category of vector spaces as
\begin{equation}
\xymatrix@1{
\ldots\ar[r]^(0.3)0&\Ext^{-1}(M,A)\ar[r]^0&\Ext^{0}(M,A)\ar[r]^0&
\Ext^{1}(M,A)\ar[r]^(0.66)0&\ldots,
}
\end{equation}
where the $\Ext$'s are vector spaces.
The triangle
\begin{equation}
\xymatrix@C=-3mm{
&\Hom(M,C)\ar[dl]|{[1]}&\\
\Hom(M,A)\ar[rr]&&\Hom(M,B)\ar[ul],\\
} 
\end{equation}
then becomes the usual long exact sequence of vector spaces
\begin{equation}
\xymatrix@1{
\ldots\ar[r]&\Ext^0(M,A)\ar[r]&\Ext^0(M,B)\ar[r]&\Ext^0(M,C)\ar[r]&
  \Ext^1(M,A)\ar[r]&\ldots
}
\end{equation}

\subsubsection{Categorical mirror symmetry at last}  \label{sss:mirN}

Of course, there is a some vagueness in the word ``potentially''
whenever we refer to binding or decay in section \ref{sss:trian}. We
have stated explicitly above that if there is an open string from $A$
to $B$ then we regard $A+B$ as a potential bound state. In order for
this to actually happen there must be some region of moduli space
where $A$ and $B$ are both themselves stable and the open string from
$A$ to $B$ is tachyonic. This is not guaranteed. Thus, the
triangulated structure appears when one has an optimistic view (which
is as much as the topological field theory can know) about what can
bind to what.

Our discussion of A-brane stability in section \ref{ss:Astab} was
approached directly rather than using the topological field theory
language. Because of this the Fukaya category need not have a
triangulated structure --- it certainly knows about the A-branes which
really are stable but it need not include the potentially stable
branes in the topological field theory which never actually make it to
stability. In particular there is no reason to suppose that the Fukaya
category is actually triangulated. That is, it may well violate
axiom {\bf TR1:} c).\footnote{I thank R.~Thomas for discussions on
this point.}

If the Fukaya category is not triangulated then the mirror symmetry
proposal in section \ref{ss:m2} cannot possibly be correct. The
derived category $\DC(X)$ is triangulated and thus cannot be
equivalent to a category which is not triangulated. The solution, of
course, is to add the extra ``potentially stable'' A-branes to the
Fukaya category so that the result is triangulated. This can be done
in a precise mathematical way by following the procedure of Bondal and
Kapranov \cite{BK:frm}.

If $\mathcal{F}(Y)$ is the Fukaya category of $Y$, then let
$\Tr\mathcal{F}(Y)$ be the triangulated category produced by the
method of Bondal and Kapranov.\footnote{This is often called the
``derived Fukaya category'' but it's not derived in the sense of
complexes etc.} The current state-of-the-art conjecture for mirror
symmetry which follows from our topological field theory constructions
is then:
\begin{center}
\shabox{\parbox{.80\hsize}{If $X$ and $Y$ are mirror \CY\ threefolds
then the category $\DC(X)$ is equivalent to the category
$\Tr\mathcal{F}(Y)$.}}
\end{center}
We should warn that even this statement is subject to corrections when
we go outside the class of \CY\ threefolds with zero $b_1$ because of
the appearance of extra coisotropic A-branes \cite{KO:Azum}.

This ``homological mirror symmetry'' statement has been demonstrated
for 2-tori \cite{AP:Cmir} and quartic K3 surfaces \cite{Seidel:quart}.

%%%%%

\subsubsection{$\Pi$-Stability}  \label{sss:Pi}

Assuming mirror symmetry to be true we may now copy the description of
the stability of A-branes in section \ref{ss:Astab} over to the case
of B-branes. 

The first ingredient we need is how to compute the central charge $Z$ of a
given B-brane. This author is not aware of a complete argument in the
literature for how to compute this, but we may proceed a little less
than rigorously as follows.

The first step is to note that $Z$ is given in the mirror by a period
$\int_L \Omega$ for some 3-cycle $L$. This varies with the complex
structure of $Y$. Thus, $Z$ must depend upon $B+iJ$ for $X$.  We saw
in section \ref{ss:mir} that we can derive the periods exactly from the
Picard--Fuchs equations. These differential equations are written in
terms of parameters that specify the complex structure
algebraically, such as the ``$\psi$'' in section
\ref{ss:mir}. However, we also saw how to relate such parameters to
the complexified K\"ahler form $B+iJ$ of $X$. Thus, a knowledge of the
Picard--Fuchs equations is sufficient to obtain exact (but
transcendental) expressions for the set of $Z$'s for objects in
$\DC(X)$.

Now, given a particular object $\cE^\bullet$ in $\DC(X)$, the next
step is to find which particular period computed above should be
associated to the central charge $Z(\cE^\bullet)$. In the case of the
A-model we may choose some basis $\gamma_k$ of $H_3(Y)$ and compute a
basis of periods $\varpi_k=\int_{\gamma_k}\Omega$. The charge, $Q$, of an
A-brane is given by its homology class in $H_3$. Thus, if
\begin{equation}
  Q = \sum b_k\gamma_k,
\end{equation}
then the central charge is given by $\sum b_k\varpi_k$. We saw in
section \ref{ss:K} that the D-brane charge of a B-brane is given by
$\ch(\cE^\bullet)\sqrt{\td(X)}$. Thus, the formula for the central
charge of a B-brane must contain this expression linearly.

We saw in section \ref{sss:Ab} that the curvature, $F$, of a bundle
associated to a D-brane is not really a physical quantity by itself since it is
not invariant under the gauge symmetry we introduced. Instead we must
always have the combination $B-F$. Since $\ch(E)=\Tr e^F$, the Chern
character by itself it not gauge invariant and must always appear in
the combination $e^{-B}\ch(E)$ in any physical quantity. Furthermore,
holomorphy of supersymmetric theories demands that $B$ always appears
in the combination $B+iJ$. All this suggests that the simplest
expression for the central charge should be
\begin{equation}
  Z(\cE^\bullet) = \int_X e^{-(B+iJ)}\ch(\cE^\bullet)\sqrt{\td(X)}.
    \label{eq:ZB0}
\end{equation}
Unfortunately this is not, in general, a solution to the Picard--Fuchs
equations! However, it is a familiar situation in mirror symmetry that
any expression depending on $B+iJ$ is subject to quantum $\alpha'$
corrections. Thus we should regard (\ref{eq:ZB0}) as the asymptotic
form of $Z$ near the large radius limit. If we know exactly just how
asymptotic this formula is, we have enough information to determine
exactly which combination of periods give $Z$ exactly.

The periods are associated to the prepotential in special geometry and
so one expects them to receive corrections in the same way. Such
corrections were discussed in \cite{CDGP:} and appear in two ways:
\begin{enumerate}
\item The perturbative corrections will be due to the 4-loop
  correction in the \nlsm\ as analyzed in \cite{GVZ:4loop}. These
  corrections will be three powers of $B+iJ$ less than the leading
  term in (\ref{eq:ZB0}). Thus, in the case of \CY\ threefolds this
  produces at most a constant term.
\item The nonperturbative corrections will produce a power series,
  with no constant term, in $q_i=\exp 2\pi i(\int_{C_i}B+iJ)$ for some basis
  $\{C_i\}$ of $H_2(X)$.
\end{enumerate}
This finally provides enough information to determine the precise
expression for $Z(\cE^\bullet)$.

Having found the central charge we may now proceed with the next stage
in copying the stability condition from the A-brane model.  Given
$\cE^\bullet$ we may choose a grading $\xi(\cE^\bullet)$ such that
\begin{equation}
\xi(\cE^\bullet) = \frac1{\pi}\arg Z(\cE^\bullet) \pmod2
\end{equation}
and demand that $\xi(\cE^\bullet)$ vary continuously with $B+iJ$ so
long as $\cE^\bullet$ is stable. Following (\ref{eq:xisft}) we have
\begin{equation}
  \xi(\cE^\bullet[n])=\xi(\cE^\bullet) + n.
     \label{eq:Bsh0}
\end{equation}

Finally we copy the picture in figure \ref{f:Adecay} by asserting that
if we have a distinguished triangle in $\DC(X)$ of the form
\begin{equation}
\xymatrix{
&C\ar[dl]|{[1]}_(0.3)c&\\
A\ar[rr]^a&&B\ar[ul]_b,\\
} \label{eq:tri2}
\end{equation}
with $A$ and $B$ stable, then $C$ is stable with respect to the decay
represented by this triangle if and only if
$\xi(B)<\xi(A)+1$. Also, if $\xi(B)=\xi(A)+1$ then $C$ is marginally
stable and we may state that
\begin{equation}
  \xi(C)=\xi(B)=\xi(A)+1.
\end{equation}
We may use axiom {\bf TR2} in section \ref{sss:trian} to rephrase this
as follows. If $A$ and $C$ are stable then $B$ is stable with respect
to decay into $A$ and $C$ so long as $\xi(A)<\xi(C)$.  

These criteria for stability are known as $\Pi$-stability and were
studied in \cite{DFR:stab,DFR:orbifold,Fl:quiv,Doug:DC,AD:Dstab}.

As stated these rules are not sufficient to determine the set of
stable B-branes at a given point in the moduli space of $B+iJ$. If we
happen to know the set of stable B-branes (including their gradings) at
some basepoint in the moduli space then these rules {\em are\/}
sufficient to determine how the stable spectrum changes as we move
along some path in the moduli space. The following rules are applied
\begin{itemize}
\item We begin with a stable set of B-branes together with a value of
  the grading $\xi$ for each B-brane. This set must be consistent with
  the rules of $\Pi$-stability. That is, no distinguished triangle may
  allow a stable B-brane to decay into two other stable B-branes.
\item As we move along a path in moduli space the gradings will change
  continuously.
\item Two stable B-branes may bind to form a new stable state.
\item A stable B-brane may decay into other (marginally) stable
  states.
\end{itemize}
Note in the last case that a brane may decay into another state which
becomes unstable at exactly the same point in moduli space. This
certainly can happen. We also emphasize that we never make any
reference to a value of $\xi$ for an unstable object. This is probably
not defined.

These rules certainly do not imply that the stable set of B-branes is
uniquely determined by a point in the moduli space of $B+iJ$. In order
to determine the stable set we explicitly specified a path from the
basepoint to the desired path. We will see in section \ref{sss:qmon} very
explicitly that there is nontrivial monodromy in $\DC(X)$ which
changes the set of stable objects as we go around loops in the moduli
space. 

The monodromy occurs because of monodromy in the gradings. This corresponds
to B-branes acquiring zero $Z$, i.e., zero {\em mass}. If a stable
B-brane becomes massless it necessarily induces a singularity in the
conformal field theory following the arguments of
\cite{Str:con,GMS:con}. Let us define the Teichm\"uller space $\cT$ as
the universal cover of the moduli space of $B+iJ$ with these singular
CFT points deleted. Thus there should be no monodromy in $\cT$ and we
expect the set of stable B-branes to be well-defined at any point in
$\cT$.

The tachyon condensation picture sets the rather discontinuous
behaviour of algebraic geometry that we saw in section \ref{ss:mdefs}
in a more natural setting. There we found that the cone of any map
$\Cone(f:A\to B)$ is invariant under a rescaling of $f$ by a nonzero
complex number. In the tachyon condensation picture, the scale of $f$
is determined by minimizing the tachyon potential. Thus, when the cone
is unstable, $f$ is fixed at zero and when the cone becomes stable $f$
acquires a definite value depending on the modulus $B+iJ$. Thus, the
only discontinuity appears as the cone decays --- as one should expect.

There is a consistency condition that any set of stable B-branes must
obey. In section \ref{sss:tachy} we found a condition (\ref{eq:unit})
equivalent to the unitarity constraint on the conformal field
theory. In B-model language this amounts to
\begin{equation}
  \xi(A) > \xi(B) \;\Rightarrow\; \Hom(A,B)=0.
   \label{eq:unit2}
\end{equation}
This removes any states of conformal weight $h<0$. By Serre duality
(\ref{eq:Serre}) it also removes states with $h>m/2$ for a \CY\
$m$-fold.

\subsubsection{Multiple decays}   \label{sss:mult}

Every object in $\DC(X)$ is either stable or unstable for a given
point in the Teichm\"uller space of $B+iJ$. A particle which is
unstable must be unstable because it decays into a set of stable
objects. Thus the set of stable objects must be big enough to account
for this property. This puts a stronger constraint on stability than
the previous section. For example, having no stable objects at all
would have been consistent with our earlier definition of
$\Pi$-stability.

If an unstable object decays into 2 stable objects we know how to
describe the decay by a distinguished triangle. We now want to
describe a decay of an object into 3 stable objects.

We use the following octahedron to describe the process:
\begin{equation} 
\xymatrix@C=16mm{
&E_2\ar[lddd]\ar[rd]&\\
A_3\ar[dd]|(0.53){[1]}_(0.40){f_2}\ar[ur]|{[1]}&&E_3
        \ar[ll]\ar[dddl]\\
&&\\
A_2\ar[rr]|(0.53){[1]}^(0.45){f_1}\ar[dr]&&A_1\ar[uuul]\ar[uu]\\
&{F}\ar[ur]|{[1]}\ar[uuul]&
} \label{eq:oct2}
\end{equation}
Suppose that we begin at a point $p_0$ in the Teichm\"uller space where
$\xi(A_1)<\xi(A_2)<\xi(A_3)$ and end at a point $p_1$ where
$\xi(A_1)>\xi(A_2)>\xi(A_3)$. Thus the open strings corresponding to
$f_1$ and $f_2$ in (\ref{eq:oct2}) go from tachyonic to massive as we
pass from $p_0$ to $p_1$. 

\iffigs
\begin{figure}
\begin{center}
\setlength{\unitlength}{0.6mm}%
\begin{picture}(100,100)
\thinlines
\put(0,50){\line(1,0){100}}
\put(50,0){\line(0,1){100}}
\put(50,105){\makebox(0,0){$W_1$}}
\put(105,50){\makebox(0,0){$W_2$}}
\put(10,90){\circle*{1}}
\put(10,86){\makebox(0,0){$p_0$}}
\put(90,10){\circle*{1}}
\put(90,6){\makebox(0,0){$p_1$}}
\put(20,70){\makebox(0,0){$\scriptstyle \xi(A_1)<\xi(A_2)<\xi(A_3)$}}
\put(80,20){\makebox(0,0){$\scriptstyle \xi(A_3)<\xi(A_2)<\xi(A_1)$}}
\put(80,70){\makebox(0,0){$\scriptstyle \xi(A_2)<\xi(A_1)$}}
\put(80,60){\makebox(0,0){$\scriptstyle \xi(A_2)<\xi(A_3)$}}
\put(20,30){\makebox(0,0){$\scriptstyle \xi(A_1)<\xi(A_2)$}}
\put(20,20){\makebox(0,0){$\scriptstyle \xi(A_3)<\xi(A_2)$}}
\end{picture}
\end{center}
  \caption{Walls of marginal stability for a decay into 3 objects.}
  \label{f:3dec}
\end{figure}
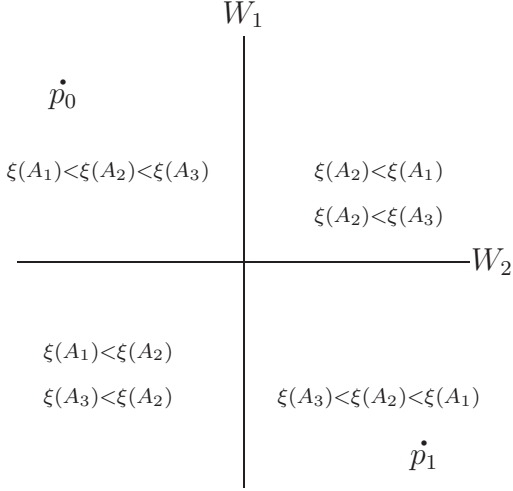
\fi

At $p_0$, with respect to the triangles in this octahedron, $E_2$ and
$F$ are stable. We may also declare that $E_3$ is stable (but this
isn't really necessary). Suppose there are two walls $W_1$ and $W_2$
between $p_0$ and $p_1$ such that $\xi(A_i)-\xi(A_{i+1})$ is negative
on the $p_0$ side of $W_i$ and positive on the $p_1$ side of $W_i$. We
depict this in figure \ref{f:3dec}.
Then there are two possibilities to consider as we move from $p_0$ to
$p_1$:
\begin{enumerate}
\item We cross $W_1$ and then $W_2$. As we cross $W_2$ the object $F$ will
  decay into $A_2$ and $A_3$. At this instant
  $\xi(F)=\xi(A_2)<\xi(A_1)$ so we know that $E_3$ must have already
  decayed into $F$ and $A_1$. Thus $E_3$ decays into $A_1$, $A_2$ and
  $A_3$ by the time we reach $p_1$.
\item We cross $W_2$ and then $W_1$. As we cross $W_1$ the object $E_2$ will
  decay into $A_1$ and $A_2$. At this instant
  $\xi(E_2)=\xi(A_2)>\xi(A_3)$ so we know that $E_3$ must have already
  decayed into $E_2$ and $A_3$. Thus $E_3$ decays into $A_1$, $A_2$ and
  $A_3$ by the time we reach $p_1$.
\end{enumerate}
Either way, the condition for $E_3$ to decay into $A_1$, $A_2$ and
$A_3$ is that
\begin{equation}
  \xi(A_1)>\xi(A_2)>\xi(A_3).
\end{equation}

We may generalize this to the case of decays into any number of
objects. For any object $E$ we define the following set of
distinguished triangles
\begin{equation}
\xymatrix@!C=3.5mm{
**[l]0=E_0\ar[rr]&&E_1\ar[rr]\ar[dl]&&E_2
\ar[rr]\ar[dl]&&\cdots\ar[rr]\ar[dl]&&E_{n-1}\ar[dl]
\ar[rr]&&**[r]E_n=E\ar[dl]\\
&A_1\ar[ul]|{[1]}&&A_2\ar[ul]|{[1]}&&A_3\ar[ul]|{[1]}
&&A_{n-1}\ar[ul]|{[1]}&&A_n\ar[ul]|{[1]} 
} \label{eq:bchain}
\end{equation}
Then $E$ decays into $A_1, A_2, \ldots, A_n$ so long as 
\begin{equation}
  \xi(A_1)>\xi(A_2)>\ldots>\xi(A_n).
    \label{eq:xich}
\end{equation}
Thus we motivate the following
\begin{conjecture}
At every point in the
Teichm\"uller space of $B+iJ$ there is a set of stable objects in
$\DC(X)$ such that every object $E$ can be written in the form
(\ref{eq:bchain}) for some $n$ (meaning it decays into $n$ stable
objects) and for stable objects $A_k$ satisfying (\ref{eq:xich}).
\end{conjecture}
Together with the unitarity constraint (\ref{eq:unit2}), this is the
form of $\Pi$-stability proposed by Bridgeland \cite{Brg:stab}.

Note that we can't claim to have proven this conjecture since there
are many objects in $\DC(X)$ which are never stable. The argument at
the start of the subsection cannot then be used to follow the
decay. We would also like to include a finiteness condition on $n$ in
(\ref{eq:bchain}). This appears to be an additional assumption too.

All that said, Bridgeland's form of the stability conditions does seem
to work very nicely \cite{Brg:stab,Brid:K3} although many aspects are
still poorly-understood in the case of \CY\ threefolds.

\subsubsection{$\mu$-stability}  \label{sss:mu}

In order to determine the set of $\Pi$-stable objects it is best if
we have a basepoint in the moduli space of $B+iJ$ from which we may
follow paths as in section \ref{sss:Pi}. The obvious choice for such a
basepoint is near the large radius limit of the \CY\ threefold $X$. 

At the large radius limit we should expect that the classical analysis
of D-branes is valid and therefore that D-branes correspond to vector
bundles supported over subspaces $S\subset X$. Furthermore, we may assume
that the world-volume approach to D-branes should be accurate. We
refer to the general literature on D-branes such as \cite{John:Dbok}
for more details.

For simplicity let us assume that we have a 6-brane wrapping $X$
associated to a holomorphic vector bundle $E\to X$ with curvature
(1,1)-form $F$. At large radius, the BPS condition reduces to the
Hermitian--Yang--Mills condition \cite{MMMS:BPS}. That is, the
curvature tensor obeys the relation 
\begin{equation}
  g^{j\bar k}F^\beta_{\alpha j\bar k} = \mu(E)\cdot\delta^\beta_\alpha,
    \label{eq:slo1}
\end{equation}
where $\alpha,\beta$ are indices in the fibre of $E$ and $\mu(E)$ is a
real number called the ``slope'' of $E$. Following the analysis of
\cite{UY:YM}, for example, one can integrate (\ref{eq:slo1}) and
obtain\footnote{Here we are following the conventions of \cite{UY:YM}
but $\mu$ is also often defined to remove the factor of $\Vol(X)$ in
(\ref{eq:mudef}).}
\begin{equation}
  \mu(E) = \frac{\deg(E)}{k\cdot\Vol(X)},
     \label{eq:mudef}
\end{equation}
where $k$ is the rank of $E$ and
\begin{equation}
  \deg(E) = \int_X J\wedge J\wedge c_1(E),
\end{equation}
is the {\em degree\/} of the bundle $E$.

As usual, this condition for a supersymmetric vacuum is a first-order
differential equation (in the connection) and is a sufficient
condition for a solution of the equations of motion, i.e., the
Yang--Mills condition, which is a second order differential equation.

The Hermitian--Yang--Mills condition (\ref{eq:slo1}) depends
explicitly on the metric and thus the K\"ahler form $J$. As such,
whether we have a BPS solution can depend upon $J$.  The existence of
Hermitian--Yang--Mills connections has been studied by Donaldson
\cite{Don:YM}, and Uhlenbeck and Yau \cite{UY:YM}, who proved the
following theorem. Let $\cE$ be the locally-free sheaf associated to
$E$. The bundle $E$ is said to be {\em $\mu$-stable\/} if every
subsheaf $\cF$ of $\cE$ satisfies $\mu(\cF)<\mu(\cE)$. We then have:
\begin{theorem}
A bundle is $\mu$-stable if and only if it admits an irreducible
Hermitian--Yang--Mills connection.
\end{theorem}
The stability of the BPS B-brane is thus equivalent to $\mu$-stability
in the classical limit where $\alpha'$ corrections are ignored.

If $\cF$ is a subsheaf of $\cE$ then we have a short exact sequence
\begin{equation}
\xymatrix@1{
  0\ar[r]&\cF\ar[r]&\cE\ar[r]&\cG\ar[r]&0,
}    \label{eq:mude}
\end{equation}
for some sheaf $\cG$. If $\cE$ is unstable then it decays into $\cF$
and $\cG$. 

The lower-dimensional branes can be analyzed similarly. We can focus
on a subspace $S\subset X$ and look for stable vector bundles (or
twisted bundles if $S$ is not spin) within the class of bundles on
$S$. The subspaces do not interfere with each other in the following
sense. A bundle on $X$ associated to $\cE$ cannot decay into a
subsheaf $\cF$ supported only on $S$ since
there is no homomorphism $\cF\to\cE$. Equally, a bundle on $S$ cannot
decay into a subsheaf on $X$ since the quotient sheaf $\cG$ in
(\ref{eq:mude}) would have negative rank.

It follows that $\mu$-stability establishes a set of stable B-branes
at the large radius limit. We should now check that this set of stable
B-branes is consistent with $\Pi$-stability.

From (\ref{eq:ZB0}) we see that, for large $J$, the leading
contribution to the central charge $Z(A)$ will be given by the lowest
degree differential form in $\ch(A)$. The Grothendieck--Riemann--Roch
formula (\ref{eq:GRR}) can be used to show that the lowest component
of $\ch(i_*\cF)$ is given by the $(6-2\dim(S))$-form $s$ which is
Poincar\'e dual to $S$.\footnote{Note that $\dim(S)$ is the {\em
complex\/}  dimension of $S$.} Thus, for large $J$, 
\begin{equation}
\begin{split}
Z&\sim\int_X(-iJ)^{\dim(S)}\wedge s\\
 &\sim\int_S(-iJ|_S)^{\dim(S)}\\
 &\sim (-i)^{\dim(S)}\Vol(S),
\end{split}
\end{equation}
yielding
\begin{equation}
  \xi(i_*\cE) = -\ff12\dim(S) \pmod2.
\end{equation}

If we choose the values of $\xi$ to fix the mod 2 ambiguity
arbitrarily we will violate the unitarity condition
(\ref{eq:unit}). For example, let $\O_X$ be the 6-brane wrapping $X$
and let $\O_p$ be the 0-brane (skyscraper sheaf) at a point $p\in
X$. Thus $\xi(\O_X)=-\ff32\pmod2$ and $\xi(\O_p)=0\pmod2$. By
restricting the value of a function on $X$ to its value at $p$ we see
that $\Hom(\O_X,\O_p)=\C$ and so we must insist $\xi(\O_X)<\xi(\O_p)$
if these B-branes are stable. Furthermore, by Serre duality,
$\Hom(\O_p,\O_X[3])=\C$ and so
$\xi(\O_p)<\xi(\O_X)+3$.\footnote{$\O_X[3]$ is the complex with $\O_X$
in position $-3$ and zero elsewhere.} So the only
possibility at large radius is that $\xi(\O_X)=\xi(\O_p)-\ff32$.

A consistent choice is to set
\begin{equation}
  \xi(i_*\cE) = -\ff12\dim(S).
\end{equation}

Let us see what happens at the subleading order in $J$. We restrict
attention to the case of 6-branes, i.e., locally-free sheaves. Now
$\ch(\cE)=k+c_1+\ldots$, where $k$ is the rank of the associated vector
bundle. Applying (\ref{eq:ZB0}) we now obtain
\begin{equation}
  \xi(\cE) =-\frac32 + \frac1\pi\tan^{-1}\frac{\mu(\cE)}{2}+\ldots,
   \label{eq:Pimu}
\end{equation}
where
$\mu(\cE)$ is, again, the slope of the sheaf $\cE$ introduced
above. Note from (\ref{eq:mudef}) that $|\mu(\cE)|\ll1$ in the large
radius limit.

The short exact sequence (\ref{eq:mude}) induces the triangle
\begin{equation}
\xymatrix{
  &\cG\ar[dl]|{[1]}&\\
  \cE\ar[rr]&&\cF,\ar[ul]
}
\end{equation}
in $\DC(X)$ as explained in section \ref{sss:trian}. If we are near
the large radius limit, then $\xi(\cE)$ and $\xi(\cF)$ are both
very close to $-\ff32$ since they are locally-free sheaves. 
$\cG$ is either locally free or supported on a complex codimension one
subspace. Thus $\xi(\cG)$ is close to either $-\ff32$ or $-1$.
Since the D-brane charges add
according to section \ref{ss:K}, we have $Z(\cF)=Z(\cE)+Z(\cG)$, which
implies that $\xi(\cF)$ lies between $\xi(\cE)$ and $\xi(\cG)$. The
$\Pi$-stability condition for $\cF$ is $\xi(\cE)<\xi(\cG)$, which is
therefore equivalent to $\xi(\cE)<\xi(\cF)$. By (\ref{eq:Pimu}) this,
in turn, is equivalent to $\mu(\cE)<\mu(\cF)$. Thus $\Pi$-stability
reduces to $\mu$-stability as first observed in \cite{DFR:stab}.

In \cite{MMMS:BPS} some $\alpha'$ corrections to the
Hermitian--Yang--Mills condition were computed. This brings the
D-brane stability condition closer to $\Pi$-stability in a sense near
the radius limit but there is a very important qualitative difference
between $\Pi$-stability and any version of $\mu$-stability as follows.

$\mu$-stability is defined for the abelian category of
coherent sheaves whereas $\Pi$-stability is defined for the
triangulated category $\DC(X)$. When checking for $\mu$-stability one
looks for decay into {\em subobjects\/} $\cE\subset\cF$. The notion of
subobjects leads to a definite hierarchy within the category of coherent
sheaves. If $\cF$ can decay into the subobject $\cE$ (and some other
decay product $\cG$) then there is no way $\cE$ can decay into
something including $\cF$. The notion of a subobject is given by the
injective map $\cE\to\cF$, i.e., a map with zero kernel. This is
well-defined since kernels always exist in an abelian category.

This hierarchy does not extend to the derived category since we no
longer have an abelian category. Each vertex of a distinguished
category may decay into the other two vertices. This essentially
arises due to the possibility of anti-branes as discussed in section
\ref{ss:K}. Thus, generically within the moduli space of $B+iJ$ we are
forced to use the more difficult notion of $\Pi$-stability on a
triangulated category. At the large radius limit, however, the
stability structure simplifies and we may use an abelian category
instead.

Does this simplification to a useful abelian structure exist elsewhere in
the moduli space? In section \ref{ss:orb} we will see that this indeed happens
for orbifolds. Douglas \cite{Doug:DC} has suggested that this might be
a key ingredient in a full understanding of $\Pi$-stability and some
progress in this direction has been made by Bridgeland
\cite{Brg:stab}. We will not pursue this idea in these lectures.

%%%%%%%%%%%%%%%%%%%%%%%%%%%%%%%%%%%%%%%%%%%%%%%%%%%%%%%%%%%%%%%%%%

\section{Applications}  \label{s:app}

%%%%%%%

\subsection{The Quintic Threefold}  \label{ss:quin}

We are now in a position to give some examples of B-branes and, in
particular, how $\alpha'$ corrections modify the na\"\i ve picture of
a B-brane as simply a holomorphic submanifold of $X$. The obvious
place to start is the quintic threefold, as introduced in section
\ref{ss:mir}, since the moduli space of $B+iJ$ is one-dimensional. As
emphasized in table \ref{tab:mir2}, since we are focusing on
stability, rather than the structure of the topological field theory, we
need to exchange the r\^oles of $X$ and $Y$ relative to section
\ref{ss:mir}. That is, in this section $X$ is the quintic threefold and
$Y$ is its mirror.

\subsubsection{Periods}   \label{sss:qper}

The first thing we analyze is how to compute the exact form of the central
charge $Z$ for any object $\cE^\bullet$ in $\DC(X)$. We argued in
section \ref{sss:Pi} that $Z$ was always a period of $\Omega$ on the
mirror $Y$. 

Since $\dim H_3(Y)=4$, we have 4 independent periods $\int_\Gamma
\Omega$ for the holomorphic 3-form $\Omega$. These periods satisfy the
Picard-Fuchs equation (\ref{eq:PFq}). In terms of the moduli space
coordinate $z$ introduced in section \ref{ss:mir} we may use the following
solutions (as in \cite{CDGP:}):
\begin{equation}
 \varpi_j = -\ff15\sum_{m=1}^\infty\frac{\alpha^{(2+j)m}\Gamma(\frac m5)}
  {\Gamma(m)\Gamma(1-\ff m5)^4}z^{-\frac m5}.  \label{eq:varpi}
\end{equation}
Since $\varpi_0+\varpi_1+\ldots+\varpi_4=0$, we may use
$\varpi_0,\ldots,\varpi_3$ as a basis.

These solutions may be analytically continued by using the Barnes'
integral method (see \cite{AGM:sd,me:min-d} for the precise method we have
used here) to match this basis with the series around the
large radius limit. We use the following basis for solutions near $z=0$:
\begin{equation}
\begin{split}
\Phi_0 &= \frac1{2\pi i}\int\frac{\Gamma(5s+1)\Gamma(-s)}
  {\Gamma(s+1)^4}(e^{\pi i}z)^s\,ds\\
 &= \sum_{n=0}^\infty\frac{(5n)!}{n!^5}z^n\\
 &= 1+O(e^{2\pi it})\\
 &= \varpi_0,
\end{split} \label{eq:Phi0}
\end{equation}

\begin{equation}
\begin{split}
\Phi_1 &= -\frac1{2\pi i}\cdot\frac1{2\pi i}
  \int\frac{\Gamma(5s+1)\Gamma(-s)^2}
  {\Gamma(s+1)^3}z^s\,ds\\
  &= \frac1{2\pi i}\log z + O(z) \\
  &= t+O(e^{2\pi it})\\
  &= -\ff15(\varpi_0-3\varpi_1-2\varpi_2-\varpi_3),
\end{split} \label{eq:Phi1}
\end{equation}

\begin{equation}
\begin{split}
\Phi_2 &= -\frac1{2\pi^2}\cdot\frac1{2\pi i}
  \int\frac{\Gamma(5s+1)\Gamma(-s)^3}
  {\Gamma(s+1)^2}(e^{\pi i}z)^s\,ds\\
 &= -\frac1{4\pi^2}(\log z)^2+\frac1{2\pi i}\log z-\ff56+O(z)\\
 &= t^2+t-\ff56+O(e^{2\pi it})\\
 &= \ff25(-2\varpi_0+\varpi_2+\varpi_3),
\end{split}
\end{equation}

\begin{equation}
\begin{split}
\Phi_3 &= \frac1{(2\pi i)^3}\cdot(-6)\cdot\frac1{2\pi i}
  \int\frac{\Gamma(5s+1)\Gamma(-s)^4}
  {\Gamma(s+1)}z^s\,ds\\
 &= \frac i{8\pi^3}(\log z)^3 + \frac{7i}{4\pi}\log z-
  \frac{30i\zeta(3)}{\pi^3} + O(z)\\
 &= t^3-\ff72t - \frac{30i\zeta(3)}{\pi^3} + O(e^{2\pi it})\\
 &= -\ff65(2\varpi_1+2\varpi_2+\varpi_3),
\end{split} \label{eq:Phi3}
\end{equation}
where $t=B+iJ=\Phi_1/\Phi_0$ by the mirror map (and abuse of notation)
of section \ref{ss:mir}. In each case the contour integral is along
the line $s=\epsilon+iy$, where $\epsilon$ is a fixed real number such
that $-\ff15<\epsilon<0$ and $y$ goes from $-\infty$ to
$+\infty$. This contour is completed to the left or to the right in
order to get the analytic continuation. These contour integrals
converge (and thus the analytic continuation is valid) for
\begin{equation}
  -2\pi < \arg z < 0.  \label{eq:argz}
\end{equation}

Now consider the sheaves $\O(m)$ on $\P^4$ as discussed in section
\ref{sss:locfree}. These restrict to $X$ to form locally-free sheaves
of rank one which we denote $\O_X(m)$. Denoting the generator of
$H^2(X,\Z)$ by $e$,
\begin{equation}
\begin{split}
  \int_X \exp(-te)\ch(\O_X(m))\sqrt{\td(X)}
   &= \int_X \exp(-te+me) \sqrt{1+\ff56 e^2}\\
   &= \ff5{12}(m-t)(2m^2+2t^2-4mt+5).
\end{split}
\end{equation}
Assuming that the term proportional to $\zeta(3)$ in (\ref{eq:Phi3})
arises from the 4-loop correction in section \ref{sss:Pi}, we may
determine the period associated to $\O_X(m)$ exactly from this leading
behaviour. The result is that
\begin{equation}
\begin{split}
Z(\O_X(m)) &= \ff56m(m^2+5)\Phi_0 - \ff52(m^2+m+2)\Phi_1
       +\ff52m\Phi_2 -\ff56\Phi_3\\
  &= \ff16(5m^3+3m^2+16m+6)\varpi_0 - \ff12(3m^2+3m+2)\varpi_1
     -m^2\varpi_2 -\ff12m(m-1)\varpi_3,
\end{split} \label{eq:Zex}
\end{equation}
and, in particular, $Z(\O_X)=\varpi_0-\varpi_1$. The conifold point,
$z=5^{-5}$ lies at the edge of the radius of convergence of the
various power series above. One can show that the power series
(\ref{eq:Phi0}) for $\Phi_0=\varpi_0$ in terms of $z$ is convergent at
the conifold point and is clearly real. For real $z^{\frac15}$,
(\ref{eq:varpi}) tells us that $\varpi_1$ is the complex conjugate of
$\varpi_0$. Thus, at the conifold point, $\varpi_0=\varpi_1$ and so
$Z(\O_X)=0$, i.e., $\O_X$ becomes massless.

We know that the conifold point corresponds to a singular conformal
field theory since a soliton, i.e., D-brane, becomes massless
\cite{Str:con,GMS:con}. Thus it seems natural to assume that the
B-brane $\O_X$ is the one in question. Note that we have not quite proven
this since we haven't proven that $\O_X$ is stable at the conifold
point. We will assume this stability in order to proceed.

We will have much to say about monodromy in section \ref{sss:qmon} but
for now let us note the following simple observation. Around the large
radius limit of the quintic, monodromy corresponds to $B\to B+1$. From
the gauge invariance discussed in section \ref{sss:Ab}, such a shift
is accompanied by a shift $F\to F+1$ in the curvature of the bundle
over any D-brane. Thus $\O_X(m)$ becomes $\O_X(m+1)$. This means that
the D-brane corresponding to $\O_X(m)$ becomes massless at the
conifold point if we circle the large radius limit $m$ times before
proceeding towards the conifold point.

\subsubsection{4-branes}  \label{sss:q4}

We can now study the stability of 4-branes on the quintic. Consider
the following short exact sequence of sheaves:
\begin{equation}
\xymatrix@1{
  0\ar[r]&\O_X(a)\ar[r]^f&\O_X(b)\ar[r]&\O_S(b)\ar[r]&0,
} \label{eq:D4}
\end{equation}
where $b>a$ and $f$ is a function of homogeneous degree $b-a$ as
discussed in section \ref{sss:olo}. $S$ is a divisor, i.e., a subspace
of complex codimension one and $\O_S(b)$ is the sheaf corresponding to
a line bundle of degree $b$ over $S$. All 4-branes corresponding to
B-branes on the quintic correspond to $\O_S(b)$ for some $f$ and some
$b$.

To leading order at large radius we have
\begin{equation}
\begin{split}
\xi(\O_X(m)) &= \frac1\pi\arg\int_X \exp((m-B-iJ)e)\\
  &= \frac1\pi\arg\bigl(5(m-B-iJ)^3\bigr)\\
  &= \frac3\pi\theta_m-3,
\end{split} \label{eq:xim0}
\end{equation}
where $\theta_m$ is the angle in the complex $(B+iJ)$-plane made
between the positive real axis and the line from $m$ to
$B+iJ$. Applying the $\Pi$-stability criterion to the distinguished
triangle associated to (\ref{eq:D4}) gives
$\xi(\O_X(b))-\xi(\O_X(a))<1$ which yields
\begin{equation}
  \theta_b-\theta_a <\frac\pi3.
\end{equation}
When this is satisfied, the open string corresponding to $f$ in
(\ref{eq:D4}) is tachyonic. Simple geometry yields that this
corresponds to the points above a circular arc in the upper
$(B+iJ)$-plane with centre $\ff12(a+b)+\ff1{2\sqrt3}(b-a)i$ and radius
$\ff1{\sqrt{3}}(b-a)$.

As expected, these 4-branes are stable in the large radius limit.
Below this arc of marginal stability the 4-brane decays into $\O_X(b)$
and $\O_X(a)[-1]$, that is a 6-brane and anti-6-brane with some
4-brane charges.  

The classical $\mu$-stability criterion of section \ref{sss:mu} would
imply that a 4-brane is always stable since it has no subobjects. We
emphasize that $\mu$-stability does {\em not\/} fail because we have
not taken enough $\alpha'$ corrections into account --- after all
(\ref{eq:xim0}) was only an approximation anyway. Rather
$\mu$-stability fails because of the qualitative aspect that it is
defined for abelian categories. This means it can never see decays
caused by anti-branes.

We also draw attention to the fact that it is misleading to think that
$\Pi$-stability corrections can always be ignored at large radius.
For very large values of $b-a$ this line of marginal stability can
extend to large values of $J$.

\iffigs
\begin{figure}[t]
  \centerline{\epsfxsize=15cm\epsfbox{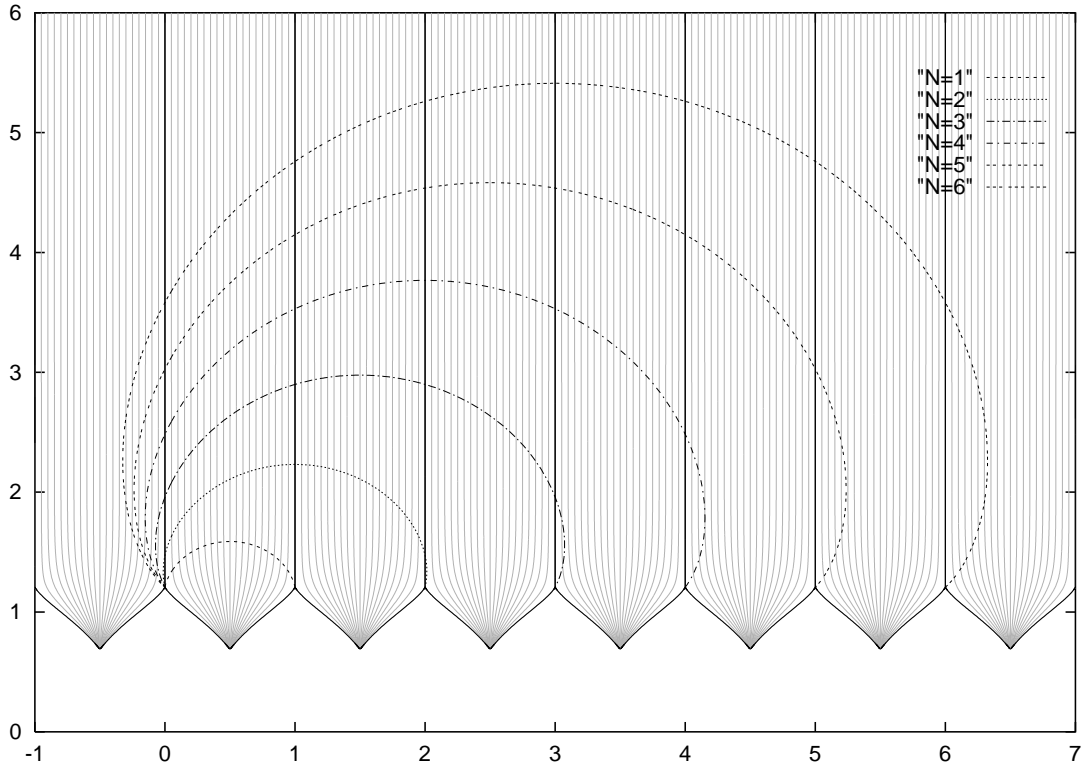}}
  \caption{Stability of various D4-branes in the $B+iJ$-plane.}
  \label{f:D4}
\end{figure}
\fi

Of course, we do not have the precise form of the line of marginal
stability since we used the large radius approximation in
(\ref{eq:xim0}). We may use numerical computation techniques and the
exact form of the periods from (\ref{eq:Zex}) to plot these curves
more precisely as was done in \cite{AD:Dstab}. The result is shown in
figure \ref{f:D4} for $a=0$ and $b=N$. The shaded areas of this figure
represent fundamental regions of the moduli space as in figure
\ref{f:scorp}.

We see that the lines of marginal stability in figure \ref{f:D4} are
not so far from being circular arcs. Note that the lines of marginal
stability end on the conifold points precisely where one of the decay
products becomes massless, and thus the grading becomes poorly-defined.

It should be emphasized that we have not strictly proven that either
the 4-branes are stable above the lines in figure \ref{f:D4} or that
the 4-branes are unstable below the lines. The 4-branes might decay by
other channels before these lines are reached, and the decay products
might also be unstable by the time we reach a line of marginal
stability. Having said that, it seems hard to imagine that figure
\ref{f:D4} is incorrect. We know the B-branes $\O_X(m)$ are stable
near the conifold points since they are supposed to be responsible for
the singularities at the conifold. Other more contrived modes of decay
of 4-branes always seem to decay at smaller radii.

\subsubsection{Exotic B-branes}  \label{sss:qex}

Having spent most of the lectures extolling the virtues of the derived
category, we have yet to see an example of a B-brane that doesn't
correspond to a single term complex --- i.e., a coherent sheaf. It is
time to rectify this situation.

The idea, as suggested in \cite{Doug:S01}, is to apply Serre duality
to the 4-brane decay of section \ref{sss:q4}. The potential tachyons
of that discussion lie in $\Hom(\O_X(a),\O_X(b))$ which is nonzero for
$b>a$. By the Serre duality of section \ref{sss:shco}, this Hilbert
space of open strings is isomorphic to
$\Hom(\O_X(b),\O_X(a)[3])$. Thus, we may consider the distinguished
triangle
\begin{equation}
\xymatrix{
  &\cX_{a,b}\ar[dl]|{[1]}&\\
  \O_X(b)\ar[rr]^g&&\O_X(a)[3],\ar[ul]
} \label{eq:exot1}
\end{equation}
where $\cX_{a,b}$ is defined as the cone of the map $g$.

The stability of $\cX_{a,b}$ can now be determined (at least relative
to the triangle (\ref{eq:exot1})) from the analysis above together
with the relation (\ref{eq:Bsh0}). For stability we require
$\xi(\O_X(a)[3])-\xi(\O_X(b))=3+\xi(\O_X(a))-\xi(\O_X(b))<1$. Using
the approximation (\ref{eq:xim0}) this yields that $\cX_{a,b}$ is
stable {\em below\/} a circular arc in the upper-half plane with
centre $\ff12(a+b)-\ff1{2\sqrt3}(b-a)i$ and radius
$\ff1{\sqrt{3}}(b-a)$. In particular, $\cX_{a,b}$ is always unstable
in the large radius limit, although we may make it stable at any given
arbitrarily large radius by choosing a large enough value of $b-a$.

\iffigs
\begin{figure}[t]
  \centerline{\epsfxsize=15cm\epsfbox{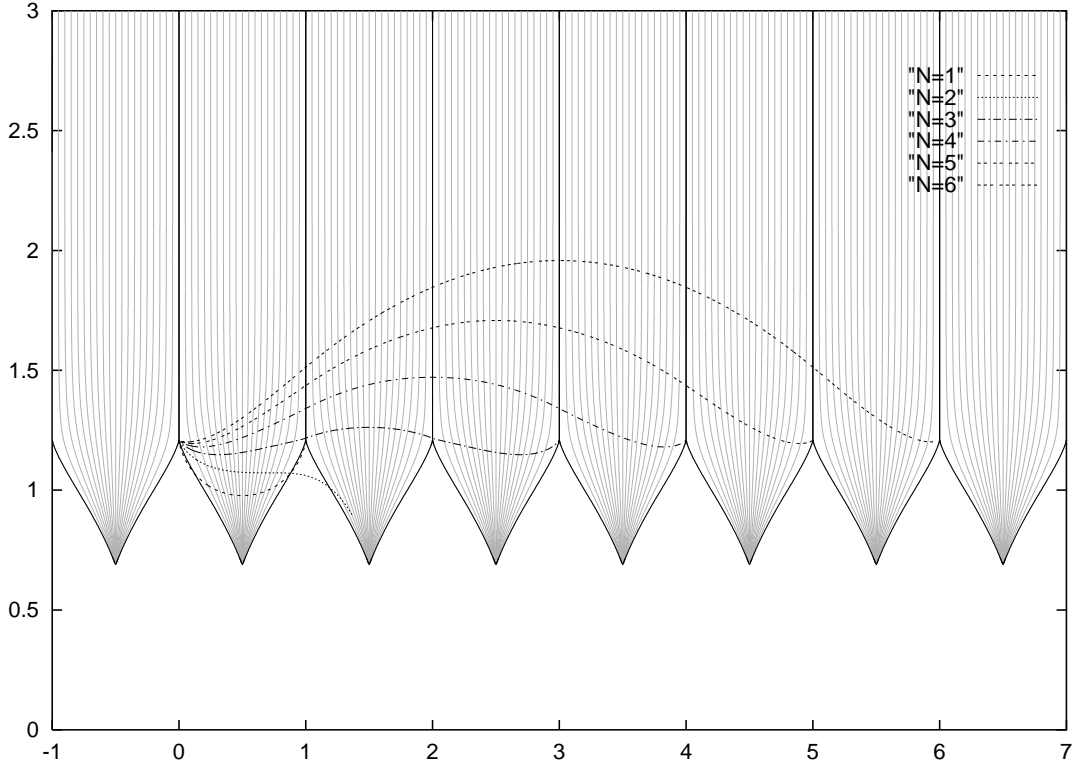}}
  \caption{Stability of the exotic objects $\cX_{0,N}$ in the $B+iJ$-plane.}
  \label{f:XN}
\end{figure}
\fi

Again we may use numerical techniques to plot a more precise version
of the lines of marginal stability. In figure \ref{f:XN} (again taken
from \cite{AD:Dstab}) we plot some examples. We should note that
something very interesting happens for $\cX_{0,2}$. It is an example
of a case where a decay product can itself decay forcing the line of
marginal stability to end on another line of marginal stability
(rather that a conifold point). We refer to \cite{AD:Dstab} for a full
discussion of this case.

So, what, exactly is $\cX_{a,b}$? It is a B-brane that exists purely
because of $\alpha'$ effects. It does not exist at large radius limit
and does not have a world-volume interpretation. Let's get a feel of
these exotic B-branes by examining $\cX_{0,1}$, say, a little more
closely. Suppose we have an injective resolution for $\O_X$:
\begin{equation}
\xymatrix@1{
  0\ar[r]&\O_X\ar[r]&\cI^0\ar[r]^{i_0}&\cI^1\ar[r]^{i_1}&\cI^2
  \ar[r]^{i_2}&\cI^3\ar[r]^{i_3}&\ldots
}
\end{equation}
Referring to section \ref{sss:shco}, an element of
$\Ext^3(\O_X(1),\O_X)$ corresponds to a map $g:\O_X(1)\to\cI_3$ such
that $i_3g=0$. Thus $\cX_{0,1}=\Cone(g)$ corresponds to the complex
\begin{equation}
\xymatrix@C=12mm{
  \ldots\ar[r]&0\ar[r]&\cI^0\ar[r]^{i_0}&\cI^1\ar[r]^(0.45){
  \left(\begin{smallmatrix}0\\i_1\end{smallmatrix}\right)}&
  {\begin{matrix}\O_X(1)\\ \oplus\\\cI^2\end{matrix}}
  \ar[r]^(0.55){\left(\begin{smallmatrix}g&i_2\end{smallmatrix}\right)}
  &\poso{\cI^3}\ar[r]^{i_3}&\ldots,
} \label{eq:excom}
\end{equation}
where we denote the zero position with a dotted underline.

The cohomology sheaves of this complex look like $\cH^{-3}=\O_X$ and
$\cH^{-1}=\O_X(1)$, so na\"\i vely this B-brane looks like an
anti-6-brane added to another anti-6-brane with a 4-brane charge. This
is certainly true if $g$ is zero. More precisely, 
(\ref{eq:excom}) is quasi-isomorphic to 
\begin{equation}
\xymatrix@1{
  \ldots\ar[r]&\O_X\ar[r]&0\ar[r]&\O_X(1)\ar[r]&\poso{0}
  \ar[r]&\ldots,
}
\end{equation}
if and only if $g=0$. To prove this assertion we may use the
discussion at the end of section \ref{sss:trian} to compute
$\Hom(\cX_{0,1},\cX_{0,1})$. After wading through several long
sequences and using the cohomology groups of $\O_X(m)$ as defined in
section \ref{sss:cech} we obtain
\begin{equation}
\Hom(\cX_{0,1},\cX_{0,1}) = \begin{cases}
  \C^2&\text{if $g=0$},\\
  \C&\text{if $g\neq0$}. \end{cases}
\end{equation}
This ties in nicely with the comments at the end of section
\ref{sss:tachy}. If $g\neq0$ we have a single irreducible D-brane. When
$g=0$ we have a direct sum of two D-branes and thus a gauge group
$\GU(1)\times\GU(1)$.

$\cX_{0,1}$ thus becomes a distinct object when the
tachyon is turned on. All this shows that $\cX_{0,1}$ is a truly
exotic object from the old point of view of D-branes. It cannot be
written as a complex with a single coherent sheaf and so it cannot be
viewed as a vector bundle supported on some subspace of $X$.

The reader should note that we were able to build these exotic
D-branes because we were able to use $\Ext^n$'s for $n>1$. This
separated terms in the complexes far enough to avoid everything
collapsing back to a single term complex. That such $\Ext$'s can ever
be physically relevant is traced back to fact that the grading $\xi$
was defined to live in $\R$ rather than on a circle. {\em If we didn't
extend the definition of $\xi$ in this way, we would never see the
exotic D-branes.}

We should note that one can directly attack the question of
identifying D-brane states at the Gepner point itself using the notion
of boundary conformal field theories. This was done in
\cite{RS:DGep,BDLR:Dq,FSW:DGep}. We will not review this method here
as it requires quite a bit of technology. However, this method is not
guaranteed to obtain {\em all\/} of the D-branes at the Gepner point.
It is possible to show that some of the charges of the exotic D-branes
we found above can coincide with charges found using this direct
method. We should add that there has also been recent progress in more
advanced methods of analyzing the Gepner model
\cite{KL:TLG1,KL:TLG2,AADia:GepD} which appear to fill in the missing
states of the boundary conformal field theory method.

\subsubsection{Monodromy}  \label{sss:qmon}

One of the easiest ways of seeing that B-branes as subspaces is an
inadequate picture is to think about monodromy in D-branes as we move
around loops in the moduli space.

In the case of A-branes, such monodromy is purely classical. The
periods of the holomorphic 3-form over integral 3-cycles undergo
non-trivial monodromy as we go around non-contractible loops in the
moduli space of complex structure. Such monodromy may be interpreted
as an automorphism of $H_3(Y,\Z)$ which preserves the intersection
form between 3-cycles. Thus the homology classes of A-branes undergo
monodromy and thus the A-branes themselves undergo monodromy. 

How is monodromy seen in the B-branes? In this case, the D-brane
charge is an element of $H^{\textrm{even}}(X,\Z)$. In the case of the
quintic we know exactly how to map between $H_3(Y,\Z)$ and
$H^{\textrm{even}}(X,\Z)$ thanks to the analysis of section
\ref{sss:qper}. Thus we can copy the monodromy action from the A-model
into the B-model. If one begins with a particular sheaf $\cF$, one can
then compute the monodromy action on $\ch(\cF)$ to get some idea of
what this B-brane becomes under monodromy. In this way, it looks as if
2-cycles may becomes mixtures of 2-cycles and 0-cycles etc., which is
certainly not classical! Furthermore, it is not hard to find examples
where the resulting element of $H^{\textrm{even}}(X,\Z)$ cannot
correspond to the Chern character of any sheaf (for example, the rank
of the bundle might be negative). Fortunately going to the derived
category solves these problems.

Upon going around a loop in the moduli space the physics must be
unchanged of course. This means that monodromy may shuffle the objects
in $\DC(X)$ around but, at the end of the day, we should be able to
relabel everything to restore the original spectra of open strings
etc. In category language this means that monodromy induces an
``autoequivalence'' on $\DC(X)$. That is, we have a functor from 
$\DC(X)$ to $\DC(X)$ that is a bijection on the (isomorphism classes
of) objects and that preserves the corresponding morphisms.

Such an autoequivalence on $\DC(X)$ is always induced by a so-called
``Fourier--Mukai transform'. To define a Fourier--Mukai transform one
begins with a fixed object $\mathcal{Z}$ (called the ``kernel'') in the
derived category of $X\times X$. We define the projection maps from
$X\times X$ to its first and second factor:
\begin{equation}
\xymatrix@C=5mm{
  &X\times X\ar[dl]_{p_1}\ar[dr]^{p_2}&\\
 X&&X\\
}
\end{equation}

Given any object $A\in\DC(X)$ we then
define the transformed object (or morphism)
\begin{equation}
  T_\mathcal{Z}(A) = \mathbf{R}p_{2*}(\mathbf{L}p_1^*A\Lotimes\mathcal{Z}).
    \label{eq:FM1}
\end{equation}
Some explanation of notation is required here. We sequentially
apply three functors in (\ref{eq:FM1}) which have been either
``left-derived'' or ``right-derived'' hence the ``$\mathbf{L}$'' or
``$\mathbf{R}$''.  This part is taken care of automatically by the
derived category machinery as follows. In each case, we take an object
in the derived category and choose a representative complex which
satisfies some nice property required by the functor. The functor can
then be applied to the complex. The three functors are:
\begin{enumerate}
\item $\mathbf{L}p_1^*$: We take the complex to be a complex of
  locally-free sheaves or, equivalently, vector bundles. $p_1^*$ is
  then the usual pull-back map on vector bundles.\footnote{Technically
  speaking $p_1$ is flat so we don't need to left-derive this functor.}
\item $\displaystyle\Lotimes\mathcal{Z}$: Again the complex must be a complex
  of locally-free sheaves. $\otimes$ is then the usual tensor product
  of sheaves. Acting on complexes, $\otimes$ produces a double complex
  which can then be collapsed back to a single complex in the usual
  way.
 \item $\mathbf{R}p_{2*}$: In this case the complex must be a complex
  of injective objects. $p_{2*}$ is then the push-forward map for
  sheaves as defined in section \ref{ss:cohB}.
\end{enumerate}

Since the derived functor machinery is built into the derived
category, we will usually drop the $\mathbf{L}$'s and $\mathbf{R}$'s from the
notation from now on.

Note that if $p:X\to\{q\}$ is map to a point, then $p_*$ is the global
section functor. Thus, its right derived functors $\mathbf{R}p_*$
amount to sheaf cohomology as explained in section \ref{sss:shco}. In
the above, $p_2:X\times X\to X$ is a fibration and therefore, roughly
speaking, its right-derived functors amount to taking cohomology in
the fibres.

It was proven in \cite{Orl:FM} that any autoequivalence of $\DC(X)$
could be written as (\ref{eq:FM1}) for a suitable choice of
$\mathcal{Z}$. Conversely, Bridgeland \cite{Brid:faith} has established the
conditions on $\mathcal{Z}$ for (\ref{eq:FM1}) to yield an autoequivalence of
$\DC(X)$. 

As a warm-up exercise let us consider how we produce the identity
Fourier--Mukai transform. Let the ``diagonal'' map $\Delta_X:X\to X\times
X$ be given by $\Delta_X(x)=(x,x)$, and let us denote $\Delta_{X*}\O_X$ by
$\O_{\Delta X}$. Suppose $\cF$ is a sheaf on $X$. Then
$p_1^*\cF\otimes\O_{\Delta X}$ is a sheaf supported on the image of
$\Delta_X$. In fact, it is precisely $\cF$ on the image of $X$ under
the map $\Delta_X$. This image is mapped identically back to $X$ by
$p_2$ and so $T_{\O_{\Delta X}}(\cF)=\cF$. Thus $\O_{\Delta X}$ generates the
identity Fourier--Mukai transform of sheaves and therefore the derived
category.

The (stringy) moduli space of $B+iJ$ on the quintic is given by a
2-sphere with 3 punctures corresponding to large radius limit, the
Gepner point and the conifold point. Thus we just need to find the
monodromy around two of these points and the third may then be
deduced.

The monodromy around the large radius limit is easy. We know that this
corresponds to $B\to B+1$ which is equivalent to a shift $F\to F+1$ in
the curvature of the bundles. This shift in curvature can be achieved
by a transformation $\cF\to\cF\otimes\O_X(1)$ on sheaves. In terms of a
Fourier--Mukai transform, this may be achieved by a kernel
$\mathcal{L}=\Delta_*\O_X(1)=\O_{\Delta X}(1)$ in a similar way to the
identity transform discussed above. We will denote this transform
$T_{\mathcal{L}}$ by $L$.

Let us next consider the Fourier--Mukai transform associated to
monodromy around the conifold point. We will denote this transform
$K$. Unlike the large radius limit, there is no direct argument that
yields $K$ in a concrete way. Instead, let us first consider the
action of $K$ on the D-brane charges. This may be deduced from the
periods listed in section \ref{sss:qper}. We may compose a loop around
the large radius limit (given by $z\mapsto e^{2\pi i}z$ for small $z$)
by a loop around the Gepner point (given by $z\mapsto e^{-2\pi i/5}z$
for large $z$) to obtain monodromy around the conifold point. For
details of the explicit monodromy matrices see \cite{CDGP:} and also
\cite{HIV:D-mir,GJ:McK,Tomas:McK,Mayr:McK} for further details about
this monodromy. The result may be expressed in the following concise
form:
\begin{equation}
  \ch(K(\cF)) = \ch(\cF) - \langle \O_X,\cF\rangle\ch(\O_X),
    \label{eq:monch}
\end{equation}
where $\langle,\rangle$ is the natural inner product given by
(\ref{eq:Bint}). We would like to do better than this however. We want
to know the monodromy transform $K$ itself, rather than just the
action on charges.

Clearly, since $Z(\O_X)$ has a simple zero at the conifold point,
monodromy around this point will shift $\xi(\O_X)$ by 2. This means
that stability conditions on any decay involving $\O_X$ may well be
affected by this monodromy. For example, consider the 0-brane
$i_*\O_p$ given by the inclusion of a point $i:p\hookrightarrow
X$. For brevity, we will refer to this skyscraper sheaf as $\O_p$ as
in section \ref{sss:coh}. From (\ref{eq:ideal0}) we obtain the
following distinguished triangle
\begin{equation}
\xymatrix{
  &\cI_p[1]\ar[dl]|{[1]}_(0.4)b&\\
  \O_X\ar[rr]^a&&\O_p.\ar[ul]_c
} \label{eq:0dec}
\end{equation}
At large radius limit we expect $\O_X$ and $\O_p$ to be stable (as
they are the basic 6-brane and 0-brane respectively) and $\cI_p$ (and
thus $\cI_p[1]$) to be unstable (since it doesn't correspond to a
vector bundle over a subspace). The latter can be seen directly from
the fact that $\xi(\O_p)-\xi(\O_X)=\ff32>1$ at large radius, making the
open string $a$ massive in (\ref{eq:0dec}).

Near the conifold point $Z(\O_p)$ and $Z(\cI_p)$ are non-zero and so the
$\xi$'s are roughly fixed for these B-branes in a small neighbourhood
of the conifold point when they are defined, i.e., when the B-branes
are stable. Thus, as we orbit the conifold point in the direction of
increasing $\xi(\O_X)$, we can make $\xi(\O_p)-\xi(\O_X)$ smaller and
thus $a$ tachyonic. That is, $\cI_p[1]$ becomes stable. At this
instant $\xi(\cI_p[1])=\xi(\O_p)$ from our $\Pi$-stability rules. Thus
when these B-branes are stable, they have approximately the same value
of $\xi$ in a small neighbourhood of the conifold point. This implies
that the open string $c$ in (\ref{eq:0dec}) is always tachyonic as
expected. 

As we continue around the conifold point increasing $\xi(\O_X)$
further, we can make $\xi(\O_X)-\xi(\cI_p[1])>0$ and thus $\O_p$
unstable. It is easy to see that this will happen an angle $\pi$ later
than the formation of $\cI_p[1]$.  The diagram (\ref{eq:0dec})
therefore encodes both the formation of $\cI_p[1]$ and the decay of
$\O_p$ as we encircle the conifold point. We refer to \cite{AD:Dstab}
for more analysis of precisely where these events occur.

The monodromy action of $\DC(X)$ should be seen as relabeling of the
B-branes as objects in the category so that the physics remains
unchanged. Thus, since $\O_p$ was stable to begin with but has now become
unstable, it must be replaced by something. The natural choice would
seem to be $\cI_p[1]$. Thus we conjecture $K(\O_p)=\cI_p[1]$. This is
consistent with the charges (\ref{eq:monch}). Furthermore $\xi(\O_X)$
has increased by 2 upon looping once around the conifold. Thus, to
restore physics, we should assert $K(\O_X)=\O_X[-2]$ to compensate for
this.

Let us define the notation $A\boxtimes B$ to mean $p_1^*A\otimes
p_2^*B$ for $A,B\in\DC(X)$. Now, consider the Fourier--Mukai transform
induced by 
\begin{equation}
   \mathcal{K} =
   \bigl(\xymatrix@1{\O_X\boxtimes\O_X\ar[r]^-r&\poso{\O_{\Delta X}}}
   \bigr), \label{eq:PsiK}
\end{equation}
where $r$ is the obvious restriction map.  Let us apply this transform
to a sheaf $\cF$. We know the $\O_{\Delta X}$ in position zero of
(\ref{eq:PsiK}) acts as the identity. As
$\O_X\boxtimes\O_X=\O_{X\times X}$, then
$p_1^*\cF\otimes\mathcal{K}$ is simply $p_1^*\cF$. Pushing this down
via $p_2$ will produce a sum of copies of $\O_X$ since $p_1^*\cF$ is
trivial (i.e., free) in this direction. As mentioned above, the act of
pushing down takes sheaf cohomology in the fibre direction which in
turn is equivalent to the functor $\Hom(\O_X,-)$. This means that, in
the derived category,
\begin{equation}
  p_{2*}p_1^*\cF = \Hom(\O_X,\cF)\otimes\O_X,
\end{equation}
which implies
\begin{equation}
  T_{\mathcal{K}}(\cF) = \Cone\bigl(\xymatrix@1{\Hom(\O_X,\cF)\otimes\O_X
      \ar[r]^-r&\cF\bigr)}, 
  \label{eq:TKq}
\end{equation}
where $r$ is now an obvious ``evaluation'' map.
This Fourier--Mukai transform reproduces the desired monodromy above
on $\O_p$ and $\O_X$. For example put $\cF=\O_X$. Then, from section
\ref{sss:cech} we may compute
\begin{equation}
H^k(X,\O_X) = \begin{cases}
  \C&\text{if $k=0$ or 3}\\
  0&\text{otherwise,} \end{cases}
\end{equation}
which implies $\Hom(\O_X,\cF)$ can be represented by a complex of
vector spaces
\begin{equation}
\xymatrix@1{
  \ldots\ar[r]&0\ar[r]&\poso{\C}\ar[r]&0\ar[r]&0\ar[r]&
     \C\ar[r]&0\ar[r]&\ldots.
}
\end{equation}
Tensoring by $\O_X$ simply replaces the $\C$'s by $\O_X$'s. The cone
construction then shifts this one place left and the first $\O_X$
cancels with the $\O_X$ on the right of the cone (since the evaluation
map is the identity) leaving $\O_X[-2]$.

A result of Bridgeland and Maciocia \cite{BM:FMq} may now be used to
show that (\ref{eq:PsiK}) is actually the {\em only\/} Fourier--Mukai
transform which produces the desired monodromy on $\O_p$ (for all $p\in
X$) and $\O_X$. We refer to \cite{AD:Dstab} for more details. Thus we
will assume that this gives the correct monodromy on $\DC(X)$ for
loops around the conifold point. Note that this particular
transformation is well-known in the study of ``helices and mutations''
(see, for example, \cite{KO:excdP} and references therein). It was
first conjectured to be applicable to monodromy in the context we are
considering by Kontsevich \cite{Kont:mon} and
was subsequently studied extensively by Seidel and Thomas
\cite{ST:braid}. It is also a special case of Horja's model of monodromy
\cite{Horj:DX,Horj:EZ,AKH:m0}.

The monodromy around the Gepner point can then be constructed by
composing the monodromies around the large radius limit and the
conifold point. In general, two Fourier--Mukai transforms based on
kernels $\mathcal{A}$ and $\mathcal{B}$ may be composed to form a
transform with a kernel
\begin{equation}
\mathcal{B}\star \mathcal{A}= p_{13*} \big( p_{12}^*\mathcal{A}\otimes
p_{23}^*\mathcal{B}\big), \label{eq:FMC}
\end{equation}
where $p_{ij}$ are the obvious projection maps from $X\times X\times
X$ to $X\times X$. Therefore, if $G$ is the monodromy around the
Gepner point, and $\mathcal{G}$ is the corresponding kernel, we may
compute
\begin{equation}
\begin{split}
\mathcal{G} &= \mathcal{K}\star \mathcal{L}\\
  &= \bigl(\xymatrix@1{\O_X(1)\boxtimes\O_X\ar[r]&
     \poso{\O_{\Delta X}(1)}}\bigr).
\end{split}
\end{equation}

Consider the result of monodromy $G^5$, i.e., five times around the
Gepner point. Given that the Gepner model is a $\Z_5$-orbifold one
might be forgiven that thinking that $G^5$ induces a trivial
monodromy, i.e., $\mathcal{G}^{\star5}=\O_{\Delta X}$. Surprisingly this
is not the case as we now show.\footnote{I thank S.~Katz for guiding
me in this computation. He in turn thanks A.~Bondal for a related
conversation.} We show the full details of this computation here so
that the reader can get a good feel for manipulations in the derived
category. Anyone not interested in these details should, of course,
skip ahead to the answer!

Let $\mathcal{S}=\O_X(1)\boxtimes\O_X$ so that
$\mathcal{G}=\Cone(\mathcal{S}\to\O_{\Delta X}(1))$. The functors in
(\ref{eq:FMC}) preserve distinguished triangles in $\DC(X\times X)$,
so $\mathcal{S}\star\mathcal{G}=\Cone(\mathcal{S}\star\mathcal{S}
\to\mathcal{S}\star\O_{\Delta X}(1))$. Now
\begin{equation}
\begin{split}
  \mathcal{S}\star\mathcal{S} &= 
             p_{13*}(\O_X(1)\boxtimes\O_X(1)\boxtimes\O_X)\\
     &= \O_X(1)\boxtimes\O_X^{\oplus5},
\end{split}
\end{equation}
since,
\begin{equation}
  H^k(X,\O_X(1)) = \begin{cases} \C^5&\text{if $k=0$,}\\
                    0&\text{otherwise.} \end{cases}
\end{equation}
Also, $\mathcal{S}\star\O_{\Delta X}(1)$ is easily seen to be
$\O_X(1)\boxtimes\O_X(1)$. Thus
\begin{equation}
\begin{split}
  \mathcal{S}\star\mathcal{G}&=\bigl(\xymatrix@1{
    \O_X(1)\boxtimes\O_X^{\oplus5}
    \ar[r]&\poso{\O_X(1)\boxtimes\O_X(1)}}\bigr)\\
  &= \O_X(1)\boxtimes\Omega(1)|_X[1].
\end{split} \label{eq:SG0}
\end{equation}
This follows from the Euler exact sequence\footnote{In this and
subsequent computations we should really keep track of the precise
forms of the morphisms. We omit this for brevity. In each case the
morphisms form representations of the $\PGl(5,\C)$ symmetry acting on
the homogeneous coordinates of $\P^4$. These representations can be
handled conveniently using Young diagrams and Schur functors as in
chapter 6 of \cite{FulHar:rep}.}
\begin{equation}
\xymatrix@1{
0\ar[r]&\Omega(1)\ar[r]&\O^{\oplus5}\ar[r]&\O(1)\ar[r]&0,
} \label{eq:Eul}
\end{equation}
on $\P^4$ restricted to $X$. Here $\Omega$ is the cotangent sheaf of
$\P^4$ and $\Omega(1)=\Omega\otimes\O(1)$. The latter sheaf is
restricted to $X$ in (\ref{eq:SG0}).\footnote{Do not confuse
$\Omega|_X$ with $\Omega_X$, the cotangent bundle of $X$!} Also
\begin{equation}
\O_{\Delta X}(1)\star\mathcal{G}=\bigl(\xymatrix@1{
  \O_X(2)\boxtimes\O_X\ar[r]&
     \poso{\O_{\Delta X}(2)}}\bigr),
\end{equation}
and so 
\begin{equation}
\begin{split}
  \mathcal{G}\star\mathcal{G}&=\Cone\bigl(\mathcal{S}\star\mathcal{G}\to
     \O_{\Delta X}(1)\star\mathcal{G}\bigr)\\
  &= \bigl(\xymatrix@1{
   \O_X(1)\boxtimes\Omega(1)|_X\ar[r]&\O_X(2)\boxtimes\O_X\ar[r]&
     \poso{\O_{\Delta X}(2)}}\bigr).
\end{split}
\end{equation}
Similarly we may prove that
\begin{equation}
\mathcal{G}\star\mathcal{G}\star\mathcal{G}=\bigl(\xymatrix@1{
   \O_X(1)\boxtimes\Omega^2(2)|_X\ar[r]&
   \O_X(2)\boxtimes\Omega(1)|_X\ar[r]&\O_X(3)\boxtimes\O_X\ar[r]&
     \poso{\O_{\Delta X}(3)}}\bigr),
\end{equation}
where $\Omega^2$ is the second exterior power of $\Omega$ which, from
(\ref{eq:Eul}), fits into the following exact sequence:
\begin{equation}
\xymatrix@1{
0\ar[r]&\Omega^2(2)\ar[r]&\O^{\oplus10}\ar[r]&\O(1)^{\oplus5}\ar[r]
    &\O(2)\ar[r]&0.
}
\end{equation}
Continuing this process yields the desired kernel
\begin{multline}
\mathcal{G}^{\star5}=\bigl(\xymatrix@1{
  \O_X(1)\boxtimes\Omega^4(4)|_X\ar[r]&
  \O_X(2)\boxtimes\Omega^3(3)|_X\ar[r]&}\\
\xymatrix@1{
   \O_X(3)\boxtimes\Omega^2(2)|_X\ar[r]&
   \O_X(4)\boxtimes\Omega(1)|_X\ar[r]&\O_X(5)\boxtimes\O_X\ar[r]&
     \poso{\O_{\Delta X}(5)}}\bigr).
  \label{eq:G5}
\end{multline}
This looks very similar to an exact sequence due to Beilinson \cite{Bei:res}
for sheaves on $\P^4\times\P^4$:
\begin{equation}
\xymatrix@1{
  0\ar[r]&\O(-4)\boxtimes\Omega^4(4)\ar[r]&\ldots\ar[r]&
   \O(-1)\boxtimes\Omega(1)\ar[r]&\O\boxtimes\O
  \ar[r]&\O_\Delta\ar[r]&0,
} \label{eq:Bres}
\end{equation}
where $\O_\Delta$ is the structure sheaf of the diagonally embedded
$\P^4$. If we tensor (\ref{eq:Bres}) by
$\O(5)\boxtimes\O$ and restrict to $X$ we obtain the complex
$\mathcal{G}^{\star5}$. This process does not preserve the exactness
of (\ref{eq:Bres}) so we need to be a little careful. We claim
\begin{equation}
  \mathcal{G}^{\star5} = \Cone\Bigl(\O_\Delta(5)\Lotimes\O_{X\times
  X}\to\O_{\Delta X}(5)\Bigr). \label{eq:G5c}
\end{equation}
This is seen as follows. As stated earlier in this section, the
operation $\displaystyle\Lotimes$ requires the object on the left or
right of this symbol to be expressed in terms of locally free
sheaves. Then the usual $\otimes$ operator may be applied. The exact
sequence (\ref{eq:Bres}) tensored by $\O(5)\boxtimes\O$ gives the
desired locally free resolution of $\O_\Delta(5)$. The functor
$\displaystyle\Lotimes\O_{X\times X}$ is then precisely restriction
$X\times X$ thus yielding the complex (\ref{eq:G5}).

Alternatively we can compute (\ref{eq:G5c}) using a locally-free
resolution of $\O_{X\times X}$ in the form
\begin{equation}
\xymatrix@1{
  0\ar[r]&\O(-5)\boxtimes\O(-5)\ar[r]&
  \bigl(\O\boxtimes\O(-5)\bigr)\oplus\bigl(\O(-5)\boxtimes\O\bigr)\ar[r]&
  \O\boxtimes\O\ar[r]&\O_{X\times X}\ar[r]&0.
}
\end{equation}
This yields
\begin{equation}
\begin{split}
  \mathcal{G}^{\star5} &= \bigl(\xymatrix@1{
    \O_\Delta(-5)\ar[r]&\O_\Delta\oplus\O_\Delta\ar[r]&
    \O_\Delta(5)\ar[r]&\poso{\O_{\Delta X}(5)}}\bigr)\\
  &= \bigl(\xymatrix@1{
    \O_\Delta(-5)\ar[r]&\O_\Delta\ar[r]&
    0\ar[r]&\poso{0}}\bigr)\\
  &= \O_{\Delta X}[2].
\end{split}
\end{equation}
This is, of course, the identity Fourier--Mukai transform shifted left
2 places. This means that we have proven, in general, that
\begin{center}
\shabox{\parbox{.85\hsize}{The action of the monodromy on $\DC(X)$
associated to looping five times around the Gepner point corresponds
to shifting the complexes two places to the left.}}
\end{center}

So how worried should we be that this is not the identity? It doesn't
contradict any statement about physics. Monodromy five times around
the Gepner point does not affect physics --- but then again monodromy
once around the Gepner point is an invariance of physics too!

What this computation is telling us is that any form of the topological
field theory associated to the Gepner model, or equivalently a \LG\
orbifold theory, which explicitly exhibits the derived category
language for D-branes must not have a $\Z_5$ quantum symmetry. This
might make such a model rather awkward to construct. 

While these notes were being completed, the paper \cite{AADia:GepD}
appeared which is based on the work of \cite{KL:TLG1,KL:TLG2} which,
in turn, is based upon the ideas of \cite{Orlov:LG}. These interesting
papers apply the derived category description directly to
Landau--Ginzburg theories and so should provide exactly the model we
are looking for. Unfortunately the shift-of-two ambiguity above
appears to be evaded in these papers by identifying such a shift with
the identity. The complete understanding of how these models fit into
the full picture has yet to be completely elucidated and so we will
not attempt to review these ideas here, although clearly they have
much to offer.

We will see in section \ref{sss:mon2} that the geometrical orbifolds
do not appear to be plagued by this shift of two anomaly. It therefore
seems to be intrinsic to the \LG\ theories in some sense.

\subsection{Flops}  \label{ss:flop}

While the quintic \CY\ threefold provides the simplest example of a
compact \CY\ threefold, we may consider simpler examples of stability
by going to noncompact cases. Perhaps the easiest, and most
interesting, is provided by the ``flop''.

The geometry of flops has been extensively reviewed elsewhere. We
refer to \cite{Greene:TASI} for example for more details. The general
idea is that there is a singular algebraic variety $X_0$ that contains
a conifold point. This singular space may be resolved by replacing the
conifold point by a $\P^1$ in two different ways to form smooth \CY\
manifolds $X$ or $X'$. Generally $X$ and $X'$ are topologically
distinct.

Geometrically the process of blowing down $X$ or $X'$ back to $X_0$
may be viewed as a deformation of the K\"ahler form $J$. Indeed, in
the space of K\"ahler forms, $X$ and $X'$ may be considered to live on
the two sides of a wall corresponding to $X_0$. Let $C$ be the
$\P^1$ inside $X$. As we approach the wall from the $X$ side, the area
of $C$ shrinks down to zero. Continuing past this wall would give $C$
negative area but by reinterpreting the geometry in terms of $X'$ we
give positive area to the new $C'\subset X'$. The process of passing
from $X$ to $X'$ is called a ``flop''.

In the context of the \nlsm, we have the $B$-field in addition to
$J$. This has a profound effect as described in \cite{AGM:II}. The
conformal field theory associated to the singular target space $X_0$
is perfectly well-defined and finite so long as the component of $B$
associated to $C$ (or $C'$) is nonzero. Thus, rather than having a
real codimension one wall of singular spaces in the moduli space of
$J$, we have a {\em complex\/} codimension one subspace of singular
conformal field theories in the moduli space of $B+iJ$. It follows
that one can pass from the $X_1$ ``phase'' to the $X_2$ ``phase''
smoothly by going around this singular subset. The conformal field
theory does not witness any ``jump'' as we move between these phases.

Once we bring D-branes into the picture we see that we must have a jump in
some sense. At least to some degree of approximation, the \CY\ target
space is the moduli space of 0-branes. Thus the moduli space of
0-branes must undergo some transition during the flop even if we avoid
the singular conformal field theory at $B=0$. This discontinuity is
provided by 0-brane stability considerations as we now show. This
calculation was first suggested in \cite{Doug:DC} and performed in
\cite{me:point} (modulo sign conventions). The flop was studied by
Bridgeland \cite{Brig:flop} in the context of the derived category and
many of the observations in that paper are relevant here.

\iffigs
\begin{figure}
  \centerline{\epsfxsize=4cm\epsfbox{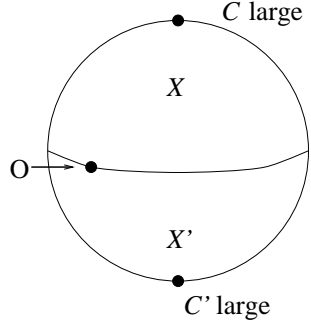}}
  \caption{The moduli space of $B+iJ$ for the flop.}
  \label{f:BJs}
\end{figure}
\fi

We imagine that we are in a \CY\ threefold $X$ where everything has a very
large area except for the flopping $\P^1$. The periods on the mirror
of this may be analyzed simply in this limit as explained in
\cite{AGM:sd}. Essentially the only component of $B+iJ$ of interest is
given by
\begin{equation}
  t = \int_C B+iJ. 
\end{equation}
This has a moduli space given by $\P^1$ as shown in figure
\ref{f:BJs}. The flop takes place on the equator and the singular
conformal field theory exists at the point labeled $O$.  Let $z$ be
the affine coordinate on this $\P^1$ so that $z=0$ in the large $C$
limit, $z=1$ corresponds to $O$ and $z=\infty$ gives the large $C'$
limit after the flop transition. The analysis of \cite{AGM:sd} then
yields the simple result that the periods on the mirror have a general
form
\begin{equation}
  \Phi=A_1 + A_2\log(z).
\end{equation}
Thus we have the {\em exact\/} relation given by the mirror map
\begin{equation}
  t = \frac1{2\pi i}\log(z). 
     \label{eq:tlog}
\end{equation}
Let $i:C\to X$ be the inclusion map and let $\O_C(m)$ denote
$i_*\O(m)$. We will use $\O_x$ to denote the skyscraper sheaf, i.e.,
0-brane, associated with the point $x\in X$.
Using the Grothendieck--Riemann--Roch theorem of section \ref{ss:K} we
see 
\begin{equation}
  \int_Xe^{-(B+iJ)}\ch(\O_C(m))\sqrt{\td(T_X)} = -t+m+1.
\end{equation}
Therefore $Z(\O_C(m))=-t+m+1$ exactly since $t$ and 1 are periods.  The
most natural statement would seem to be therefore that the brane
corresponding to $\O_C(-1)$ becomes massless at $O$.  Actually we are
free to choose the branch of the logarithm however we feel and we
could, instead, say that $\O_C$ becomes massless for simplicity. This
is equivalent to choosing a basepoint near the large radius limit but
then going once around this large radius limit before heading towards
$O$.  With this choice, we focus on this neighbourhood of $O$ by
putting $t=1+\epsilon e^{i\theta}$ for a small real and positive
$\epsilon$. We sketch this neighbourhood in figure \ref{f:theta}.

\iffigs
\begin{figure}[t]
  \centerline{\epsfxsize=6cm\epsfbox{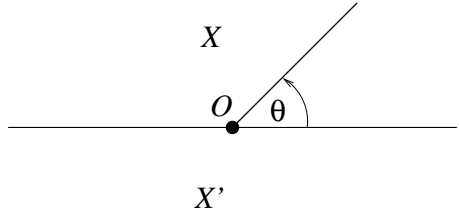}}
  \caption{The neighbourhood of $O$.}
  \label{f:theta}
\end{figure}
\fi

Suppose $x\in C$. Then we have a short exact sequence
\begin{equation}
\xymatrix@1{
  0\ar[r]&\O_C(-1)\ar[r]&\O_C\ar[r]&\O_x\ar[r]&0.
}
\end{equation}
This leads to a distinguished triangle we write in the form
\begin{equation}
\xymatrix@C=1mm{
  &\O_C(-1)[1]\ar[dl]|{[1]}_(0.6){\ff{\theta}{\pi}}&\\
  \O_C\ar[rr]_{1-\ff{\theta}{\pi}}&&\O_x.\ar[ul]_(0.45)0
} \label{eq:flopT}
\end{equation}
Near $O$, $Z(\O_C)$ is very small and $Z(\O_x)=1$. It follows that
$\xi(\O_x)=\xi(\O_C(-1)[1])=0$ and $\xi(\O_C)=\theta/\pi-1$. The
morphisms in (\ref{eq:flopT}) are labeled by the differences in the
$\xi$'s between the head and tail of the arrow. Thus a vertex is
$\Pi$-stable (with respect to that triangle) when and only when the
label on the opposite edge is $<1$.  Therefore, the 0-brane $\O_x$
decays into $\O_C$ and $\O_C(-1)[1]$ as $\theta$ increases beyond
$\pi$.

We also have a distinguished triangle
\begin{equation}
\xymatrix@!C=7mm{
  &\O_C[1]\ar[dl]|{[1]}_(0.6){1-\ff{\theta}{\pi}}&\\
  \O_C(1)\ar[rr]_{\ff{\theta}{\pi}}&&\O_x,\ar[ul]_(0.45)0
} \label{eq:flop2}
\end{equation}
which shows that $\O_x$ decays into $\O_C(1)$ and $\O_C[1]$ for $\theta<0$.
Either way, we see that $\O_x$ decays as we move from the $X$
phase into the $X'$ phase in figure \ref{f:theta}.

Suppose $y\not\in C$. Then, even though $\O_y$ has exactly the same
D-brane charge as $\O_x$ it does not decay by (\ref{eq:flopT}) or
(\ref{eq:flop2}) since there are no morphisms from $\O_C$ or $\O_C(1)$
to $\O_y$. Indeed, we would not expect 0-branes away from $C$ to
be affected by the flop transition. Note again the superiority of the
derived category approach to D-brane physics. Any method based solely
on the notion of D-brane charges would not be able to distinguish
between $\O_x$ and $\O_y$ even though their properties are quite
different with respect to a flop.

We should also be able to see the objects in the derived category of
$X$ which play the r\^ ole of points on $C'$ {\em after\/} we do the
flop. Before we do this we need to introduce a method of computing
some of the relevant $\Ext$'s. Suppose we are given sheaves $\cE$ and
$\cF$ on $C$. Given the embedding $i:C\to X$ with a normal bundle $N$
on $C$, there is a spectral sequence with
\begin{equation}
  E_2^{p,q} = \Ext_C^p(\cE,\cF\otimes\wedge^q N),
\end{equation}
converging to $E_\infty^{p,q}$ with $\bigoplus_{p+q=n}E_\infty^{p,q}=
\Ext^n_X(i_*\cE,i_*\cF)$.\footnote{A quick way of proving this is to
use the ``right adjoint'' functor of $i_*$ which is written $i^!$ and
whose properties are well-understood \cite{Hart:dC} in the derived
category.}

For example, consider $\Ext^n_X(\O_C,\O_C(-1))$. Since
$N=\O(-1)\oplus\O(-1)$, we have an $E_2$ stage of the spectral
sequence given by $E_2^{p,q}=H^p(C,\O_C(-1)\otimes\wedge^q N)$:
\begin{equation}
\begin{xy}
\xymatrix@!C=0mm@R=7mm{
0&\C^2\\
0&\C^2\\
0&0
} 
\save="x"!LD+<-3mm,0pt>;"x"!RD+<0pt,0pt>**\dir{-}?>*\dir{>}\restore
\save="x"!LD+<0pt,-3mm>;"x"!LU+<0pt,-2mm>**\dir{-}?>*\dir{>}\restore
\save!CD+<0mm,-3mm>*{p}\restore
\save!CL+<-2.5mm,0mm>*{q}\restore
\end{xy}  \label{eq:Cs1}
\end{equation}
and thus
\begin{equation}
  \Ext^n_X(\O_C,\O_C(-1))=\begin{cases}
   \C^2&\text{if $n=2$ or 3,}\\
   0&\text{otherwise.}\end{cases}
\end{equation}
Thus we may use morphisms $f\in\Hom(\O_C[-1],\O_C(-1)[1])\cong\C^2$
to form new objects $D_f$:
\begin{equation}
\xymatrix@!C=7mm{
  &D_f\ar[dl]|{[1]}_(0.6){\ff{\theta}{\pi}-1}&\\
  \O_C[-1]\ar[rr]_{2-\ff{\theta}{\pi}}^f&&\O_C(-1)[1],\ar[ul]_0
} \label{eq:flop3}
\end{equation}
which become stable for $\theta>\pi$. As noted in section
\ref{ss:mdefs}, rescaling $f$ by a complex number has no effect on
$D_f$, so we have a $\P^1$'s worth of $D_f$'s. These indeed represent
the points on $C'$ which become stable as we flop into $X'$ by
increasing $\theta$ through $\pi$ as argued in \cite{Brig:flop}. We
leave it to the reader to find the objects corresponding to points on
$C'$ which become stable as we flop by decreasing $\theta$ through 0.

The objects $D_f$ are exotic in the same sense as those in section
\ref{sss:qex}. They were described as ``perverse sheaves'' in
\cite{Brig:flop}. Note that it is a trivial observation that the
derived category of $X$ is equal to that of $X'$ in the context of the
B-model. The B-branes yield $\DC(X)$ and $X$ is converted into $X'$ by a
shift in $B+iJ$ --- but such a shift has no affect on the B-model. The
equivalence of $\DC(X)$ and $\DC(X')$ was established rigorously in
\cite{BO:flop,Brig:flop}.

%%%%%%%

\subsection{Orbifolds}  \label{ss:orb}

Let $G$ be a finite subgroup of $\SU(d)$. In this section we are
interested in a string propagating on the orbifold $\C^d/G$. Of
course, beginning with the seminal work of \cite{DHVW:}, orbifolds
have played an enormously important r\^ole in the understanding of
stringy geometry. It should therefore be no surprise that the subject
of B-branes on orbifolds provides a rich laboratory for further
insight into the properties of D-branes. Much of what follows is a
review of the works \cite{DM:qiv,DDG:wrap,DFR:orbifold,
DD:stringy,DG:fracM,Fl:quiv,DFR:stab}. We also refer to
\cite{LMW:McK,Mayr:McK} for related work.

One of the key concepts in studying the geometry of orbifolds is that
these singularities can be blown-up (at least in the case $d\leq3$) to
produce a smooth manifold. This blow-up process introduces an
``exceptional divisor'' to replace the orbifold singularity.  We will
assume the reader is familiar with the basic ideas of this process.
We refer to \cite{me:orb2}, for example, for a review of this process
in the context of conformal field theory.

\subsubsection{The McKay correspondence}  \label{sss:McK}

We begin with the purely mathematical and beautiful observation of
McKay \cite{McKay:} in the case $d=2$. Let $G\subset\SU(2)$ be a
finite group. It is well-known that such groups have an A-D-E
classification and that the resulting exceptional divisor in the
blow-up is given by a collections of $\P^1$'s intersecting according
to the corresponding Dynkin diagram. We refer to
section 2.6 of \cite{me:lK3} for a review of these facts. 

Let
$V_i$, $i=1,\ldots,r$ be the irreducible representations of $G$. In
addition let $Q$ be the fundamental 2-dimensional representation of $G$
induced by the embedding $G\subset\SU(2)$. Consider the following
decompositions
\begin{equation}
  Q\otimes V_i = \bigoplus_{j=1}^r a_{ji}V_j,
    \label{eq:McD}
\end{equation}
for non-negative integers $a_{ij}$. The ``McKay quiver'' is defined by
drawing $r$ nodes, one for each irreducible representation, and then
drawing $a_{ij}$ arrows from the node associated to $V_i$ to the node
associated to $V_j$. In the case of $d=2$, $a_{ij}=a_{ji}$ but this is
not necessarily true for $d>2$.

McKay's observation was that the resulting quiver is precisely the
{\em extended\/} Dynkin diagram associated to $G$. For example, if $G$
is the binary icosahedral group, it is associated to $E_8$ and the
McKay quiver is:
\begin{equation}
\begin{xy}
  (0,0)*{\circ};(10,0)*{\circ}**\dir2{-}
       ?(0.58)*_!/1pt/^\dir{>}?(0.42)*^!/1pt/_\dir{<},
  (0,-3)*{\scriptstyle2},
  (10,0)*i{\circ};(20,0)*{\circ}**\dir2{-}
       ?(0.58)*_!/1pt/^\dir{>}?(0.42)*^!/1pt/_\dir{<},
  (10,-3)*{\scriptstyle4},
  (20,0)*i{\circ};(30,0)*{\circ}**\dir2{-}
       ?(0.58)*_!/1pt/^\dir{>}?(0.42)*^!/1pt/_\dir{<},
  (20,-3)*{\scriptstyle6},
  (30,0)*i{\circ};(40,0)*{\circ}**\dir2{-}
       ?(0.58)*_!/1pt/^\dir{>}?(0.42)*^!/1pt/_\dir{<},
  (30,-3)*{\scriptstyle5},
  (40,0)*i{\circ};(50,0)*{\circ}**\dir2{-}
       ?(0.58)*_!/1pt/^\dir{>}?(0.42)*^!/1pt/_\dir{<},
  (40,-3)*{\scriptstyle4},
  (50,0)*i{\circ};(60,0)*{\circ}**\dir2{-}
       ?(0.58)*_!/1pt/^\dir{>}?(0.42)*^!/1pt/_\dir{<},
  (50,-3)*{\scriptstyle3},
  (60,0)*i{\circ};(70,0)*{\circ}**\dir2{-}
       ?(0.58)*_!/1pt/^\dir{>}?(0.42)*^!/1pt/_\dir{<},
  (60,-3)*{\scriptstyle2},
  (70,-3)*{\scriptstyle1},
  (20,0)*i{\circ};(20,10)*{\circ}**\dir2{-}
       ?(0.58)*_!/1pt/^\dir{>}?(0.42)*^!/1pt/_\dir{<},
  (17,10)*{\scriptstyle3},
  (72,-1)*{,}
\end{xy} \label{eq:E8}
\end{equation}
where the numbers refer to the dimensions of the irreducible
representations. Thus, except for the extra node present in the
extended diagram, the McKay quiver represents exactly the
configuration of $\P^1$'s in the exceptional divisor upon blowing-up
the orbifold singularity. In this latter context, the numbers in
(\ref{eq:E8}) refer to the multiplicity of the $\P^1$'s in the
exceptional divisor.

Within algebraic geometry we may give a quite different interpretation
of the McKay quiver which turns out to be very relevant for
D-branes. We follow the description in \cite{CI:flopM}.
Let $V$ be a finite-dimensional complex vector space. Let us
denote $\C^d$ by $Q$ and, as usual, $Q^*=\Hom(Q,\C)$. Now let
$S\in\Hom(Q^*,\End(V))=\Hom(V,Q\otimes V)$, i.e., $S$ defines a linear
action of the coordinates of $Q$ on $V$. If $x$ and $y$ are any two
coordinates in $Q$ we demand that $S(x)$ commutes with $S(y)$. We
denote this condition $S\wedge S=0$. 

$\C[Q]$ denotes the polynomial ring of functions on $Q$ and it is
equal to $\oplus_k \Sym^k Q^*$. Thus $S$ defines an action of $\C[Q]$
on $V$. In other words it gives $V$ the structure of a $\C[Q]$-module.

Now let $G\subset\SU(d)$ act on $V$ and on $Q$. $V$ is an arbitrary
representation of $G$ and $Q$ is the standard $d$-dimensional
representation. We now demand that $S$ commutes with the $G$-action by
putting\footnote{The notation
$f\in\Hom_G(A,B)$ means $fg=gf$ for all $g\in G$.}
\begin{equation}
S\in\Hom_G(V,Q\otimes V),\qquad S\wedge S=0.
  \label{eq:qhom}
\end{equation}
Such an $S$ defines a ``G-equivariant'' $\C[Q]$-module structure on
$V$. In fact, we have defined a $G$-equivariant {\em sheaf\/} on
$Q\cong\C^d$. For a sheaf one would normally define a $\O_X(U)$-module
structure for all open sets $U$. We have only done this for
$U=\C^d$, i.e., we have only considered global sections. However,
since $\C^d$ is contractible, there is no information content beyond
these global sections.

Now decompose $V$ into irreducible representations, $V=\bigoplus_k
m_kV_k$. Schur's lemma states $\Hom(V_i,V_j)=\C\delta_{ij}$. Thus,
using (\ref{eq:McD}), it follows that $S$ is represented by a
collection of matrices of complex numbers. To be precise, there are
$a_{ij}$ matrices representing a map from $\C^{m_i}$ to $\C^{m_j}$ for
each $i,j$. In other words, we associate a number $m_i$ to each node
in the McKay quiver and a matrix of dimension $m_j\times m_i$ to each
arrow from the $i$th node to the $j$th node. This collection of linear
maps associated to the arrows on a McKay quiver is called a {\em
  representation\/} of the quiver. We refer to \cite{Ben:quiv} for
more background in this subject.

The condition $S\wedge S=0$ puts constraints on these matrices. For
example, suppose $G$ is abelian. Then every irreducible representation
is one-dimensional and thus $Q^*=q_1\oplus q_2\oplus\ldots\oplus
q_d$ for suitable irreducible representations $q_\alpha$. It
follows that there are $d$ arrows leaving each node in the McKay
quiver. Let $M^i_\alpha$ represent the matrix associated to the
$\alpha$'th arrow leaving node $i$. Let $\alpha(i)$ be the node at the head
of this arrow. Then the $S\wedge S=0$ relations read
\begin{equation}
  M^{\alpha(i)}_\beta M^i_\alpha = M^{\beta(i)}_\alpha M^i_\beta. 
    \label{eq:rel}
\end{equation}
These are said to be {\em relations\/} on the representation of the
quiver.

\def\biS#1#2{
\begin{xy} <0.8mm,0mm>:
  (0,0)*{\circ};(0,10)*{\circ}**\dir{-}
    ?(0.58)*\dir{>}, (0,-3)*{\scriptstyle#1}, (0,13)*{\scriptstyle#2},
  (-5,0)*i{x},(5,0)*i{x}%pad
\end{xy}}
Now we consider morphisms of $G$-equivariant sheaves. In the quiver
language this amounts to a morphism of $\C[Q]$-modules respecting the
$G$-action. It is not hard to find the explicit form of these
morphisms. Let $W=\bigoplus_k n_kV_k$. A $G$-invariant morphism from a
quiver representation associated to $V$ to one associated to $W$ will,
by Schur's lemma, be a choice of a matrix in $\Hom(\C^{m_k},\C^{n_k})$
for each node of the quiver. If these maps commute with the arrows
within each quiver then we preserve the $\C[Q]$-module structure. 
For example, suppose we have a quiver with two nodes and a single
arrow connected them. A morphism
\begin{equation}
\xymatrix@1{
\biS{m_1}{m_2}\ar[r]&\biS{n_1}{n_2}}
\end{equation}
is equivalent to the following commutative diagram
\begin{equation}
\xymatrix{
\C^{m_2}\ar[r]&\C^{n_2}\\
\C^{m_1}\ar[r]\ar[u]&\C^{n_1}\ar[u]
}
\end{equation}

We have constructed a $G$-invariant morphism of $\C[Q]$-modules. We can use
this construction to define a morphism of quiver representations and
thus the category of quiver representations.

Our definitions show that {\em the category of $G$-equivariant
sheaves on $\C^d$ is equivalent to the category of representations of
the McKay quiver with the relations $S\wedge S=0$.}

The category of representations of the McKay quiver (with relations)
is an abelian category. Each quiver representation is associated to a
vector space $V$ as defined above. Kernels and cokernels may easily be
computed purely from this vector space structure. When we start to
write down exact sequences we see a new interpretation of the arrows
in a quiver as follows. Note that
\begin{equation}
\xymatrix@1{
0\ar[r]&\biS01\ar[r]&\biS11\ar[r]&\biS10\ar[r]&0,\POS(36.5,0)*{\scriptstyle
  f}} \label{eq:extQ}
\end{equation}
is a short exact sequence of quiver representations where the map
$f\in\Hom(\C,\C)=\C$ can be multiplication by any complex number. Let
$F_i$ be the quiver representation with $n_j=\delta_{ij}$, i.e., $V$
is simply the irreducible representation $V_i$.  Assuming there are no
arrows beginning and ending on the same node, all the maps in $F_i$
are obviously zero and so $F_i$ specifies a unique object. The short
exact sequence (\ref{eq:extQ}) represents an extension of $F_1$ by
$F_2$. That is $\Ext^1(F_1,F_2)=\C$. This easily generalizes to an
arbitrary quiver to give the following result (even if relations are
imposed): {\em the number of arrows from node $i$ to node $j$ equals
the dimension of $\Ext^1(F_i,F_j)$.} Thus the quiver is often called
an ``Ext quiver''.

The most impressive generalization of the McKay correspondence was
given by Bridgeland, King and Reid (BKR) \cite{BKM:MisM} who said the
following. 
\begin{theorem}
Suppose $X$ is a smooth resolution (i.e., blow-up) of the
orbifold $\C^d/G$ with $G$ a finite subgroup of $\,\SU(d)$ and
$d\leq3$.\footnote{The only reason why this should fail for $d>3$ is that
smooth resolutions need not exist.} Then the derived category $\DC(X)$
is equivalent to the derived category of $G$-equivariant sheaves on
$\C^d$.
\end{theorem}
In other words, $\DC(X)$ is equivalent to the derived category of
McKay quiver representations associated to $G$ with relations $S\wedge S=0$.
Thus, whereas the original McKay correspondence viewed the arrows in the
Dynkin diagram (\ref{eq:E8}) as something to do with the intersection
theory of the components of the exceptional divisor, in the
generalization to three dimensions it is best to view the arrows as a
statement about $\Ext$'s in the derived category.

Of course, the appearance of the derived category in the latter
version of the McKay correspondence is excellent news for B-brane
physics! The blow-up of an orbifold singularity is viewed as a
deformation of $B+iJ$ and so should not affect the B-model. It
immediately follows that
\begin{center}
\shabox{\parbox{.85\hsize}{B-branes on the orbifold $\C^d/G$ and open
strings between them are described by the derived category of
McKay quiver representations (with relations).}}
\end{center}

It follows that we have a distinguished set of D-branes on the
orbifold associated to the {\em irreducible\/} representations of
$G$. These will be associated to the quivers representations $F_i$
above. These branes were dubbed {\bf fractional branes} in
\cite{DDG:wrap}.

The BKR result gives a precise recipe for mapping between these two
derived categories as follows. Consider the following diagram
\begin{equation}
\xymatrix{
  Z\ar[r]^q\ar[d]_p&\C^d\ar[d]^\gamma\\
  X\ar[r]^-\pi&\C^d/G,
}
\end{equation}
where $Z$ is defined as the ``fibre product'' of $X$ and $\C^d$. That
is, $Z$ is the subspace of $X\times\C^d$ given by points $(x,z)$ such
that $\pi(x)=\gamma(z)$. The map $\gamma$ is the quotient by $G$ and
$\pi$ is the blow-down. The maps $p$ and $q$ are then the obvious
projections on the fibre product. BKR then say that $q_*p^*$ gives the
desired equivalence between $\DC(X)$ and the derived category of
$G$-equivariant sheaves on $\C^d$. Note that we think of sheaves on
$X$ as trivially $G$-equivariant. The map $p^*$ then typically
introduces some non-trivial $G$-action.

As an example, consider the skyscraper sheaf $\O_x$ on $X$ for $x$ at
a point in $X$ not fixed by the $G$-action. Then $p^*\O_x$ will be a
collection of $|G|$ skyscraper sheaves on $Z$ transforming in the
regular representation of $G$. This pushes forward to $\C^d$ to give
the same collection of skyscraper sheaves. Clearly the global sections
of this sheaf on $\C^d$ form the regular representation of $G$. Thus
{\em the 0-branes on $X$ correspond to quivers for the regular
  representation of $G$.} 

This means that, for 0-branes, the integer $m_i$ attached to each node
in the quiver representation is equal to the dimension of the
corresponding irreducible representation. The location of the 0-brane
will be dictated by the matrices associated to the arrows of the
quiver. Once we study the stability of these B-branes we will see that
the moduli space of such stable quiver representations is equal to $X$
as expected.

It also follows that the 0-brane is always composed of a nontrivial
sum of {\em fractional branes\/} (hence the name). We will see that, at
the orbifold point, the 0-brane is always marginally stable against decay
into this set of underlying fractional branes. 

\subsubsection{The Douglas--Moore construction}  \label{sss:DM}

We arrived at quivers from the mathematical direction of the McKay
correspondence. While this is probably the best approach for seeing
the appearance of the full derived category of quiver representations,
there is a wonderfully direct physics way of seeing why quivers
themselves should appear in the context of D-Branes on orbifolds.
This is due to Douglas and Moore \cite{DM:qiv}.

Suppose our full ten-dimensional spacetime looks like
$\R^{1,3}\times\C^3/G$ and we have a 3-brane that fills the space-like
directions of $\R^{1,3}$ and so appears as a 0-brane in the $\C^3/G$
directions. The world-volume theory of this 3-brane yields an $N=1$
supersymmetric field theory in $\R^{1,3}$.

One should note that this theory may, for general D-branes, have
anomalies. This is because of the RR flux generated by the D-brane ---
an issue we have been able to ignore up to this point. One
can remove these anomalies by using more background RR fluxes along
the lines of \cite{GHM:inf}. See also \cite{CFIKV:} for discussion of
when these theories can be anomaly-free. Actually all our discussion
of this four-dimensional quantum field theory will essentially be
classical, so we will ignore this issue of anomalies.

The idea is that one can analyze this quantum field theory by
analyzing a collection of D-branes in $\C^3$ corresponding to the
preimage of the quotient map, and then imposing $G$-invariance.

Suppose we have $m$ D-branes at the origin of $\C^3$ transforming in the
representation $V$ of $G$. Let $\rho_V(g)$ represent the $m\times m$
matrix representing $g\in G$ in this representation.  The field theory
will have a $\GU(m)$ gauge symmetry. Let $A$ be the corresponding
gauge connection. The action of $G$ is given explicitly by
\begin{equation}
  g(A) = \rho_V(g)\cdot A\cdot\rho_V(g)^{-1} 
\end{equation}
In other words, $A$ transforms in the representation $\End(V)=V\otimes
V^*$ and the $G$-invariant part of this can be written $\Hom_G(V,V)$.
By Schur's lemma, the resulting gauge group is
$\GU(m_1)\times\GU(m_2)\times\ldots$, where $V=\bigoplus_k m_kV_k$ as
in the last section.

Before dividing by $G$, we have an $N=4$ supersymmetric gauge theory
in four dimensions. The $N=4$ gauge superfield contains three scalar
fields $Z$ which arise from the components of the ten-dimensional
connection pointing in the $\C^3$ directions. These fields transform
under $G$ similarly to $A$ except that $G$ acts on the $\C^3$
directions too. This latter 3-dimensional representation is
clearly $Q$ from the previous section. So the $Z$'s transform in
$Q\otimes\End(V)$. In other words, the $G$-invariant subspace of
invariant scalar fields is given by $\Hom_G(V,Q\otimes V)$. Thus, from
(\ref{eq:qhom}), the $Z$'s play the r\^ole of the matrices associated
to the arrows in a quiver representation.

Finally, to obtain complete agreement with the last section, we need
to find the commutation relations on the quiver. These arise from the
superpotential of the $N=4$ theory \cite{BSS:N=4}. If we write
$Z_\mu$, $\mu=1,2,3$, for the three scalar fields (each transforming
in the adjoint of $\GU(m)$) corresponding to the 3 directions in
$\C^3$, this superpotential may be written
\begin{equation}
  W = \Tr(Z_1[Z_2,Z_3]).
\end{equation}
The critical points of this superpotential impose precisely the
relations $S\wedge S=0$ from section \ref{sss:McK}.

To recap, the $N=1$ supersymmetric field theory in $\R^{1,3}$ is
described by the McKay quiver representations of section
\ref{sss:McK}. The integers $m_i$ describe the effective gauge
group. The homomorphisms associated to the arrows are associated to
``bifundamental'' scalar fields $Z$ transforming accordingly in this
gauge group. Finally, the commutation relations on the quiver
representation are given by the superpotential.

This Douglas--Moore construction of a D-brane world-volume is
identical mathematically to a problem studied by Sardo Infirri
\cite{SI:I,SI:II}. In this case one studies translation-invariant
$G$-equivariant holomorphic bundles on $Q=\C^d$. Let the fibre of a
vector bundle be given by a representation $V$ of $G$. Then, for
$G$-equivariance, the connection on this bundle transforms yet again
in $\Hom_G(V,Q\otimes V)$. Let us write this connection in the form
\begin{equation}
  \sum_{\mu=1}^d Z_\mu dz_\mu-Z^\dagger_\mu d\bar z_\mu.
\end{equation}
The $(2,0)$-part of the curvature is then\cite{SI:I}
\begin{equation}
\sum_{\mu,\nu}\left(-\frac{\partial Z_\mu}{\partial
  z_\nu}+\ff12[Z_\mu,Z_\nu]\right)dz_\mu\wedge dz_\nu.
\end{equation}
If we impose translation invariance we demand that the derivatives of
$Z_\mu$ vanish. The condition that the bundle be holomorphic then
amounts to $[Z_\mu,Z_\nu]=0$ which again imposes the relations on the
quiver representation.

Thus we have three interpretations for the quiver representation:
\begin{enumerate}
\item A $G$-equivariant sheaf.
\item The scalar fields in the world-volume of a D-brane on an orbifold.
\item A connection on a translation-invariant holomorphic
  $G$-equivariant vector bundle.
\end{enumerate}

The fact that these descriptions coincide allow us to prove a fact
about the gradings of the fractional branes at the orbifold point. The
scalar fields $Z_\mu$ in our world-volume theory must arise as open
string states in the worldsheet description. That is, they occur as
certain $\Ext$'s between the D-branes. The Ext-quiver language
immediately tells us, of course, that these scalars are actually
associated to $\Ext^1$'s between the fractional branes. The discussion
around equation (\ref{eq:FukM}) and section \ref{sss:Pi} tells us that
a scalar state associated with $\Ext^1$ is massless if and only if the
$\xi$-gradings at the end of the string are equal. Thus we have proven:
\begin{theorem}
The gradings are equal for all the fractional branes at
the orbifold point.
\end{theorem}

A consequence of this theorem is, of course, that the central charges
of fractional D-branes align to have the same arg at the orbifold
point. We will see this explicitly in an example in section
\ref{sss:Z3}.

It is worth noting that the quiver language may be used in more
general cases than orbifolds. So long as the gradings of a set of
branes are aligned, their Ext-quiver will represent the field content
of an $N=1$ theory in four dimensions. This ties the
derived category picture into quiver-related work such as
\cite{CFIKV:,Wijn:dP,HW:dib,Herz:exc} etc. Sadly we do not have the space to
expand on this relation further.

\subsubsection{$\theta$-stability}  \label{sss:theta}

We have established that the category of topological B-branes for an
orbifold $\C^d/G$ corresponds to the derived category of quiver
representations with relations. Now we want to impose the BPS
condition again, that is, we want a stability condition. We already
know what the answer is --- $\Pi$-stability. However, since we have a
relatively simple description of B-branes on an orbifold, we might
expect there to be a simpler description of the stability condition
for the orbifold for a region of the moduli space of
$B+iJ$ around the orbifold point (i.e., small blow-ups). In other
words, we want something analogous to the way $\mu$-stability was a
good description near the large radius limit. The answer is called
$\theta$-stability as introduced by King \cite{King:th}. This was
studied in the context of D-branes in
\cite{DFR:stab,DFR:orbifold,Fl:quiv}.

Both the Douglas--Moore D-brane world-volume picture and the Sardo
Infirri picture of a $G$-equivariant bundle yield the condition for a
BPS state. As one might guess, in the $G$-equivariant bundle picture
we simply impose the condition that the connection is
Hermitian--Yang--Mills. In the D-brane language we are required to
minimize the potential arising from the ``D-terms'' in the
action. Either way, the condition for a BPS state
becomes\footnote{Note that in order to define the Hermitian conjugate
$Z_\mu^\dagger$ we require an inner product on the
representation. This is intrinsic in the quantum field theory but is
``extra data'' in the abstract quiver language.}
\begin{equation}
  \sum_i[Z_\mu,Z_\mu^\dagger]=0.
     \label{eq:Dt0}
\end{equation}
In section \ref{sss:McK} we saw that the regular
representation of $G$ should correspond to the skyscraper sheaf
$\O_x$. One can show \cite{SI:I} that the moduli space of $Z$'s
associated to the regular representation that satisfy (\ref{eq:Dt0})
is indeed given by $\C^d/G$ as expected.

Now we want to resolve the orbifold a little. This means that we want
to deform our problem in such a way that the moduli space of $Z$'s
associated to the regular representation becomes $X$, a resolution of
the orbifold. In the D-brane language one may do this by adding
``Fayet--Iliopoulos'' terms to the action
\cite{DM:qiv,DGM:Dorb,GLR:nonab}. One such term may be added for each
unbroken $\GU(1)$ of the gauge group, i.e., one for each irreducible
representation of $G$. Let us call the coefficients of these terms
$\zeta_i$, where $i$ is an index running over the irreps of $G$.
The equivalent deformation is seen in the Sardo Infirri picture as a
deformation of a ``moment map'' \cite{SI:I}.

The result is that (\ref{eq:Dt0}) becomes
\begin{equation}
  \sum_\mu[Z_\mu,Z_\mu^\dagger]=\diag(\underbrace{\zeta_1,\zeta_1,\ldots,
    \zeta_1}_{\dim(V_1)},\underbrace{\zeta_2,\zeta_2,\ldots,
    \zeta_2}_{\dim(V_2)},\ldots).
     \label{eq:Dtz}
\end{equation}
In the case that $G$ is abelian and $d=3$, Sardo Infirri proved that the
resulting moduli space of the 0-brane is indeed a resolution of
$\C^d/G$ (see also \cite{DGM:Dorb}). The case $d=2$ and an arbitrary
$G$ was proven earlier by Kronheimer \cite{Kron:ALE}. 

Since the blow-up of an orbifold singularity is obtained by a
deformation of $B+iJ$, this resolution should be produced by closed
string marginal operators. To be precise, {\em twisted\/} closed
string marginal operators. Twisted operators are labeled by
conjugacy classes in $G$ \cite{DHVW:} which we denote $C$. Let
$\phi_C$ be a twisted operator present in the topological A-model for
closed strings for the conjugacy class $C$ and consider a deformation
of $B+iJ$ produced by adding a term
\begin{equation}
  a_C\int_\Sigma d^2z\, \phi_C,
\end{equation}
to the action. Turning on the $\zeta_i$'s should be equivalent to
turning on some $a_C$'s which implies that $\phi_C$ should acquire a
nonzero 1-point function in the D-brane background. This one-point
function was computed in \cite{DM:qiv,DGM:Dorb}. The result is, at
least in a linear approximation for very small blow-ups
\begin{equation}
  a_C = \sum_i \chi_i(C)\,\zeta_i,
    \label{eq:zetamarg}
\end{equation}
where $\chi_i$'s are the {\em characters\/} of the group $G$. 

The operator $\phi_1$ associated to the conjugacy class of the
identity is the closed string tachyon and is removed by the GSO
projection. Thus we require that $a_1=0$, i.e.,
\begin{equation}
  \sum_i\dim(V_i)\,\zeta_i=0.
   \label{eq:a1}
\end{equation}
Thus condition is also imposed by (\ref{eq:Dtz}) since a commutator
must be traceless.

So far we have discussed the effects of nonzero $\zeta_i$'s on the
0-brane, i.e., regular representation of $G$. What about general
representations? The set-up is essential identical except that the
$Z_\mu$'s now transform in another representation of $G$. Let us
consider the representation $V=\bigoplus_i m_iV_i$. Then
(\ref{eq:Dtz}) is modified so that the right-hand side is a diagonal
matrix with each $\zeta_i$ appearing $m_i$ times. In this more general
setting the condition (\ref{eq:a1}) can prevent (\ref{eq:Dtz}) from
having a solution.

In D-brane world-volume language this means that we cannot
find a solution which makes the contribution of the D-term to the
potential equal to zero. At first sight, this would appear to break
supersymmetry. In fact, this is not the case as pointed out in
\cite{DFR:stab}. If we simply minimize the potential then a
not-so-manifest $N=1$ supersymmetry still exists implying we do have a
BPS state. We refer to \cite{DFR:stab} for more details. Suppose we
minimize the potential by setting
\begin{equation}
  \sum_\mu[Z_\mu,Z_\mu^\dagger]=\diag(\underbrace{\theta_1,\theta_1,\ldots,
    \theta_1}_{m_1},\underbrace{\theta_2,\theta_2,\ldots,
    \theta_2}_{m_2},\ldots),
     \label{eq:Dtth}
\end{equation}
for some real numbers $\theta_i$. The potential is then given by
\begin{equation}
  \sum_i(\zeta_i-\theta_i)^2,
\end{equation}
which is minimized subject to the condition (\ref{eq:a1}) by
\begin{equation}
  \theta_i = \zeta_i - \frac{\sum_j m_j\zeta_j}{\sum_j m_j}.
     \label{eq:thdef}
\end{equation}

The equation (\ref{eq:Dtth}) may be written in a more quiver-friendly
way as follows. Let $a$ be an arrow in the quiver with head $h(a)$ and tail
$t(a)$. Let $Z_a$ be the $m_{h(a)}\times m_{t(a)}$ matrix associated
with this arrow in a given quiver representation. Then (\ref{eq:Dtth})
becomes
\begin{equation}
  \sum_{h(a)=i} Z_aZ_a^\dagger - \sum_{t(a)=i} Z_a^\dagger Z_a
     = \theta_i\id. 
   \label{eq:DKth}
\end{equation}
This is exactly the equation studied by King \cite{King:th}. Fix a
representation of the quiver associated to a representation
$V=\bigoplus_i m_iV_i$ of $G$. For any representation $W=\bigoplus_i
n_iV_i$ of $G$ we define
\begin{equation}
  \theta(W) = \sum_i \theta_in_i.
\end{equation}
Thus, by the tracelessness of (\ref{eq:Dtth}), we see
$\theta(V)=0$. We say that the quiver representation is {\em
$\theta$-stable\/} if, for any nontrivial quiver subrepresentation
associated to a representation $W$ of $G$, $\theta(W)>0$. King proved the
following
\begin{theorem}
A quiver representation satisfies (\ref{eq:DKth}) (with an inner
product unique up to obvious automorphisms) if and only if it a
direct sum of $\theta$-stable representations.
\end{theorem}

Thus, very close to the orbifold point we have a stability condition
expressed purely in terms of quivers. 

\subsubsection{Periods}  \label{sss:per2}

For the remainder of these lectures we will focus on a particular
example of an orbifold rather than attempt to prove any general
statements. The example is $\C^3/\Z_3$ where $g$ generates $\Z_3$ and
acts as $g:(z_1,z_2,z_3)\mapsto(\omega z_1,\omega z_2, \omega z_3)$
for $\omega=\exp(2\pi i/3)$. Much of the analysis of D-branes on
orbifolds has been done in this simplest example (e.g.,
\cite{DGM:Dorb,DDG:wrap,DG:fracM,DFR:orbifold}). 

Let $V_i$, $i=0,\ldots,2$ be the one-dimensional irreducible
representations of $\Z_3$ given by $\rho(g)=\omega^i$. A
representation $V=\oplus_im_iV_i$ is then associated to a quiver
representation
\begin{equation}
\begin{xy} <1mm,0mm>:
  (0,0)*{\circ};(16,0)*{\circ}**\dir3{-}
    ?(0.58)*\dir3{>}, (0,-3)*{\scriptstyle m_2},
  (16,0)*i{\circ};(8,10)*{\circ}**\dir3{-}
    ?(0.58)*\dir3{>},(16,-3)*{\scriptstyle m_1.},
  (8,10)*i{\circ};(0,0)*{\circ}**\dir3{-}
    ?(0.58)*\dir3{>},(8,13)*{\scriptstyle m_0}
\end{xy} \label{eq:Z3Q}
\end{equation}

As is well-known, this orbifold is resolved with an exceptional
divisor $E\cong\P^2$. In this case, $X$ may be viewed as the total
space of the line bundle corresponding to the sheaf $\O_E(-3)$.

The computation to compute the periods and thus the central charges
can be done in a way very similar to that of the quintic. 
In this section we perform the computations corresponding to section
\ref{sss:qper}.

Similarly to the quintic, we have a one-dimensional moduli space of
$B+iJ$ that can be viewed as a $\P^1$. One point on this $\P^1$
corresponds to the large radius limit where the exceptional divisor
$E$ is infinitely large. At the other extreme, we have the orbifold
point where we have no blow-up. At a third point on the $\P^1$, which
we denote $P_0$, we have the analogue of the ``conifold point'' where
$B+iJ$ have a special value that makes the associated conformal field
theory singular. Again, as in the quintic, we may use mirror symmetry
to analyze this moduli space and the associated periods exactly, as
was first done in \cite{AGM:sd}.

The Picard--Fuch's equation in question is given by
\begin{equation}
  \left(z\frac{d}{dz}\right)^3\Phi +27z\left(z\frac{d}{dz}\right)
\left(z\frac{d}{dz}+\ff13\right)\left(z\frac{d}{dz}+\ff23\right)\Phi =0.
   \label{eq:PFZ3}
\end{equation}
Clearly any constant solves this differential equation. Putting
$z=(3e^{-\pi i}\psi)^{-3}$ we may write a basis for the remaining
solutions near $\psi=0$ as:
\begin{equation}
  \varpi_j = \frac1{2\pi i}\sum_{n=1}^\infty
     \frac{\Gamma(\ff n3)\omega^{nj}}{\Gamma(n+1)\Gamma(1-\ff n3)^2}
     (3\psi)^n, \label{eq:perZ3}
\end{equation}
where $\omega=\exp(2\pi i/3)$. Thus 1, $\varpi_0$ and $\varpi_1$ form
a basis for the solutions of the Picard--Fuchs equation. The analytic
continuation to a basis for small $z$ (valid for $|\arg(z)|<\pi$) is
performed similarly to section \ref{sss:qper}:
\begin{equation}
\begin{split}
\Phi_0&=1,\\[2mm]
\Phi_1 &= \frac1{2\pi i}\cdot\frac3{2\pi i}
    \int\frac{\Gamma(3s)\Gamma(-s)}{\Gamma(1+s)^2}z^s\,ds\\
  &= \frac1{2\pi i}\log z + O(z) \\
  &= t\\
  &= \varpi_0,\\[2mm]
\Phi_2 &= -\frac1{4\pi^2}\cdot\frac{-6}{2\pi i}
  \int\frac{\Gamma(3s)\Gamma(-s)^2}
  {\Gamma(s+1)}(e^{-\pi i}z)^s\,ds\\
 &= -\frac1{4\pi^2}(\log z-i\pi)^2-\ff5{12}+O(z)\\
 &= t^2-t-\ff16+O(e^{2\pi it})\\
 &= -\ff23(\varpi_0-\varpi_1),
\end{split} \label{eq:Z3ac}
\end{equation}
where the mirror map is given by $t=\int_C B+iJ=\frac1{2\pi
i}\log(z)+O(z)$, and $C$ is a $\P^1$ hyperplane in $E$.
The analogue of figure \ref{f:scorp} for the $\C^3/\Z_3$ orbifold is
given in figure \ref{f:Z3scp}. Note that the orbifold point lies at
exactly $B=J=0$.\footnote{There is a false assumption in
\cite{AGM:sd} which shifts $B$ by $\ff12$. The correct argument
appears in \cite{MP:inst}.}

\iffigs
\begin{figure}[t]
$$
\begin{xy}
  \xyimport(100,100){\epsfxsize=14cm\epsfbox{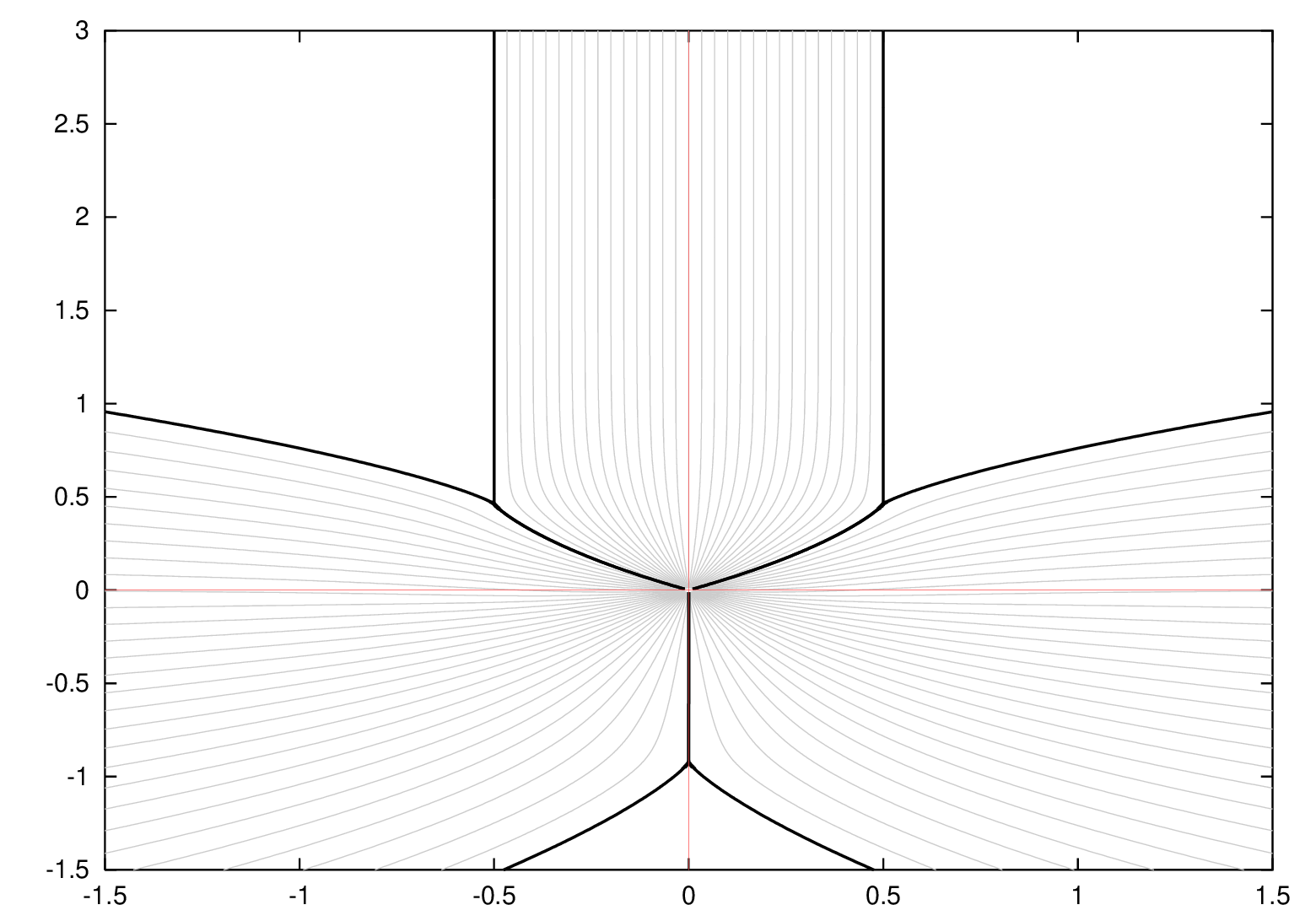}}
  ,(50,-1)*{B}
  ,(2,50)*{J}
  ,(52.4,36.0)*{\bullet}
  ,(53.8,32)*{\psi=0,\text{Orbifold}}
  ,(67.2,45.2)*{\bullet}
  ,(77,43)*{P_0,\psi=e^{2\pi i/3}}
  ,(37.6,45.2)*{\bullet}
  ,(30,43)*{\psi=e^{4\pi i/3}}
  ,(52.4,17.3)*{\bullet}
  ,(58,17.3)*{\psi=1}
\end{xy}
$$
  \vspace{-5mm}
  \caption{Fundamental regions for the moduli space of the $\Z_3$-orbifold.}
  \label{f:Z3scp}
\end{figure}
\fi

Now, using (\ref{eq:sub0}) and (\ref{eq:ZB0}), we can compute the
exact value for central charges. For example, consider the 4-branes
$\O_E(m)$ wrapping the exceptional divisor:
\begin{equation}
  Z(\O_E(m)) = -(m+\ff43)\varpi_0+\ff13\varpi_1 +
          \ff12m^2+\ff32m+\ff43.
    \label{eq:ZOEm0}
\end{equation}
Consider the point $P_0$ in the moduli space where $\psi=2\pi i/3$ and
we have a singular conformal field theory. At $P_0$ we have $\varpi_0=
t=\ff12+iJ_0$ where $J_0\approx 0.4628$. From (\ref{eq:perZ3}) the
value of $\varpi_1$ at $P_0$ will clearly be equal to the value of
$\varpi_0$ at $\psi=4\pi i/3$, namely $-\ff12+iJ_0$. Thus
$\varpi_0-\varpi_1=1$ at $P_0$. {\em It follows from (\ref{eq:ZOEm}) that
$\O_E(-1)$ becomes massless at $P_0$.} Similarly $\O_E(-2)$ becomes
massless at the point $\psi=4\pi i/3$.

We will therefore assume that the singularity in the conformal field
theory at $P_0$ is caused by the stable B-brane $\O_E(-1)$ becoming
massless. This appears to be very similar to the statement that
$\O_C(-1)$ became massless for the singular conformal field theory for
the case of the flop in section \ref{ss:flop}. In the latter case we
reset our definitions so that $\O_C$ became massless instead. We will
do the same here to make the results prettier. In effect we shift
$t\mapsto t+1$ so that (\ref{eq:ZOEm0}) becomes
\begin{equation}
  Z(\O_E(m)) = -(m+\ff13)\varpi_0+\ff13\varpi_1 +
          \ff12m^2+\ff12m+\ff13,
    \label{eq:ZOEm}
\end{equation}
and now $\O_E$ becomes massless at $P_0$.

\subsubsection{Monodromy}  \label{sss:mon2}

In general orbifolds may exhibit a ``quantum symmetry'' which acts on
a state twisted by $g\in G$ by multiplication by $q(g)$, where
\begin{equation}
  q \in \Hom(G,\C^*).
\end{equation}
The group $\Hom(G,\C^*)$ is thus the group of quantum symmetries.  It
is not hard to show that this is isomorphic to the {\em
abelianization\/} of $G$, i.e., $G/[G,G]$.  Yet another interpretation
of $\Hom(G,\C^*)$ is the group of one-dimensional representations of
$G$ where the group operation is the tensor product. Given a
one-dimensional representation $U$ of $G$ we may act on the set of
representations by $U\otimes-$. This gives a symmetry of the McKay
quiver and thus shows exactly how the quantum symmetries of an
orbifold act on the category of D-branes.

It is manifest from (\ref{eq:E8}) that the extended Dynkin diagram for
$E_8$ has no symmetries and thus the quantum symmetry group associated
to the binary icosahedral group is trivial. That is, there is no
quantum symmetry for this orbifold. Thus, we cannot possibly use
quantum symmetries as a tool for making {\em general\/} statements
about orbifolds but they can be very useful in examples. In
particular, if $G$ is abelian, the quantum symmetry group is
isomorphic to $G$ and it acts transitively on the nodes of the McKay
quiver. In our case, the quantum symmetry group $\Z_3$ acts by
rotating the McKay quiver by $2\pi/3$.

In section \ref{sss:qmon} we analyzed the monodromy around the Gepner
point and discovered that the quantum $\Z_5$ symmetry one might expect
in this context was in fact broken in the derived category. In this
section we will show that the quantum symmetry of an orbifold, at
least in our example, is {\em not\/} broken by the derived
category. 

Monodromy around the large radius limit corresponds to tensoring by
$\O_X(D)$ where $D$ is a divisor Poincar\'e dual to the component of
the K\"ahler form which is being taken to be very large. Thus we
require $D$ to intersect $C$ (the $\P^1$ hyperplane of $E$) in one
point. To fit in with the notation used in the case of the quintic we will
denote $\O_X(D)$ by $\O_X(1)$. This notation is also consistent with
the fact that $\O_X(1)\otimes\O_E=\O_E(1)$.  Note that since the
normal bundle of $E$ corresponds to $\O_E(-3)$, the analogue of the
exact sequence (\ref{eq:OX}) is
\begin{equation}
\xymatrix@1{
0\ar[r]&\O_X(3)\ar[r]&\O_X\ar[r]&\O_E\ar[r]&0.
} \label{eq:OE3}
\end{equation}

The monodromy associated to the ``conifold point'' $P_0$ must be
associated to the fact that $\O_E$ becomes massless there. If
$\mathcal{K}$ is the Fourier--Mukai transform associated to monodromy
around $P_0$, then, comparing to (\ref{eq:TKq}) we would expect a
transform
\begin{equation}
  T_{\mathcal{K}}(\cF) = \Cone\bigl(\xymatrix@1{\Hom(\cE,\cF)
  \otimes\cE\ar[r]^-r&\cF\bigr)}, 
  \label{eq:TKZ}
\end{equation}
where $\cE$ is now $\O_E$. This corresponds to
\begin{equation}
   \mathcal{K} =
   \bigl(\xymatrix@1{\cE^\vee\boxtimes\cE\ar[r]^-r&\poso{\O_{\Delta X}}}
   \bigr),
\end{equation}
where $\cE^\vee$ is the dual of $\cE$ defined in the derived category
as $\mathbf{R}\sHom(\cE,\O_X)$.
Thus, if $\mathcal{G}$ is the transform associated to the orbifold
point, we have
\begin{equation}
\begin{split}
\mathcal{G} &= \bigl(\xymatrix@1{\cE^\vee(1)\boxtimes\cE\ar[r]
    &\poso{\O_{\Delta X}(1)}}\bigr)\\
  &= \bigl(\xymatrix@1{\O_E(-1)^\vee\boxtimes\O_E\ar[r]
    &\poso{\O_{\Delta X}(1)}}\bigr).
\end{split}
\end{equation}
The computation proceeds in a way very similar to section
\ref{sss:qmon} to yield
\begin{equation}
\mathcal{G}^{\star3} = \bigl(\xymatrix@1{
    \O_E(-1)^\vee\boxtimes\Omega_E^2(2)\ar[r]&
    \O_E(-2)^\vee\boxtimes\Omega_E(1)\ar[r]&
    \O_E(-3)^\vee\boxtimes\O_E\ar[r]&
    \poso{\O_{\Delta X}(3)}}\bigr).
\end{equation}
Using (\ref{eq:OE3}) we see that $\O_E^\vee$ is $\bigl(\xymatrix@1{
\poso{\O_X}\ar[r]&\O_X(-3)}\bigr)$ which is $\O_E(-3)[-1]$. We may then
apply the Beilinson sequence (\ref{eq:Bres}) for $E\cong\P^2$
to yield
\begin{equation}
\begin{split}
\mathcal{G}^{\star3} &= \Cone\bigl(\xymatrix@1{\O_{\Delta E}[-1]
      \ar[r]&\O_{\Delta X}(3)}\bigr)\\
  &= \Cone\Bigl(\bigl(\xymatrix@1{
   \poso{\O_{\Delta X}(3)}\ar[r]& \O_{\Delta X}}\bigr)
     \longrightarrow\O_{\Delta X}(3)\Bigr)\\
  &= \O_{\Delta X}.
\end{split}
\end{equation}
Thus $\mathcal{G}^{\star3}$ is the identity on the nose with no shifts
involved. The quantum $\Z_3$ symmetry is therefore preserved even at
the level of the derived category.

\def\triZ#1#2#3{
\begin{xy} <0.6mm,0mm>:
  (0,0)*{\circ};(16,0)*{\circ}**\dir3{-}
    ?(0.58)*\dir3{>}, (0,-4)*{\scriptstyle#3},
  (16,0)*i{\circ};(8,10)*{\circ}**\dir3{-}
    ?(0.58)*\dir3{>}, (16,-4)*{\scriptstyle#2},
  (8,10)*i{\circ};(0,0)*{\circ}**\dir3{-}
    ?(0.58)*\dir3{>}, (8,14)*{\scriptstyle#1}
\end{xy}}
\def\triZz#1#2#3{
\begin{xy} <0.6mm,0mm>:
  (0,0)*{\circ};(16,0)*{\circ}**\dir3{-}
    ?(0.58)*\dir3{>}, (0,-4)*{\scriptstyle#3},
  (16,0)*i{\circ};(8,10)*{\circ}**\dir3{-}
    ?(0.58)*\dir3{>}, (16,-4)*{\scriptstyle#2},
  (8,10)*i{\circ};(0,0)*{\circ}**\dir3{-}
    ?(0.58)*\dir3{>}, (8,14)*{\scriptstyle#1}, (0,7)*{\scriptstyle0}
\end{xy}}
We may apply the BKR map of section \ref{sss:McK} to determine the
relationship between quiver representations and coherent sheaves. The
McKay equivalence may of course be composed with any autoequivalence
of $\DC(X)$ and still give an equivalence. This gives a degree of
ambiguity to how we may associate the derived category of sheaves to
the derived category of quivers. We refer to \cite{CI:flopM,KPS:McK}
for further details on how to compute the correspondence exactly. Let
$\Delta_{m_0m_1m_2}$ denote a quiver representation of the form
(\ref{eq:Z3Q}) and consider the fractional branes $F_0=\Delta_{100}$,
$F_1=\Delta_{010}$ and $F_2=\Delta_{001}$. A choice of the McKay
correspondence consistent with our conventions is then given by
\begin{equation}
\begin{split}
  F_0 &= \vbox{\triZ100} = \O_E\\
  F_1 &= \triZ010 = \Omega_E(1)[1]\\
  F_2 &= \triZ001 = \O_E(-1)[2].
\end{split} \label{eq:Fs}
\end{equation}
From the above analysis of $\mathcal{G}$ it is easy to show that
\begin{equation}
\begin{split}
  T_{\mathcal{G}}(\O_E) &= \Omega_E(1)[1]\\
  T^2_{\mathcal{G}}(\O_E) &= \Omega^2_E(2)[2] = \O_E(-1)[2].
\end{split}
\end{equation}
Thus $T_{\mathcal{G}}$ indeed corresponds to rotating the McKay quiver
clockwise by $2\pi i/3$ or, equivalently, by tensoring by the
representation $F_1$. Perhaps we should emphasize that this picture
does not work if we were to assert that $F_2=\O_E(-1)$ without the
shift of 2 as is done in much of the literature.

We may easily generalize the quintic hypersurface and the orbifold
computation above to other dimensions (i.e., degree $d$
hypersurfaces in $\P^{d-1}$ and orbifolds $\C^n/\Z_n$). In each case
the quantum symmetry of the Landau--Ginzburg orbifold is broken to
become a shift left by two and the quantum symmetry of the geometrical
orbifold is preserved. Based on this rather limited set of examples it
is tempting to conjecture that this is a general result for all Gepner
models and for all orbifolds $\C^d/G$. It would be interesting to prove
this, or at least study some more examples.

\subsubsection{Examples of stability}  \label{sss:Z3}

In the case of the quintic we used the large radius limit as our base
point for determining stability. In the case of the orbifold, we can
use the quantum symmetry to use the orbifold point as the base point.
First note that from (\ref{eq:Z3ac}) (with $t$ shifted by 1) and
(\ref{eq:Fs}) we have
\begin{equation}
\begin{split}
  Z(F_0) &= \ff13(1-\varpi_0+\varpi_1)\\
  Z(F_1) &= \ff13(1-\varpi_0-2\varpi_1)\\
  Z(F_2) &= \ff13(1+2\varpi_0+\varpi_1).
\end{split}
\end{equation}
Thus, at the orbifold point, the $F_i$'s all have central charge
$\ff13$. This is not surprising as the $\Z_3$ quantum symmetry
cyclically permutes the $F_i$'s are so their physics must be identical
at the orbifold point. In particular, they must all have the same
value of $\xi$ --- a fact we proved in general at the end of section
\ref{sss:DM}. We may thus declare at the orbifold point that
\begin{equation}
  \xi(F_0)=\xi(F_1)=\xi(F_2)=0.
\end{equation}
Given (\ref{eq:perZ3}), to a linear approximation in $\psi$ near the
orbifold we therefore have
\begin{equation}
\begin{split}
  \xi(\Delta_{m_0m_1m_2}) &= -c\frac{(-m_0-m_1+2m_2)\Re(\psi)+
              (m_0-2m_1+m_2)\Re(\omega\psi)}{m_0+m_1+m_2}\\
  &= c\frac{\sum_im_i\zeta_i}{\sum_im_i},
\end{split} \label{eq:xizet}
\end{equation}
for a positive constant $c$ and we {\em define\/} the $\zeta_k$ by
\begin{equation}
  \zeta_k = \sqrt{3}\Re(e^{\frac{\pi i}6(4k-1)}\psi),
\end{equation}
so that
\begin{equation}
\begin{split}
  \zeta_0+\zeta_1+\zeta_2 &= 0\\
  \zeta_0+\omega\zeta_1+\omega^2\zeta_2 &= 
          \ff{3\sqrt3}{2}e^{\frac{\pi i}6}\bar\psi\\
  \zeta_0+\omega^2\zeta_1+\omega\zeta_2 &=  
          \ff{3\sqrt3}{2}e^{-\frac{\pi i}6}\psi.
\end{split}
\end{equation}
Clearly this is the analogue of (\ref{eq:zetamarg}) and the $\zeta$'s
we have just introduced here correspond to those of section
\ref{sss:theta}. Indeed, we may now explicitly show that
$\theta$-stability is a limiting form of $\Pi$-stability near the
orbifold point. Suppose we have a short exact sequence of quiver
representations
\begin{equation}
\xymatrix@1{
  0\ar[r]&\triZ{n_0}{n_1}{n_2}\ar[r]&\triZ{m_0}{m_1}{m_2}\ar[r]&
   \triZ{m_0-n_0}{~~m_1-n_1}{m_2-n_2~~}\ar[r]&0.
} \label{eq:th1}
\end{equation}
Near the orbifold point, the $\xi$'s of these 3 D-branes will be very
close to zero. Thus, by the way central charges add, the $\xi$ of the
middle entry in (\ref{eq:th1}) must lie between the $\xi$'s of the
other two. For $\Pi$-stability of the middle entry we draw the
distinguished triangle
\begin{equation}
\xymatrix@!C=10mm{
  &\triZ{m_0}{m_1}{m_2}\ar[dl]&\\
  \triZ{m_0-n_0}{~~m_1-n_1}{m_2-n_2~~}\ar[rr]|{[1]}^(0.6)f&&
    \triZ{n_0}{n_1}{n_2}\ar[ul]\\
} \label{eq:th1t}
\end{equation}
and look for the condition for $f$ to be tachyonic. From
(\ref{eq:xizet}) this is precisely
\begin{equation}
 \frac{\sum_in_i\zeta_i}{\sum_in_i}<\frac{\sum_im_i\zeta_i}{\sum_im_i},
\end{equation}
which, from (\ref{eq:thdef}) is equivalent to King's $\theta$-stability
statement of section \ref{sss:theta}.

This $\theta$-stability formulation allows us to completely classify
the stable irreducible B-branes near the orbifold point. As mentioned
in section \ref{sss:tachy}, the irreducibility for an object $A$
amounts to $\Hom(A,A)=\C$. A quiver representation satisfying this
condition is known as a ``Schur representation''. The problem of
finding such representations was discussed in \cite{DFR:orbifold}.

Determining whether a quiver representation (with relations) is Schur
is a purely algebra question but turns out to be fairly awkward. In
many cases it is actually more convenient to use the BKR equivalence
and rephrase the question in terms of coherent sheaves.

As an example of a non-Schur quiver representation, consider
$\Delta_{211}$ with {\em generic\/} maps on the arrows in the quiver. With
some effort one can show that the short exact sequence
\begin{equation}
\xymatrix@1{
  0\ar[r]&\triZ100\ar[r]&\triZ211\ar[r]&
   \triZ111\ar[r]&0,
}
\end{equation}
is {\em split}. This immediately implies that
$\Hom(\Delta_{211},\Delta_{211})\supset\C^2$. This fact
becomes more obvious when written in terms of sheaves. $\Delta_{111}$
is a 0-brane which is generically nowhere near the exceptional divisor
$E$, whereas $\Delta_{100}$ is the 4-brane $\O_E$ wrapping
$E$. Thus $\Delta_{211}$ is a sum of two quite disjoint D-branes and
is obviously reducible. Note that when the maps on the arrows are not
generic this quiver presentation might actually be Schur. This would
correspond to the 0-brane being on $E$ leading to a possible
irreducible bound state with the 4-brane.

We will not attempt to explicitly provide a complete solution to the
classification problem here but some Schur representations of interest
are $\Delta_{111}$, and $\Delta_{abc}$, where $\{a,b,c\}$ is any
permutation of $\{0,1,n\}$ for $n\leq3$.

The fractional branes $F_k$ are obviously always stable near the
orbifold point since they have no nontrivial subobject in the category
of quiver representations. Let us next focus on $\Delta_{111}$, some of
which correspond to 0-branes. The quiver
\begin{equation}
\begin{xy} <0.6mm,0mm>:
  (0,0)*{\circ};(16,0)*{\circ}**\dir3{-}
    ?(0.58)*\dir3{>}, (0,-4)*{\scriptstyle1}, (8,-4)*{\scriptstyle \neq0},
  (16,0)*i{\circ};(8,10)*{\circ}**\dir3{-}
    ?(0.58)*\dir3{>}, (16,-4)*{\scriptstyle1}, (17,7)*{\scriptstyle \neq0},
  (8,10)*i{\circ};(0,0)*{\circ}**\dir3{-}
    ?(0.58)*\dir3{>}, (8,14)*{\scriptstyle1}, (-1,7)*{\scriptstyle \neq0}
\end{xy}
\end{equation}
with at least one nonzero map between each pair of nodes is stable
since there is no injective map from any possible subobject to
it. According to the explicit computations in \cite{SI:I,SI:II} such a
quiver represents a 0-brane away from the exceptional divisor $E$ (or
orbifold point if we haven't blown up). Indeed, one would expect that
the stability of such a 0-brane should not be affected by
orbifold-related matters.

Now consider the following short exact sequence:
\begin{equation}
\xymatrix@1{
0\ar[r]&\triZ100\ar[r]&\triZz111\ar[r]&\triZ011\ar[r]&0.
}
\end{equation}
This $\Delta_{111}$ is stable against decay to $\Delta_{100}$ by
$\theta$-stability if $\zeta_0<0$. We also have the sequence
\begin{equation}
\xymatrix@1{
0\ar[r]&\triZ110\ar[r]&\triZz111\ar[r]&\triZ001\ar[r]&0,
}
\end{equation}
giving a further constraint $\zeta_2>0$ on the stability of this
0-brane. Thus this 0-brane is stable in $2\pi/3$ wedge coming out of
the orbifold point. After blowing-up a little into this wedge, this
0-brane corresponds to a point on the exceptional divisor.
Obviously a cyclic permutation of the zero to another edge of the
quiver results in similar statements with the $\zeta$'s permuted
accordingly. Thus, all other quivers $\Delta_{111}$ are unstable in the
wedge $\zeta_2>0,\zeta_0<0$ and do not correspond to 0-branes at all.

The quiver representations with zero maps on the left edge have a
close connection to sheaves on $E$ as can be seen as follows. The
general representation $\Delta_{abc}$ falls into the sequence
\begin{equation}
\xymatrix@1{
0\ar[r]&\triZ100\ar[r]&\triZz abc\ar[r]&\triZz{a-1}bc\ar[r]&0,
}
\end{equation}
and thus we iterate
\begin{equation}
\begin{split}
\xymatrix@1{\triZz abc} &= \Cone\left(\xymatrix@1{
  \triZz{a-1}bc[-1]\ar[r]&\O_E}\right)\\
&= \Cone\left(\xymatrix@1{
  \triZz{a-2}bc[-1]\ar[r]&\O_E^{\oplus 2}}\right)\\
&= \Cone\left(\xymatrix@1{
  \triZ0bc[-1]\ar[r]&\O_E^{\oplus a}}\right).
\end{split}
\end{equation}
Continuing this process yields
\begin{equation}
\xymatrix@1{\triZz abc} = \Bigl(\xymatrix@1{\O_E(-1)^{\oplus c}
\ar[r]&\Omega_E(1)^{\oplus b}\ar[r]&\poso{\O_E^{\oplus a}}}\Bigr),
  \label{eq:Beilt}
\end{equation}
explicitly mapping this quiver representation into the derived
category of coherent sheaves on $X$ (or $E$). This is precisely
Beilinson's construction of sheaves on $\P^2$ \cite{Bei:res} and this
correspondence was identified in \cite{DFR:orbifold}. Thus, quiver
representations with zero maps on the left edge are seen to be
associated to D-branes on $E$. We will get a
pure sheaf of course only if the cohomology of the complex in
(\ref{eq:Beilt}) is concentrated at one term. 

\iffigs
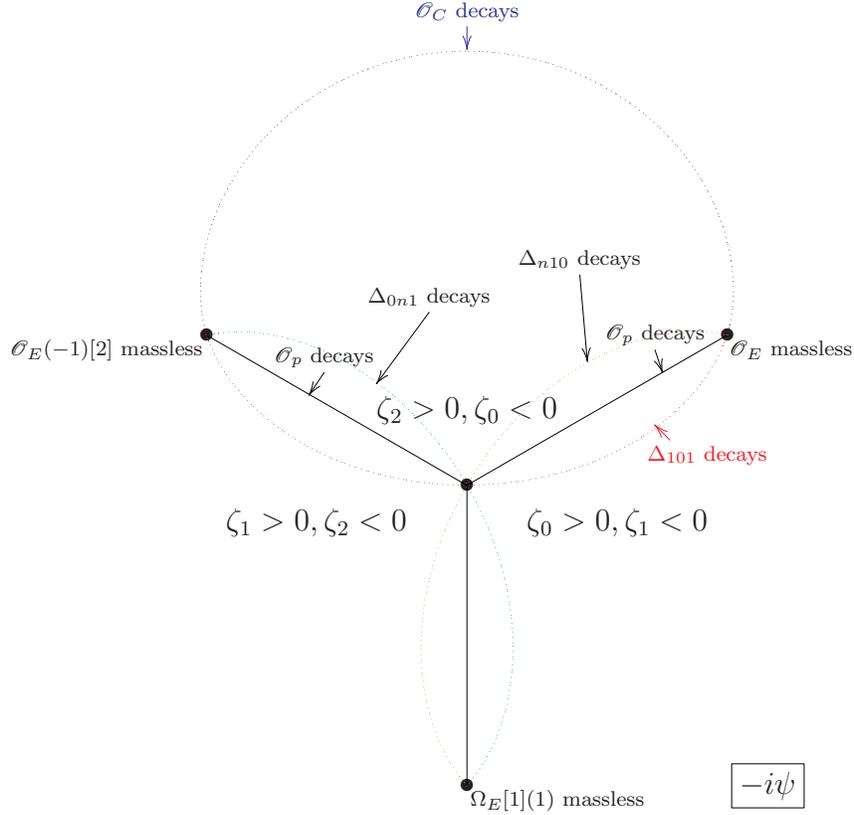
\begin{figure}[t]
$$
\begin{xy} <1mm,0mm>:
  (40,-40)*{-i\psi}*+\frm{-},
  (0,0)*{\bullet},
  (0,-40)*{\bullet}, (12,-42)*{\hbox{\scriptsize $\Omega_E[1](1)$ massless}},
  (34.64,20)*{\bullet}, (43,18)*{\hbox{\scriptsize $\O_E$ massless}},
  (-34.64,20)*{\bullet}, (-48,18)*{\hbox{\scriptsize $\O_E(-1)[2]$ massless}},
  (0,0);(0,-40)**\dir{-},
  (0,0);(34.64,20)**\dir{-},
  (0,0);(-34.64,20)**\dir{-},
  (0,10)*{\zeta_2>0,\zeta_0<0},
  (20,-5)*{\zeta_0>0,\zeta_1<0},
  (-20,-5)*{\zeta_1>0,\zeta_2<0},
  (0,63)*[Blue]{\hbox{\scriptsize $\O_C$ decays}},
  (0,61);(0,58)**[Blue]\dir{-}?>*[Blue]\dir{>},
  (25,20)*{\hbox{\scriptsize $\O_p$ decays}},
  (25,18);(26,15)**\dir{-}?>*\dir{>},
  (-19,17)*{\hbox{\scriptsize $\O_p$ decays}},
  (-19,15);(-21,12)**\dir{-}?>*\dir{>},
  (32,4)*[Red]{\hbox{\scriptsize $\Delta_{101}$ decays}},
  (27,6);(25,8)**[Red]\dir{-}?>*[Red]\dir{>},
  (15,30)*{\hbox{\scriptsize $\Delta_{n10}$ decays}},
  (15,28);(16,16.5)**\dir{-}?>*\dir{>},
  (-5,25)*{\hbox{\scriptsize $\Delta_{0n1}$ decays}},
  (-5,23);(-12,13.5)**\dir{-}?>*\dir{>},
  (34.64,20);@={
   (34.9257,21.0667),(35.2213,22.6667),(35.4243,24.8),(35.4375,26.601),
   (35.272,28.8),(34.9312,30.9969),(34.2838,33.6),(33.728,35.2604),
   (33.0574,36.9438),(32.3184,38.5229),(31.564,39.9438),(30.5603,41.6),
   (29.421,43.2356),(28.2838,44.6896),(27.2349,45.9072),(26.1243,47.0728),
   (24.6077,48.5064),(22.2993,50.3876),(20.9426,51.3564),(18.6375,52.802),
   (16.8375,53.7688),(14.6824,54.7708),(12.9684,55.448),(11.2941,56.0124),
   (9.71248,56.4792),(7.90588,56.9124),(5.82024,57.2908),(4.23528,57.5052),
   (2.57868,57.6532),(0.70478,57.7364),(-1.12941,57.7292),(-2.82353,57.6376),
   (-5.08236,57.404),(-6.77648,57.1324),(-8.4706,56.7824),(-10.2331,56.3312),
   (-11.9934,55.7908),(-14.3669,54.902),(-15.8526,54.2488),(-17.389,53.4896),
   (-19.1757,52.4876),(-20.539,51.6304),(-22.072,50.5564),(-23.4662,49.4812),
   (-24.878,48.2668),(-26.2632,46.9356),(-27.5228,45.5876),(-28.6334,44.2668),
   (-29.8831,42.6012),(-31.0544,40.8),(-31.918,39.2927),(-32.6426,37.8667),
   (-33.3397,36.2667),(-34.028,34.4),(-34.5948,32.4698),(-34.9919,30.676),
   (-35.2456,29.0667),(-35.3956,27.4667),(-35.4419,25.5302),(-35.3228,23.4531),
   (-35.0934,21.8667)},
   (-34.64,20)*[Blue]\qspline{.},@i,
  (34.64,20);@={
   (34.1537,18.6667),(33.4081,17.0667),(32.5875,15.6229),(31.6114,14.1562),
   (30.3904,12.5823),(29.3007,11.3636),(28.1526,10.2114),(26.8511,9.04064),
   (25.318,7.82188),(23.9388,6.84012),(22.4377,5.88436),(20.6994,4.90104),
   (19.1857,4.14012),(17.4744,3.3776),(15.9039,2.75938),(14.1127,2.14271),
   (12.4095,1.63984),(10.7075,1.20912),(9.03528,0.853384),(7.34116,0.559896),
   (5.7484,0.341927),(4.00656,0.165364),(1.97647,0.0402996),(0,0),
   (-2.25882,0.052474),(-4.1396,0.176302),
   (-5.76768,0.344271),(-7.51076,0.58698),
   (-9.31764,0.909116),(-10.8854,1.24974),(-12.581,1.68724),(-14.1176,2.14505),
   (-15.7666,2.70938),(-17.4604,3.37188),(-19.0641,4.08228),(-20.834,4.97032),
   (-22.3814,5.84792),(-23.8627,6.78908),(-25.2022,7.73332),(-26.6581,8.87812),
   (-28.0864,10.1479),(-29.2522,11.3062),(-30.5669,12.8),(-31.6743,14.2427),
   (-32.7948,15.9604),(-33.6772,17.6)},
   (-34.64,20)*[Red]\qspline{.},@i,
  (-34.64,20);@={
   (-33.2427,20.2446),(-31.4842,20.3989),(-29.8236,20.4102),(-28.0654,20.2982),
   (-26.0918,20.0278),(-24.4915,19.6934),(-22.9196,19.2751),(-21.255,18.7334),
   (-19.4329,18.0151),(-17.8931,17.3115),(-16.3148,16.4894),(-14.5941,15.4757),
   (-13.1783,14.5452),(-11.6623,13.4445),(-10.3416,12.3935),(-8.91198,11.1506),
   (-7.62489,9.927),(-6.40087,8.66839),(-5.25669,7.39809),(-4.15546,6.07768),
   (-3.17032,4.8073),(-2.14649,3.3871),(-1.02314,1.69153),(0,0),
   (1.08397,-1.98244),(1.91712,-3.67315),(2.58569,-5.16709),(3.24704,-6.798),
   (3.8715,-8.52387),(4.36037,-10.0519),(4.82929,-11.7391),(5.20115,-13.2988),
   (5.53689,-15.0089),(5.81007,-16.8071),(5.99668,-18.5511),(6.11258,-20.5279),
   (6.12627,-22.3069),(6.05182,-24.0602),(5.90385,-25.6924),(5.64036,-27.5256),
   (5.25484,-29.3975),(4.83461,-30.9863),(4.19833,-32.8717),(3.50258,-34.5521),
   (2.57531,-36.3814),(1.59655,-37.9653)},
   (0,-40)*[Green]\qspline{.},@i,
  (0,-40);@={
   (-0.911021,-38.9113),(-1.92386,-37.4656),
   (-2.76391,-36.0331),(-3.54604,-34.4544),
   (-4.29865,-32.61),(-4.80923,-31.0569),
   (-5.23291,-29.4866),(-5.59614,-27.7741),
   (-5.88505,-25.837),(-6.04568,-24.1517),
   (-6.12283,-22.3738),(-6.10529,-20.3768),
   (-6.00739,-18.6853),(-5.81213,-16.8221),
   (-5.56225,-15.1529),(-5.2007,-13.2933),
   (-4.78459,-11.5668),(-4.30661,-9.87751),
   (-3.77859,-8.25147),(-3.1857,-6.63758),
   (-2.57808,-5.14922),(-1.86007,-3.55246),(-0.953336,-1.73182),(0,0),
   (1.17486,1.92996),(2.22248,3.49685),(3.18199,4.82282),(4.26372,6.21102),
   (5.44614,7.61475),(6.52499,8.80213),(7.75167,10.0518),(8.91649,11.1537),
   (10.2297,12.2996),(11.6503,13.4352),(13.0674,14.4688),(14.7214,15.5576),
   (16.2552,16.4589),(17.8109,17.2711),(19.2984,17.9591),(21.0177,18.6475),
   (22.8316,19.2496),(24.4176,19.68),(26.3686,20.0717),(28.1717,20.3094),
   (30.2195,20.421),(32.0806,20.3653)},
   (34.64,20)*[Orange]\qspline{.},@i,
\end{xy}
$$
  \vspace{-5mm}
  \caption{Some lines of marginal stability for the $\Z_3$-orbifold.}
  \label{f:Z3m}
\end{figure}
\fi

We denote some lines of marginal stability for $\Pi$-stability in
figure \ref{f:Z3m}. In each case, the arrow denotes the direction you
cross the line to cause the relevant object to decay.  Naturally this
agrees with $\theta$-stability near the origin.  The figure shows the
moduli space in the form of the complex $(-i\psi)$-plane. We make this
choice so that the picture is aligned with figure \ref{f:Z3scp}, i.e.,
the large radius limit is upwards. Note that the lines of marginal
stability corresponding to $\O_p$, $p\in E$, (i.e., the corresponding
$\Delta_{111}$ quivers above) follow the lines of constant
$\arg(\psi)$ even when the non-perturbative effects of $\Pi$-stability
are taken into account.

Some decays of note are the following:
\begin{enumerate}
\item $\O_C$: This sheaf fits into the exact sequence
\begin{equation}
\xymatrix@1{
0\ar[r]&\O_E(-1)\ar[r]&\O_E\ar[r]&\O_C\ar[r]&0,
} \label{eq:OC1}
\end{equation}
and thus decays by $\Pi$-stability in a way essentially identical 
to the 4-branes in the quintic as in section \ref{sss:q4}. Thus, these
2-branes are stable at large radius but decay before the orbifold
point is reached. Note that (\ref{eq:OC1}) implies that, in the
derived category of quiver representations we have
\begin{equation}
  \O_C = \Cone(F_2[-2]\to F_0).  \label{eq:OC1a}
\end{equation}
That is, this D-brane is essentially a complex of quivers and cannot
be written in terms of a single quiver. In other words, it is not in
the abelian category of quiver representations. It is therefore
consistent with our picture that it decays before we get close to the
orbifold point.
\item $\Delta_{101}$: Following the logic of section \ref{sss:qex} we
can now look for an ``exotic'' D-brane by taking the ``Serre dual''
of (\ref{eq:OC1a}). This gives $\Cone(F_0[-1],F_2)$, i.e., an extension
of $F_0$ by $F_2$. This is precisely $\Delta_{101}$. As expected from
section \ref{sss:qex}, these objects should not be stable at large radius
but can become stable as we shrink the exceptional divisor down. The
line of marginal stability is shown in figure \ref{f:Z3m}. Note that
they do not actually become stable until we shrink down to, or beyond,
the orbifold point. These objects generically have nonzero maps along
the left edge of the triangle and so are not classified by Beilinson's
construction (\ref{eq:Beilt}).

We see a nice complementarity between the D-branes $\Delta_{101}$ and
$\O_C$. $\O_C$ is an object in the category of coherent sheaves but is
a complex in terms of quivers. $\Delta_{101}$ is an object in the
category of quiver representations but becomes an exotic complex
$\Cone(\O_E[-1],\O_E(-1)[2])$ in the derived category of sheaves.

\item $\Delta_{n10}$: This fits into the sequence
\begin{equation}
\xymatrix@1{
  0\ar[r]&\triZ100\ar[r]&\triZ n10\ar[r]&
   \triZ{n-1}10\ar[r]&0.
}
\end{equation}
The produces a decay as shown in figure \ref{f:Z3m}. The identification
(\ref{eq:Beilt}) together with the short exact sequence 
\begin{equation}
\xymatrix@1{
  0\ar[r]&\Omega_E(1)\ar[r]&\O_E^{\oplus3}\ar[r]&\O_E(1)\ar[r]&0,
}
\end{equation}
may be used to show that 
\begin{equation}
\begin{split}
  \Delta_{210} &\cong \O_C(1)\\
  \Delta_{310} &\cong \O_E(1).
\end{split}
\end{equation}
These D-branes are therefore simply objects both in terms of sheaves
and quivers. It is not surprising therefore that they are both stable
at large radius limit and near the orbifold point.

On the other hand $\Delta_{110}$ does not correspond to a simple sheaf
since the complex in (\ref{eq:Beilt}) has cohomology in more than one
place. In this case we have a sequence
\begin{equation}
\xymatrix@1{
  0\ar[r]&\triZ110\ar[r]&\triZ111\ar[r]&
   \triZ001\ar[r]&0,
}
\end{equation}
which makes $\Delta_{110}$ only marginally stable at the large radius limit.

\item $\Delta_{0n1}$: This is similar to the $\Delta_{n10}$ case and
again we plot the line of marginal stability in figure
\ref{f:Z3m}. This time $\Delta_{011}$ corresponds to the ideal sheaf
of a point $\cI_{E,p}[1]$ and is again only marginally stable at the
large radius limit.

\end{enumerate}

Clearly the study of D-brane stability on orbifolds is a very
interesting subject and we have only just begun to scratch the
surface. The results above, together with previous analysis in the
literature such as \cite{DGM:Dorb,DDG:wrap,DG:fracM,DFR:orbifold}
provide a good start to the analysis of the subject but much remains
to be done.

%%%%%%%%%%%%%%%%%%%%%%%%%%%%%%%%%%%%%%%%%%%%%%%%%%%%%%%%%%%%%%%%%%

\section{Conclusion}  \label{s:conc}

We hope that the reader is convinced that the derived category program
is essential for understanding D-branes on a \CY\ threefold and thus,
presumably, D-branes in any nontrivial spacetime. 

The essential ingredient is the extension of the na\"\i ve concept of
``branes'' and ``anti-branes'' to grading branes by arbitrary
integers. This inexorably leads one to discuss complexes and soon
the whole machinery of the derived category becomes unavoidable. While
the mathematics involved in this story might look excessive at first
sight, it is hard to imagine how one would understand B-branes without
using this language or essentially reinventing something identical.

Given this complexity of B-branes it is perhaps worrying to note that
we still made some drastic simplifications in these lectures. The most
egregious is probably the assumption that the string coupling is zero
and thus the mass of any D-brane is infinite (unless it's zero!). The
next step one might therefore wish for is the notion of ``quantizing''
the derived category story which presumably introduces many many more
complications. Clearly our current knowledge of D-brane physics
remains relatively poor.

%%%%%%%%%%%%%%%%%%%%%%%%%%%%%%%%%%%%%%%%%%%%%%%%%%%%%%%%%%%%%%%%%%

\section*{Acknowledgments}

It is a pleasure to thank M.~Douglas who provided many of the basic
ideas involved in the derived category approach to D-branes and
$\Pi$-stability, and with whom I collaborated on this subject. I also
thank my collaborators A.~Lawrence, P.~Horja and R.~Karp. In addition
I have had many useful conversations with T.~Bridgeland, S.~Katz,
I.~Melnikov, D.~Morrison, R.~Plesser, R.~Thomas, and I.~Zharkov. I
would like to thank K.~T.~Mahanthappa and J.~Maldecena for organizing
TASI-03 for which these lecture notes were primarily written. Much of
the motivation for working on a review of this subject was also
provided by the ``If I Only Had a Brane'' seminar series at Duke
organized by Owen Patashnick.

The author is supported in part by NSF grant DMS-0301476.

%\bibliographystyle{my-phys}
%\bibliography{string}

\end{document}